\documentclass[10pt,a4paper]{article}

\usepackage{graphicx}

\begin{document}

\title{Renormalization of chiral nuclear forces with multiple subtractions in peripheral channels}

\author{E. F. Batista$^1$, S. Szpigel$^2$ and V. S. Tim\'oteo$^3$ \\
{\small $^1$\em Departamento de Ciências Exatas e Naturais, Universidade Estadual do Sudoeste da Bahia} \\ 
{\small $^2$ \em Centro de R\'adio-Astronomia e Astrof\'\i sica, Universidade Presbiteriana  Mackenzie} \\ 
{\small $^3$ \em Faculdade de Tecnologia - FT, Universidade Estadual de Campinas - UNICAMP}}

\maketitle

\begin{abstract}
We analyse the renormalization of the of two-nucleon interaction with multiple subtractions 
in peripheral waves considering two chiral forces at N3LO. Phase shifts at low energies are 
then computed with several subtraction points below $\mu = 10~{\rm fm}^{-1}$. We show that 
for most peripheral waves the phase shifts have nearly no dependence on the renormalization 
scale. In two cases the phase shifts converge slowly as the renormalization scale approaches 
$\mu = 1 ~ {\rm fm}^{-1}$ and in one case the phase shifts presented oscillations with respect
to the subtraction point $\mu$.
\end{abstract}

\section{Introduction}

It is widely established in nuclear physics, that Quantum Chromodynamics (QCD) is in fact 
the underlying theory which describes the properties of strong nuclear forces. In this theory, 
the fundamental degrees of freedom are quarks which interact with each other via exchange 
of colored gluons. On the other hand, the strong nuclear force is also responsible for the 
binding of protons and neutrons in atomic nuclei. But according to QCD, nucleons are bound 
states of quarks and the nuclear force is  considered as the residual part of the quark-quark 
interaction inside of the nucleon with exchange of gluons. Due to the property of asymptotic 
freedom, the running coupling constant is small enough at high energies to allow QCD be 
handled within a perturbative approach. However, at low energies, where mesons and 
nucleons are the relevant degrees of freedom, the running coupling constant becomes large 
and QCD is no longer perturbative. This strong non-perturbative nature of QCD at low-energies 
implies in several mathematical and computational difficulties in describing nucleon properties 
at this energy level.

Since QCD cannot be treated perturbatively at low energies, a new approach was developed 
to handle nuclear forces with degrees of freedom appropriate for low-energy systems. The idea
was to use quantum field theory but replacing quarks and gluons degrees of freedom by pions 
and nucleons, keeping the fundamental properties of the underlying theory like chiral symmetry. 
This Effective Field Theory (EFT) scheme was already used in other systems to describe different
types of interactions and Weinberg proposed an EFT approach to nuclear systems based on QCD.
This idea generated a new branch in nuclear physics and allowed a deeper understanding of the
nuclear force and few-nucleon systems. In particular, the two-nucleon system requires a non-perturbative
extension of Chiral Perturbation Theory (ChPT) which works well in the case of pion-nucleon scattering.

Basically, an EFT is constructed by isolating the most relevant degrees of freedom and symmetries
for the system under consideration and applying standard quantum field theory. In nuclear physics, 
replacing quarks and gluons by pions and nucleons means moving to a different (lower) energy scale
and a connection between the symmetry properties of the underlying fundamental theory and the 
symmetries of their effective versions must be well-established. Thus, in a nuclear effective theories it is 
necessary to establish scale parameters which enable us to separate the high-energy components of the 
interaction from the low-energy part. In a series of papers \cite{wein}, Weinberg proposed an effective field 
theory scheme for nuclear forces based on the chiral symmetry of QCD.

This approach was first applied by Ord\'o\~nez, Ray and van Kolck \cite{bira} and allowed the perturbative
treatment of the $NN$ interaction.  An expansion in powers of $(Q/\Lambda_\chi)^\nu$ is performed and  
$Q$ is a generic low momentum scale and $\Lambda$ is the chiral symmetry breaking scale which is 
approximately $1 GeV$. This expansion is  controlled by a power counting scheme, called Weinberg Power 
Counting (WPC), which provides an hierarchical organization for the processes in few-nucleon systems.
Following the WPC, the $NN$ interaction at leading order (LO) consists of one-pion-exchange (OPE) plus 
a contact term. At next-to-leading order (NLO), two-pion-exchange (TPE) and $\mathcal{O}(p^2)$ contact 
interactions are added, at next-to-next-to-leading order (N2LO) there is an additional set of TPE diagrams 
and, finally, at next-to-next-to-next-to-leading order (N3LO) corrections to both OPE and TPE are included 
along with $\mathcal{O}(p^4)$ contact interactions.  

Regardless of how the chiral expansion is organized, the issue of how to renormalize the two-body interaction 
is of fundamental importance and has been subject of investigation for decades. Early works by Adhikari et al 
started by discussing the renormalization of two-body quantum hamiltonians \cite{adh}. Later, the problem was
focused in the $NN$ interaction using renormalization group analysis and a new power counting was proposed 
by Kaplan, Savage and Wise \cite{ksw}. Discussions on the renormalization of singular and one-pion-exchange 
two-body interactions, power counting and renormalization of the three-body system are detailed by van Kolck et al 
\cite{vK,NTvK}. Another renormalization group approach to two-body and nucleon-nucleon scattering was presented 
by Birse et al \cite{MCbirse} and a complete analysis of cutoff renormalization in configuration space was performed 
by the Granada group \cite{GRen}. Also, a comparison between renormalization in configuration and momentum spaces 
has been carried out in Ref. \cite{RxP}.

Another renormalization approach for the $NN$ system consists of a hybrid scheme, where the LO contribution is treated 
non-perturbatively and the higher orders are handled perturbatively \cite{mnl11}. Results for $P$-waves and
$D$-waves show that perturbative two-pion-exchange reproduces the experimental data up to 
$k_{\rm cm} \sim 300~{\rm MeV}$. Here we treat all terms non-perturbatively since in our renormalization scheme 
the pieces of the interaction are inserted as the subtractions are performed.    

Apart from the divergences due to pion loops in the irreducible diagrams, the reducible diagrams also generate 
divergences in the scattering equation. To overcome this problem the most common employed method is 
introducing a cutoff regularization scale $\Lambda$ which limits the momentum integration, in the scattering equation, 
above this scale parameter resulting in a finite phase-shifts. The cutoff scheme handles the divergences 
by modifying the potential and keeping the scattering equation intact. The regularized interaction contains 
only low-momentum components and the cutoff scale is fixed at some scale, typically $\sim 2 - 3 ~{\rm fm}^{-1}$. 
A slight discomfort with this method comes from the fact that all physics above a certain momentum scale are excluded. 
Recently, a N3LO interaction has been optimized by an improved renormalization approach in configuration space 
which maintains the analytic structure of the scattering amplitude \cite{ekm15}.

An alternative renormalization procedure referred as subtractive method or multiple subtractions, treat the divergences 
with a different perspective: instead of modifying the potential, as in cutoff method, and keeping the scattering equation 
untouched, here the interaction is kept intact and the scattering equation is modified by the introduction of subtractions 
in its kernel. The dependence of the phase shifts on the cutoff is replaced by a dependence on the subtraction point that 
can be later eliminated by using the renormalization group flow. With this procedure no components are neglected from 
the interaction and both low and high momentum components are included. Detailed descriptions of this approach can be 
found in Refs.\cite{NPA99,PLB00,PLB05,npa07,ijmp07,PRC11,JPG12}.  

In this work we employ the subtract kernel method to renormalize the Lippmann-Schwinger equation in N3LO 
and perform a detailed renormalization scale dependence analysis of the phase shifts in peripheral channels. 
The high angular momentum waves are interesting because the force in their channels contains no contact 
interactions and consists only of pion exchanges. The N3LO potential have contact contributions up to 
$D$-waves so that $F$ and higher waves have contributions purely from pion exchanges and no 
core given by contact interactions.

This paper is organized as follows: Section 2 describes the renormalization of N3LO interactions with five subtractions, 
Section 3 presents the numerical results for both uncoupled and coupled channels up to $J=6$ and our main conclusions 
are given in Section 4.

\section{Renormalization of N3LO interactions}

At any given order, the modern two-nucleon interactions can always be separated in two components:
the pion exchange interactions and the contact terms, which parametrize the short range core of the interaction 
and are determined by fitting scattering data. The $NN$ potential is then written as
\begin{equation}
V_{\rm NN} = V_{\rm pions} + V_{\rm cont} \; ,
\end{equation}
where the first term contains one-pion-exchange and two-pion exchanges 
\begin{equation}
V_{\rm pions} = V_{\pi} + V_{2\pi}   \; . 
\end{equation} 

The power counting scheme organizes which set of Feynman diagrams must be included 
in each order in the chiral expansion:
\begin{eqnarray}
V_{\rm pions} & = & V_{\rm pions}^{(0)} + V_{\rm pions}^{(2)} + V_{\rm pions}^{(3)} + V_{\rm pions}^{(4)} \cdots \; , \nonumber \\
V_{\rm cont}   & = & V_{\rm cont}^{(0)} + V_{\rm cont}^{(2)} +V_{\rm cont}^{(4)} + \cdots \; ,
\label{pi+ct}
\end{eqnarray}
where the superscript numbers in parentheses indicates the order in the chiral expansion. 
Note that in Eq. (\ref{pi+ct}) for the contact contribution, all the odd powers cancel due to symmetry requirements.
Hence, in the Weinberg power counting scheme, there is no contact interaction $V_{\rm cont}^{(3)}$ at the third order in the
chiral expansion (N2LO). This actually breaks the order-by-order improvement of the chiral expansion when going from
NLO to N2LO as shown very clearly in Ref. \cite{JPG12}. Nogga, Timmermans and van Kolck \cite{NTvK}, looked at the
interaction at leading order (LO) and found that additional counter terms that are not predicted by the Weinberg power counting
are required in order to improve the description. Later, Valderrama showed that there are also problems in higher orders and 
they could be treated perturbatively \cite{mnl11}. 

For finite cutoff, however, Epelbaum et al. \cite{egm00}) showed that a better description of the phase-shifts is obtained when 
N2LO instead of NLO interactions are used (with the same contact terms). The same conclusion can be found by studying the 
$\chi^2 / {\rm datum}$ in E. Marji et al. \cite{marji13}. 

Now let us turn to the Renormalization of N3LO interactions with five subtractions. This approach introduces a renormalization 
scale $\mu$ (subtraction point), which denotes the momentum at which the subtractions are performed. For a given energy 
$E$ the Lippmann-Schwinger (LS) equation for the $T$-matrix in operator form is written as 
\begin{eqnarray}
T(E) = V + V~G_{0}^{+}(E)~T(E)\; ,
\label{LSE}
\end{eqnarray}
where $V$ is the $NN$ potential at a given order in the chiral expansion and $G_{0}^{+}$ is the free Green's function which, 
in terms of  the free Hamiltonian $H_0$, is given by
\begin{eqnarray}
G_{0}^{+}(E)= \frac{1}{E - H_0 + i \epsilon} \; .
\end{eqnarray}

When bare potentials are introduced in the equation above, an ultraviolet divergence arises due to the 
implicit integral in the second term of the right-hand side of Eq. (\ref{LSE}), which diverges when the 
momentum goes to infinity. In the standard cutoff procedure, the $NN$ potential $V$ is multiplied by a 
regularising function,
\begin{equation}
V(p, p') \rightarrow  V_{\Lambda}(p, p') \equiv \exp[-(p/\Lambda)^{2r}]~ V(p, p')~ \exp[-(p'/\Lambda)^{2r}] \; ,
\label{SCR}
\end{equation}
where $\Lambda$ is the cutoff scale and $r \geq 1$. This function suppresses contributions from larger momenta, 
eliminating the ultraviolet divergences in the momentum integral. Non-relativistic nucleon-nucleon potentials 
based on chiral effective field theory with cutoff regularization provide a very accurate description of $NN$ scattering 
data below pion production threshold $E_{\rm lab} \sim 350~{\rm MeV}$. 

The renormalization with multiple subtractions handles this problem in a different way since the $NN$ potential is not 
modified in favour of changing the Green's function instead. The N3LO interactions require five subtractions to be 
renormalised with no cutoff and in this case the subtracted scattering equation is given by 
\begin{equation}
T^{(5)}_{\mu}(E) = V^{(5)}_{\mu}(E) + V^{(5)}_{\mu}(E)~G_{5}^{+}(E;-\mu^2)~T^{(5)}_{\mu}(E) \; ,
\label{LS5}
\end{equation}
where $\mu$ is the subtraction scale, $V^{(n)}_{\mu}(E)$ is  the driving term 
\begin{eqnarray}
V^{(n)}_{\mu}(E) = V^{(n-1)}_{\mu}(E) + V^{(n-1)}_{\mu}(E) \frac{(-\mu^2-E)^{n-1}}{(E-q^2)^{n}} V^{(n)}_{\mu}(E) \; ,
\end{eqnarray}
which has to be calculated recursively, $G_{5}^{+}(E;-\mu^2)$ is the $5$-times subtracted Green's function and 
$q$ is the the relative intermediate two nucleon momentum.
\begin{eqnarray}
G_{5}^{+}(E) &=& {\cal F}_{5}(E;-\mu^2)~G_{0}^{+}(E)  \; ,
\label{G5}
\end{eqnarray}
where
\begin{equation}
{\cal F}_{5}(E;-\mu^2) = \left(\frac{\mu^2+E}{\mu^2+H_0} \right)^{5} 
\end{equation}
is a term that arises from the recursive nature of the Renormalization process 
and works as a form factor, being responsible for providing a regular $T$-matrix. 
Detailed expressions for the integral equations in the recursive calculation with 
partial-wave basis are given in Ref. \cite{JPG12} for the case of N2LO interactions 
with four subtractions.

The LS equation with five subtractions Eq. (\ref{LS5}) has the same operator structure as the original equation 
Eq. (\ref{LSE}), with the effective $NN$ potential $V$ replaced by the driving term $V^{(5)}_{\mu}(E)$ and the 
free Green's function $G_{0}^{+}(E)$ replaced by the Green's function with five subtractions $G_{5}^{+}(E;-\mu^2)$. 
The recursive driving term encodes the physical information apparently lost due to the removal of the propagation 
through intermediate states at the subtraction point $\mu$. Then, once the driving term is determined for a particular 
subtraction point, the subtracted Lippmann-Schwinger equation provides a renormalised solution for the $T$-matrix 
at any given energy $E$. 

The driving terms $V^{(n)}_\mu$ are built recursively with the components of the nucleon-nucleon interaction.
Here we use non-regulated N3LO interactions from Entem and Machleidt (EM) \cite{mach} and from Epelbaum, 
Gl\"ockle and Meissner (EGM) \cite{epel}. The main difference between the two chiral forces is the Two-Pion Exchange 
part. The EGM potential uses Spectral Function Regularization (SFR) for the pion loop integrals resulting in a softer TPE 
component. Differences in the pionic part will then be compensated by changes in the Low-Energy Constants so that in 
the end the two forces give similar descriptions for the $NN$ system. Note that in the case of the EGM potential the 
SFR is still present for the loop integrals, but there is no smooth regulator function to suppress large momentum
contributions. The smooth cutoff is also removed from the EM potential so that the interactions we are using are 
the original EM and EGM interactions with their cutoffs removed. Once the contact interactions are determined
for both potentials, even with different off-shell behavior, the resulting on-shell scattering amplitudes are similar.
Thus, we expect comparable results for cutoff regularization and subtractive renormalization at low energies.
At high energies, $E_{\rm lab} > 200~{\rm MeV}$ the cutoff scheme is more efficient than the renormalization 
with multiple subtractions as far as describing the phase-shifts is concerned.  

\section{Numerical Results}

Here we work in a partial-wave relative momentum space basis and compute the phase 
shifts in each peripheral channel. However, a three-dimensional approach without any 
partial wave decomposition have also been employed \cite{hadi}. 

For numerical reasons, when implementing the Renormalization procedure, 
we solve the subtracted LS equation for the $K$-matrix using the principal value prescription.
We then compute the neutron-proton phase-shifts for channels with angular momentum in the 
range $3 \leq J \leq 6$ with the N3LO potentials EM and EGM. Expressions for the phases-shits
as functions of the on-shell $K$-matrix in coupled and uncoupled channels are given in Refs. \cite{batista}.

For each wave, we consider several renormalization scales up to $\mu = 10~{\rm fm}^{-1}$, 
limiting the momentum integrations at $\Lambda = 30 ~{\rm fm}^{-1}$. In practice this means 
we have an infinite cutoff and the renormalization is completely imposed by the five subtractions, 
unlike in Refs. \cite{phill} where the cutoff still plays a role since only one subtraction was performed 
for the N2LO potential which requires four subtractions to allow an infinite cutoff. 

In the case of the EGM potential, we have used an SFR cutoff of $\tilde\Lambda = 4~m_\pi (550~{\rm MeV})$, 
the most common choice. The only parameter of the EGM potential we changed was the cutoff $\Lambda\to\infty$. 
The EGM interaction depend on the SFR cutoff and the results are different if the SFR cutoff is modified, but this 
dependence is not related to the subtractions. We also believe that it is the SFR that drives the difference between 
the EM and EGM potentials as far as the renormalization scale dependence is concerned. Nevertheless, here we treated 
the SFR cutoff as an internal parameter of the EGM potential and looked only at the dependence of the phases on $\mu$.

\begin{figure*}[t]
\begin{center}
\includegraphics[scale=0.2]{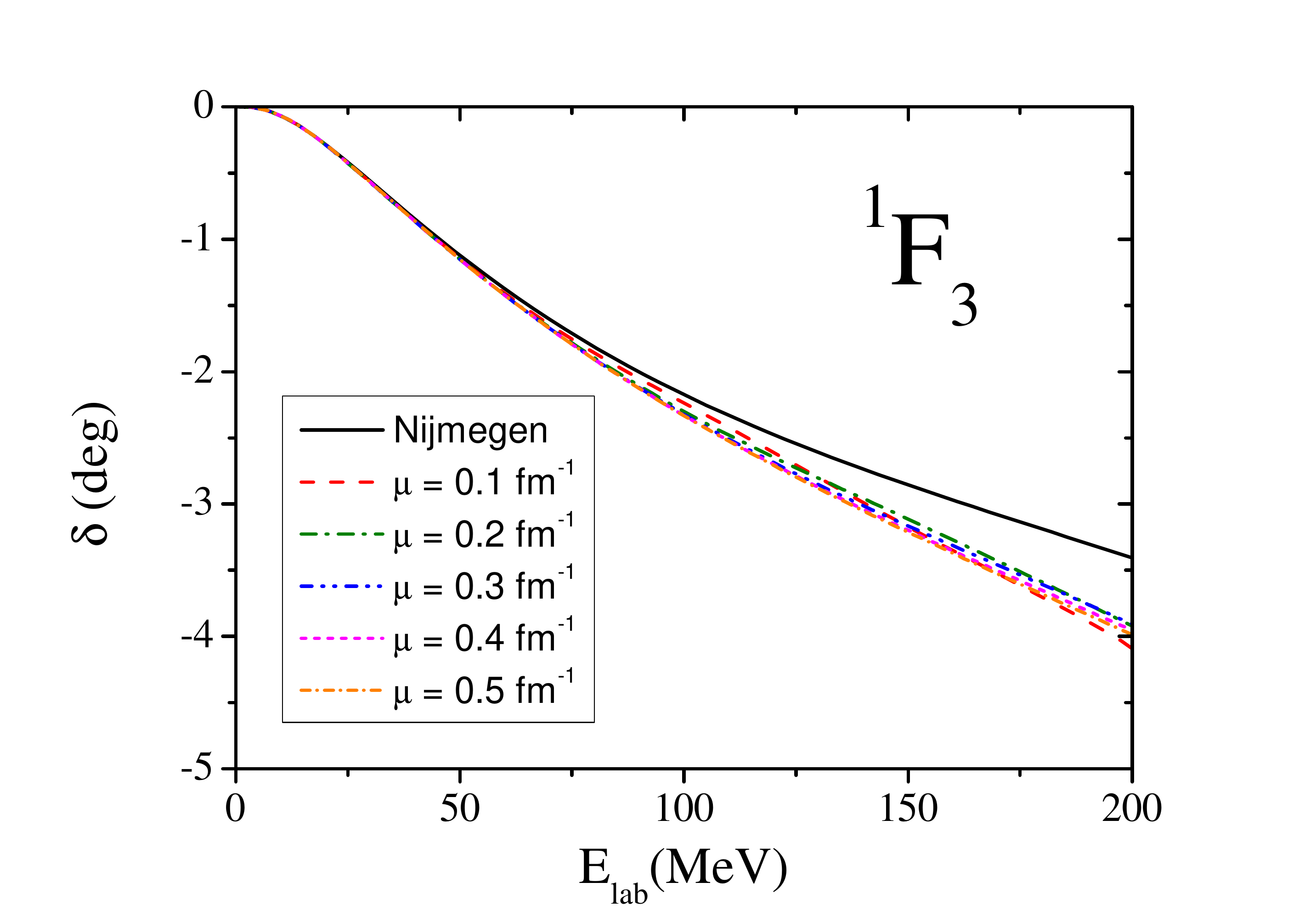}\hspace*{0.1cm}\includegraphics[scale=0.2]{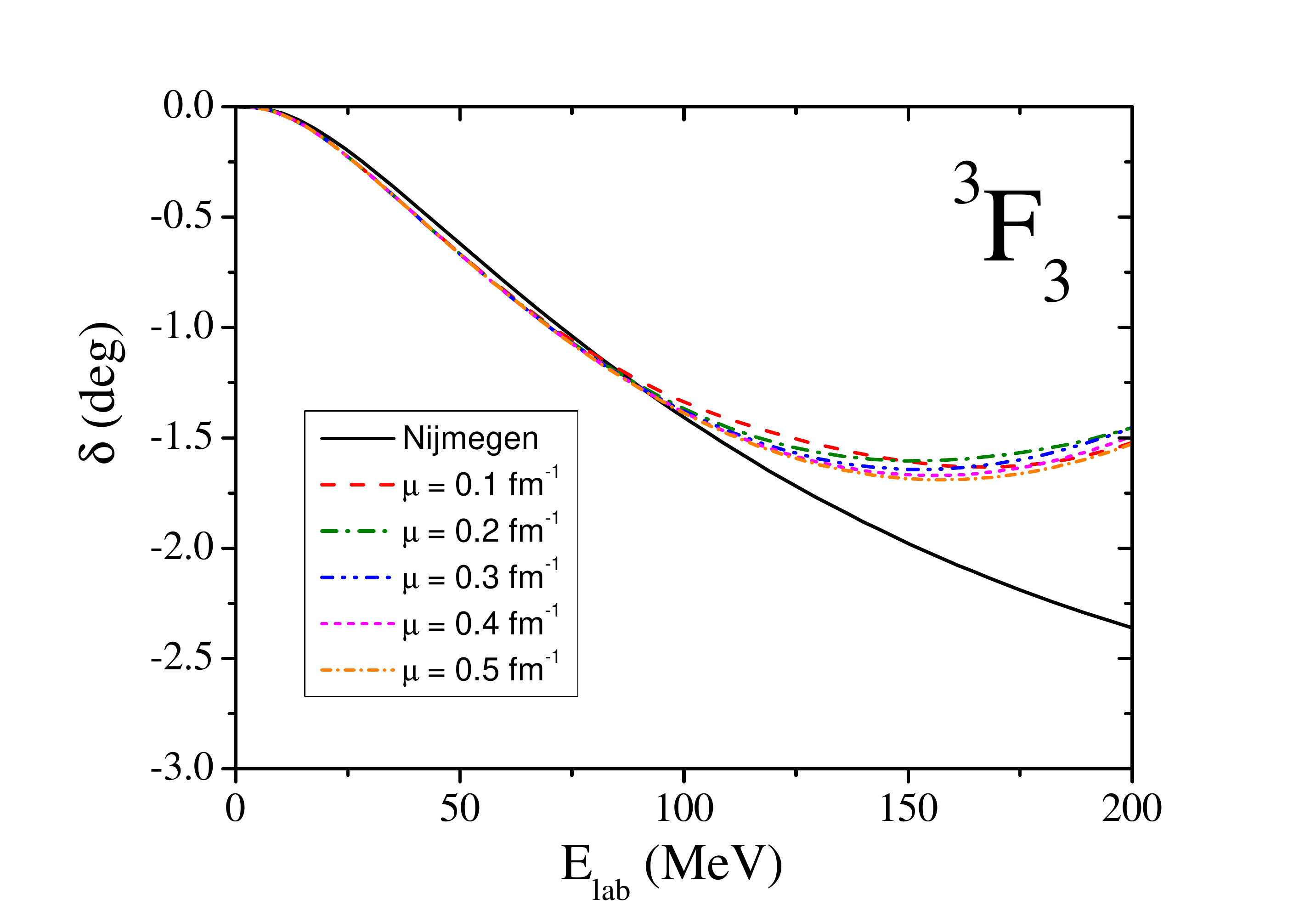} \\
\includegraphics[scale=0.2]{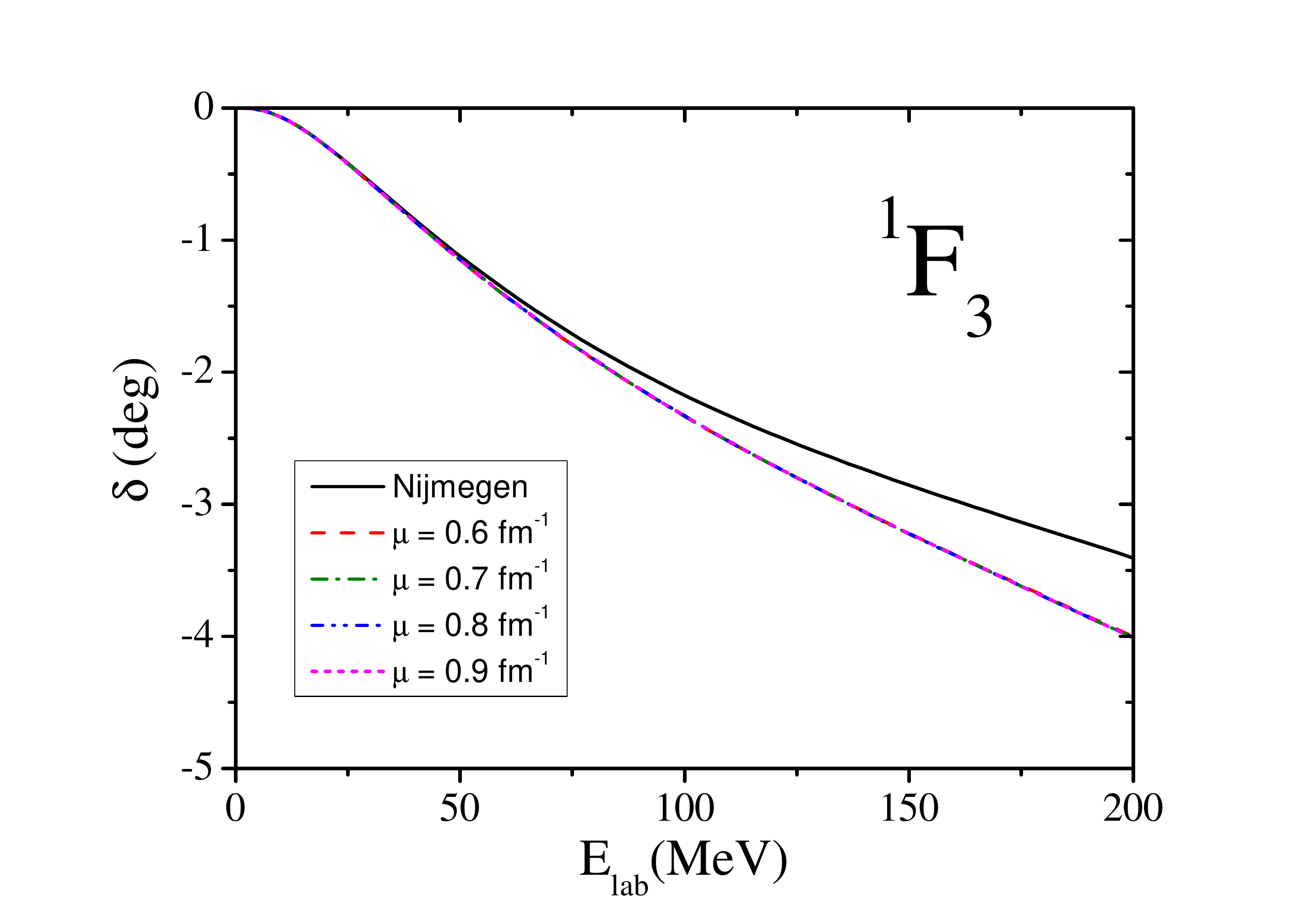}\hspace*{0.1cm}\includegraphics[scale=0.2]{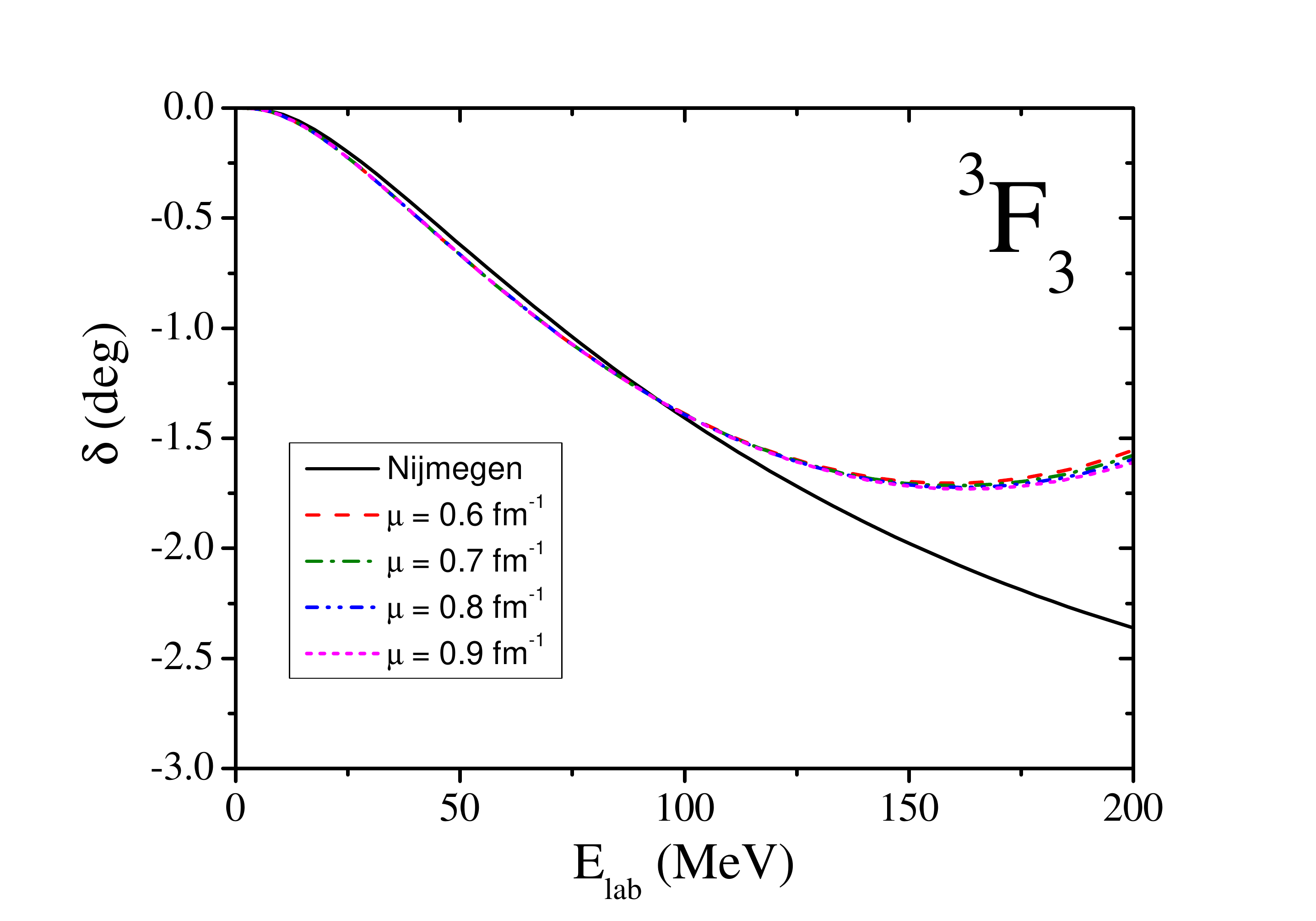} \\
\includegraphics[scale=0.2]{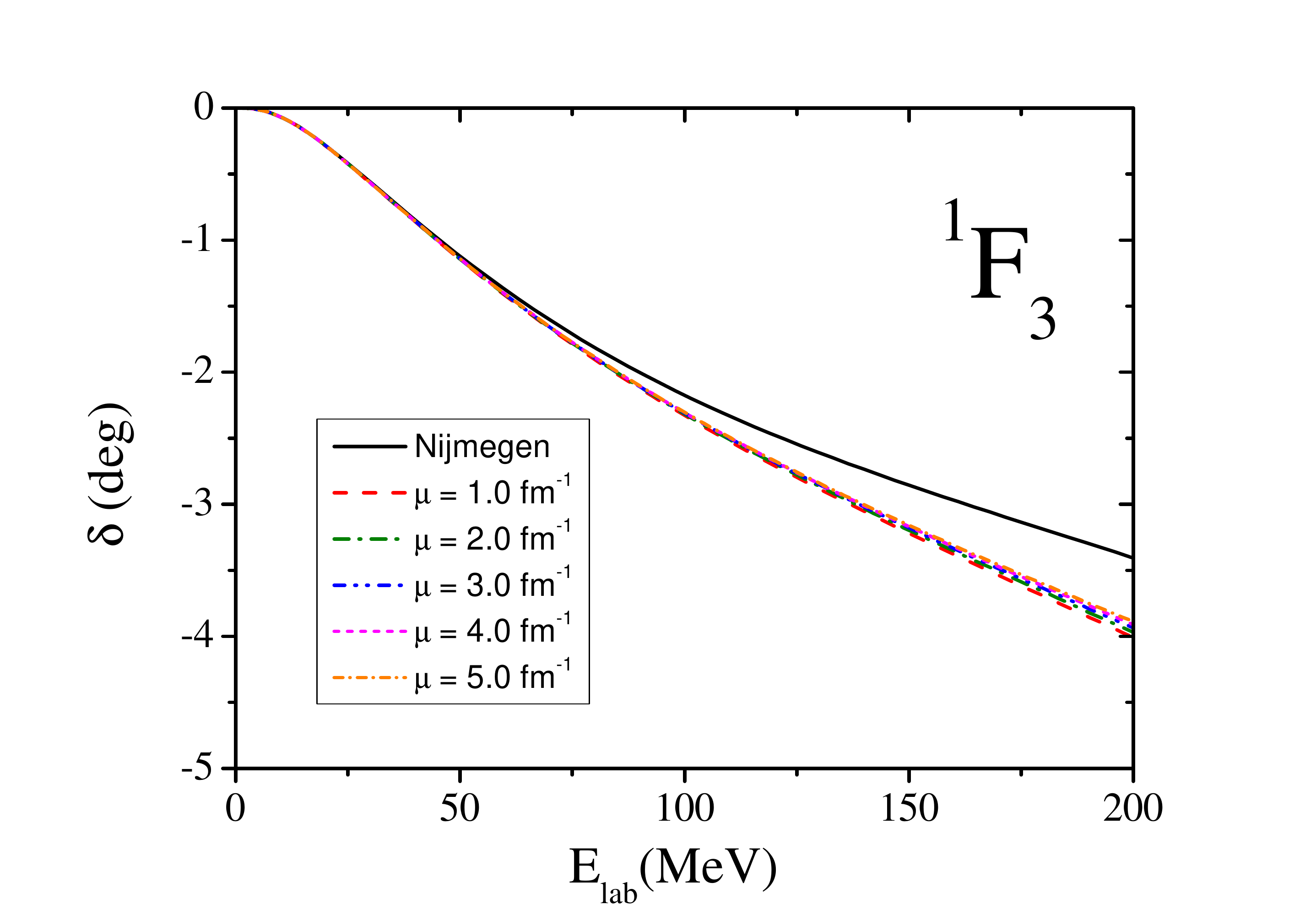}\hspace*{0.1cm}\includegraphics[scale=0.2]{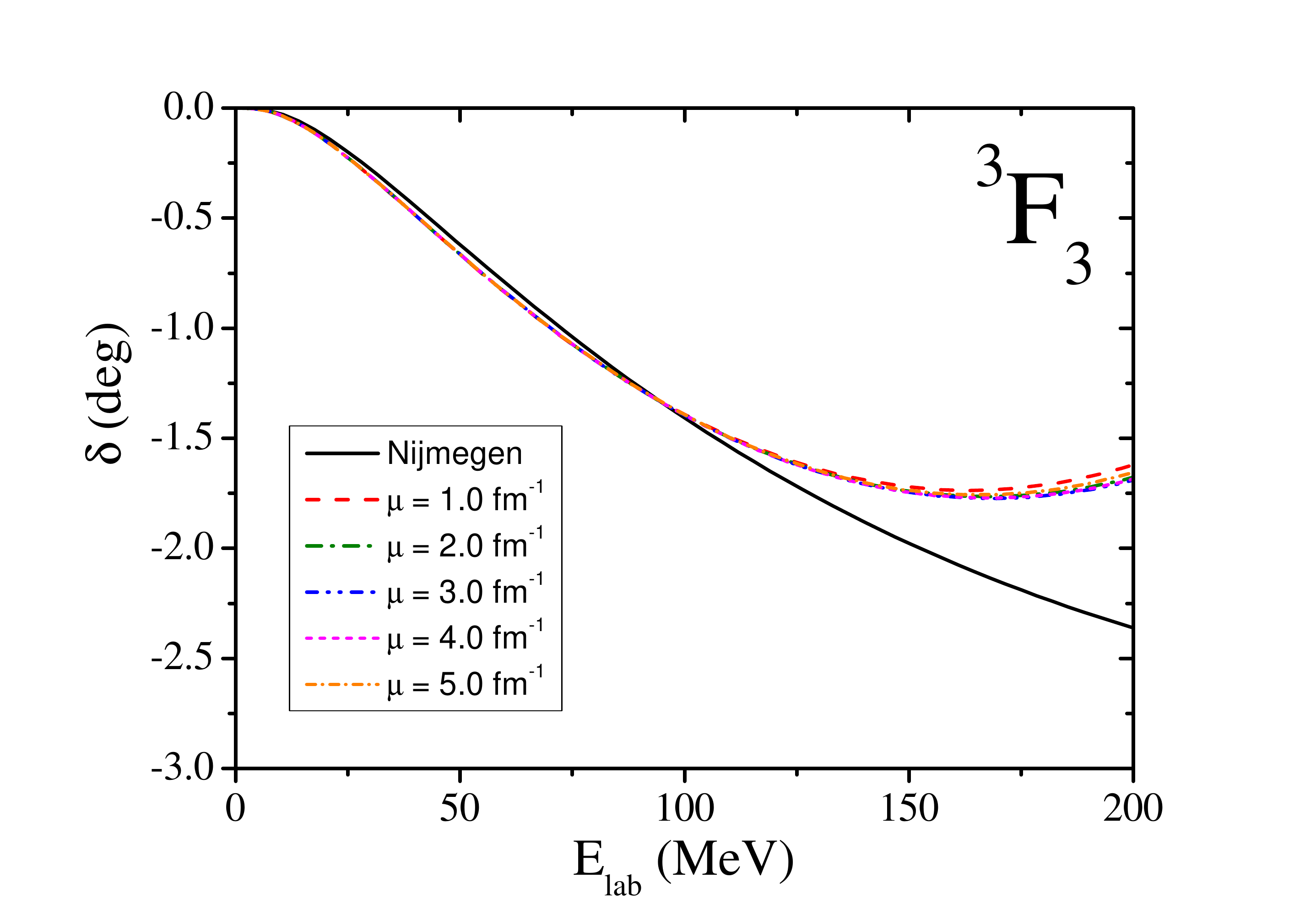} \\
\includegraphics[scale=0.2]{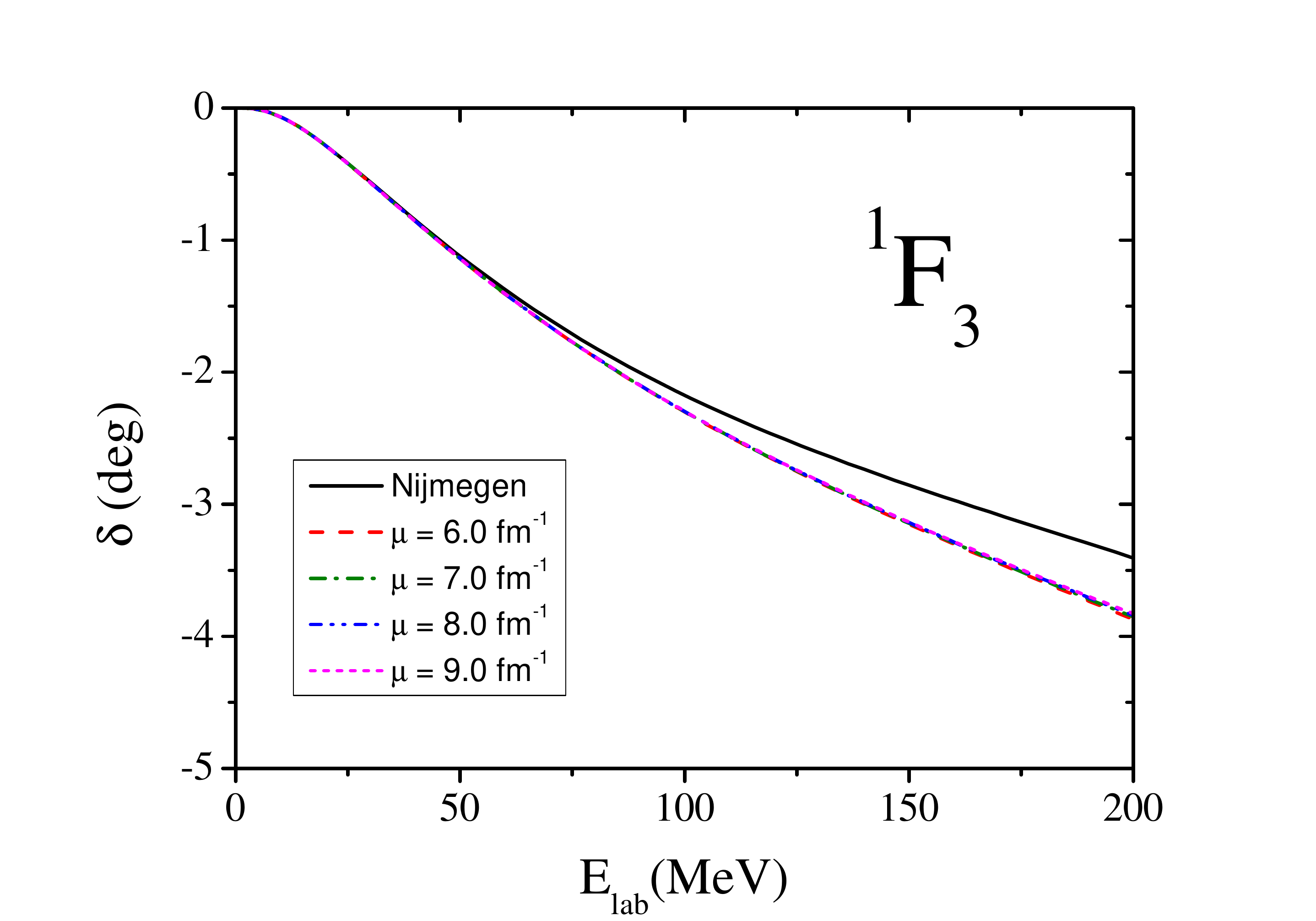}\hspace*{0.1cm}\includegraphics[scale=0.2]{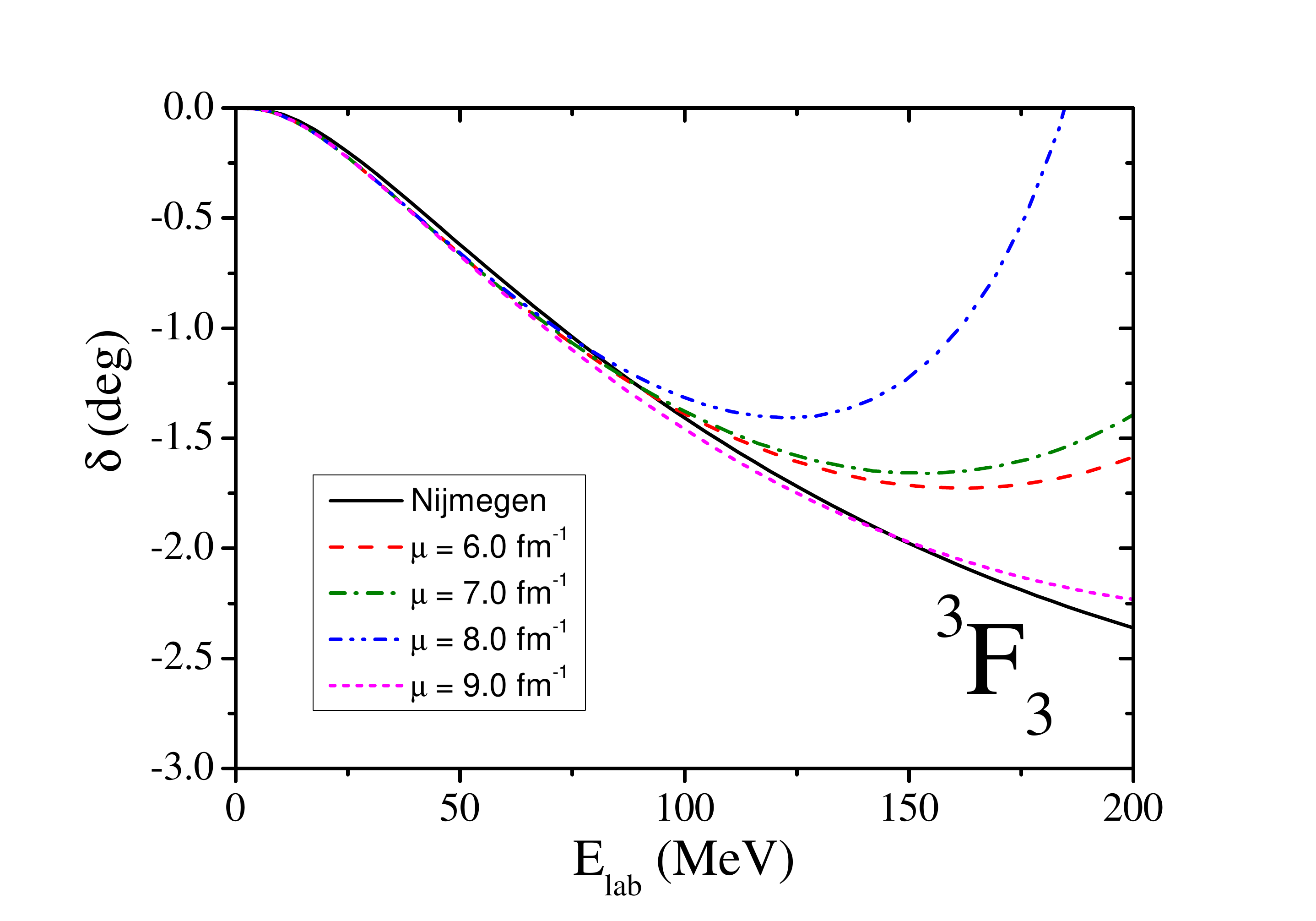} 
\end{center}
\caption{(Color on-line) Phase-shifts in the $^1F_3$ and $^3F_3$ uncoupled channels calculated from the solution of the subtracted LS equation for the $K$-matrix with five subtractions for the N3LO-EM potential for several values of the renormalization scale compared to the Nijmegen partial wave analysis.}
\label{fig1}
\end{figure*}

\begin{figure*}[t]
\begin{center}
\includegraphics[scale=0.2]{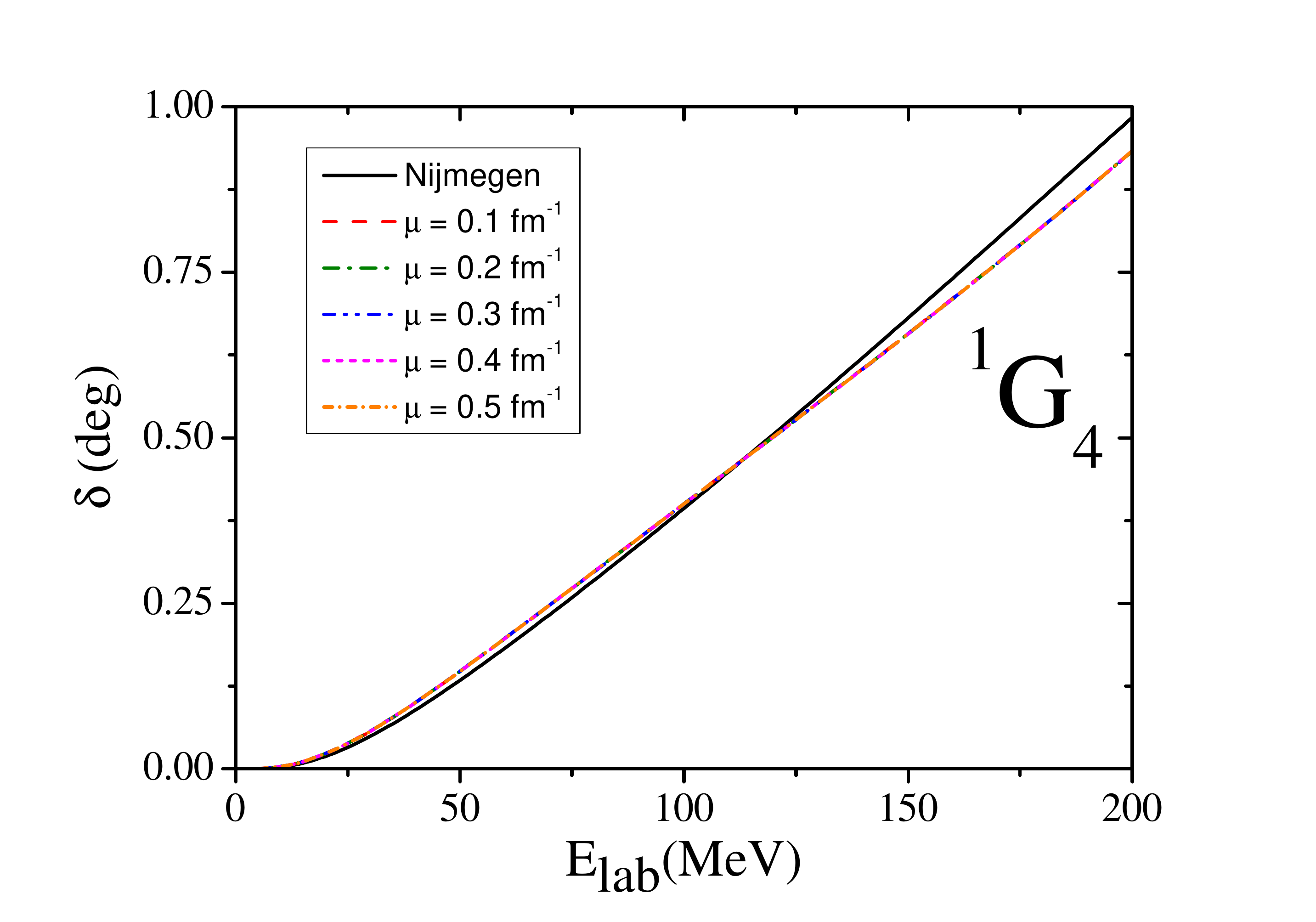}\hspace*{0.1cm}\includegraphics[scale=0.2]{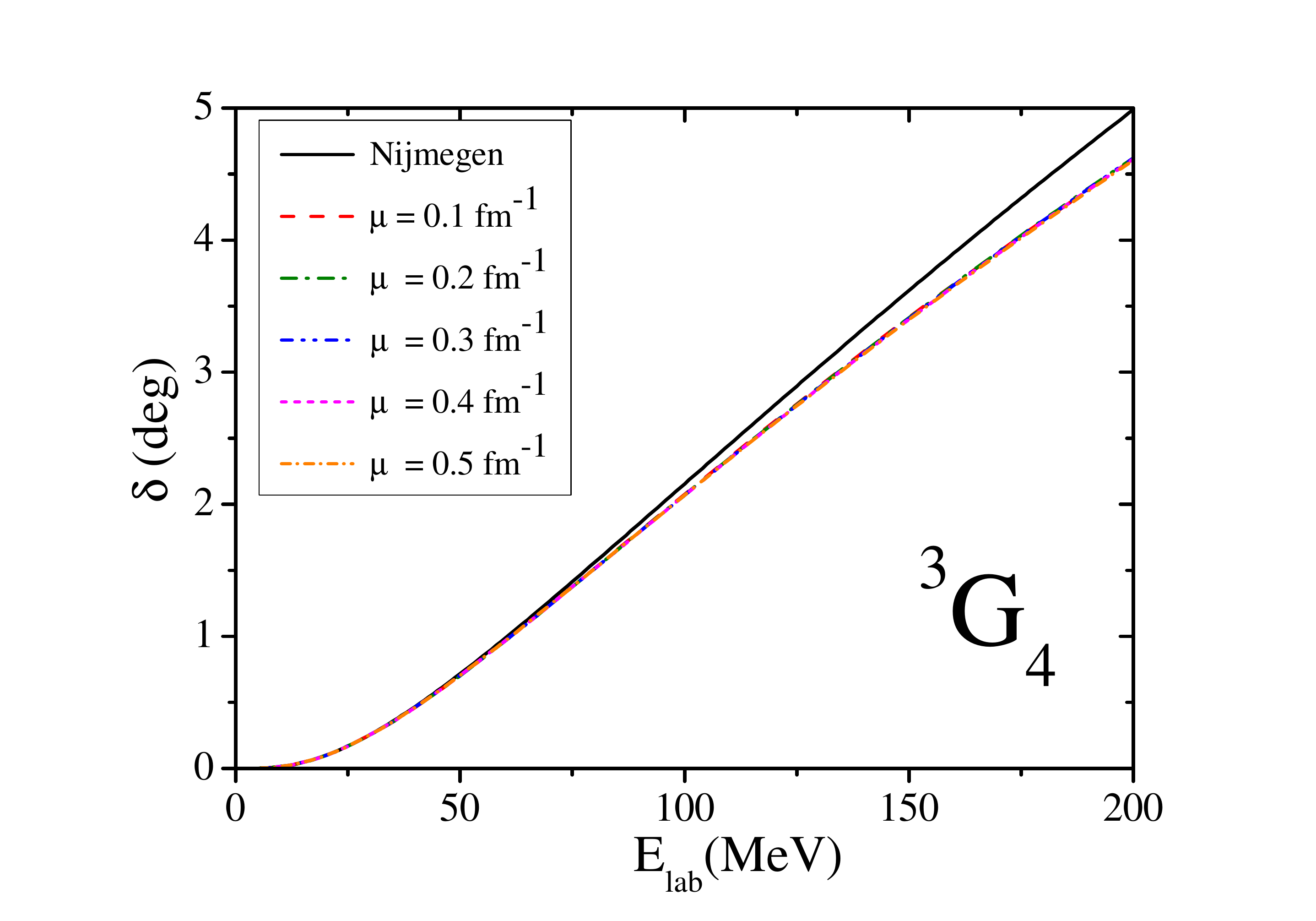} \\
\includegraphics[scale=0.2]{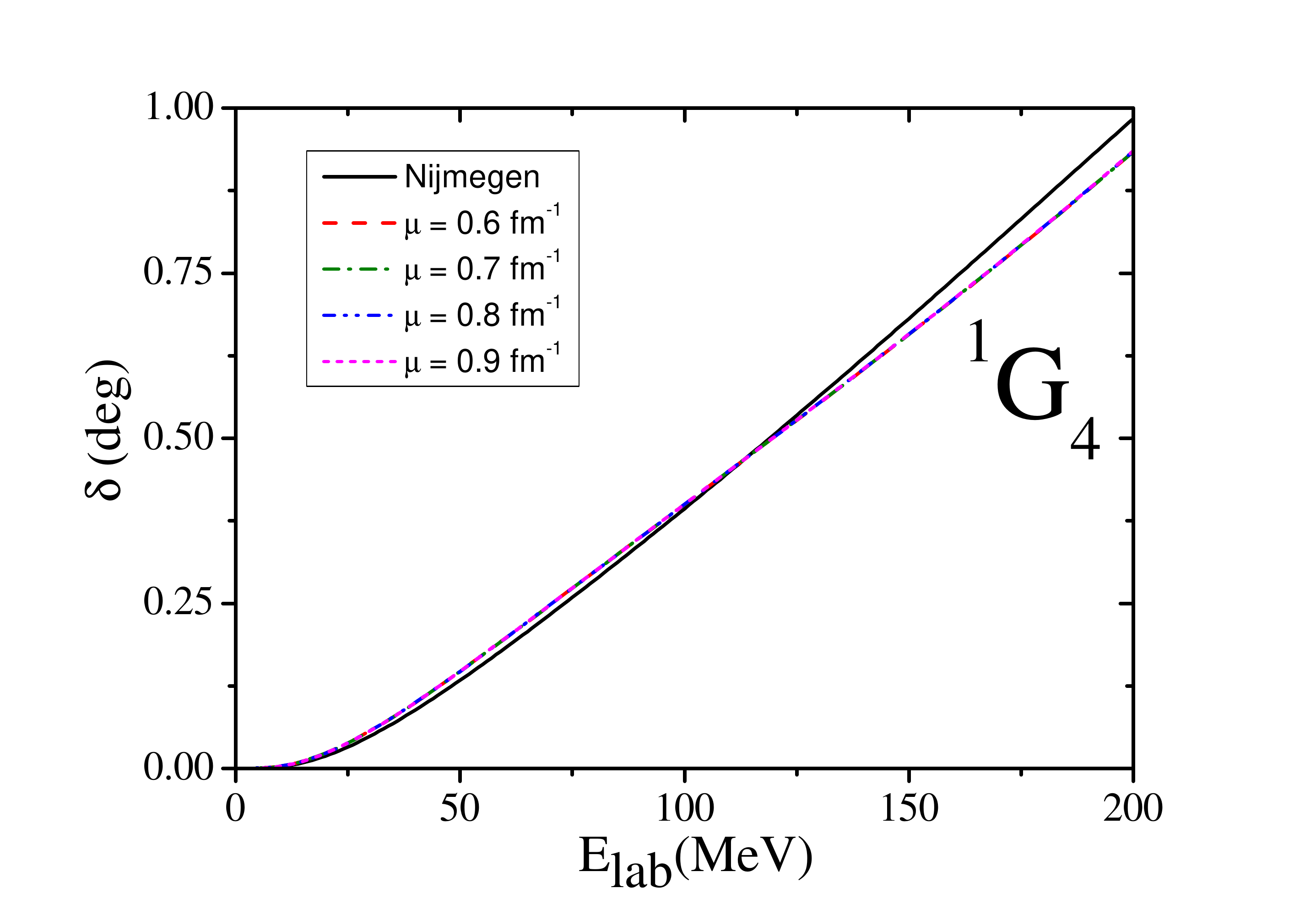}\hspace*{0.1cm}\includegraphics[scale=0.2]{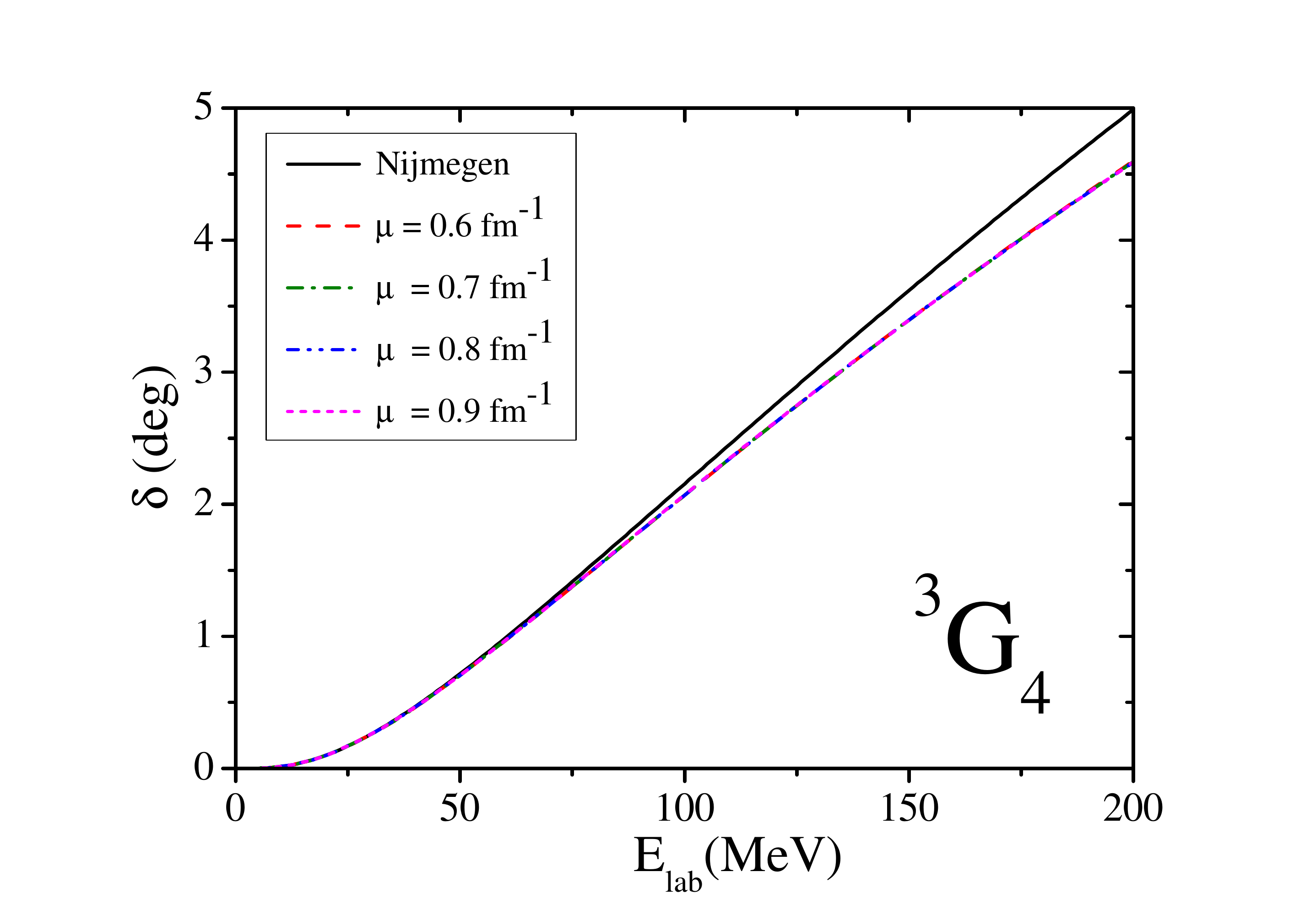} \\
\includegraphics[scale=0.2]{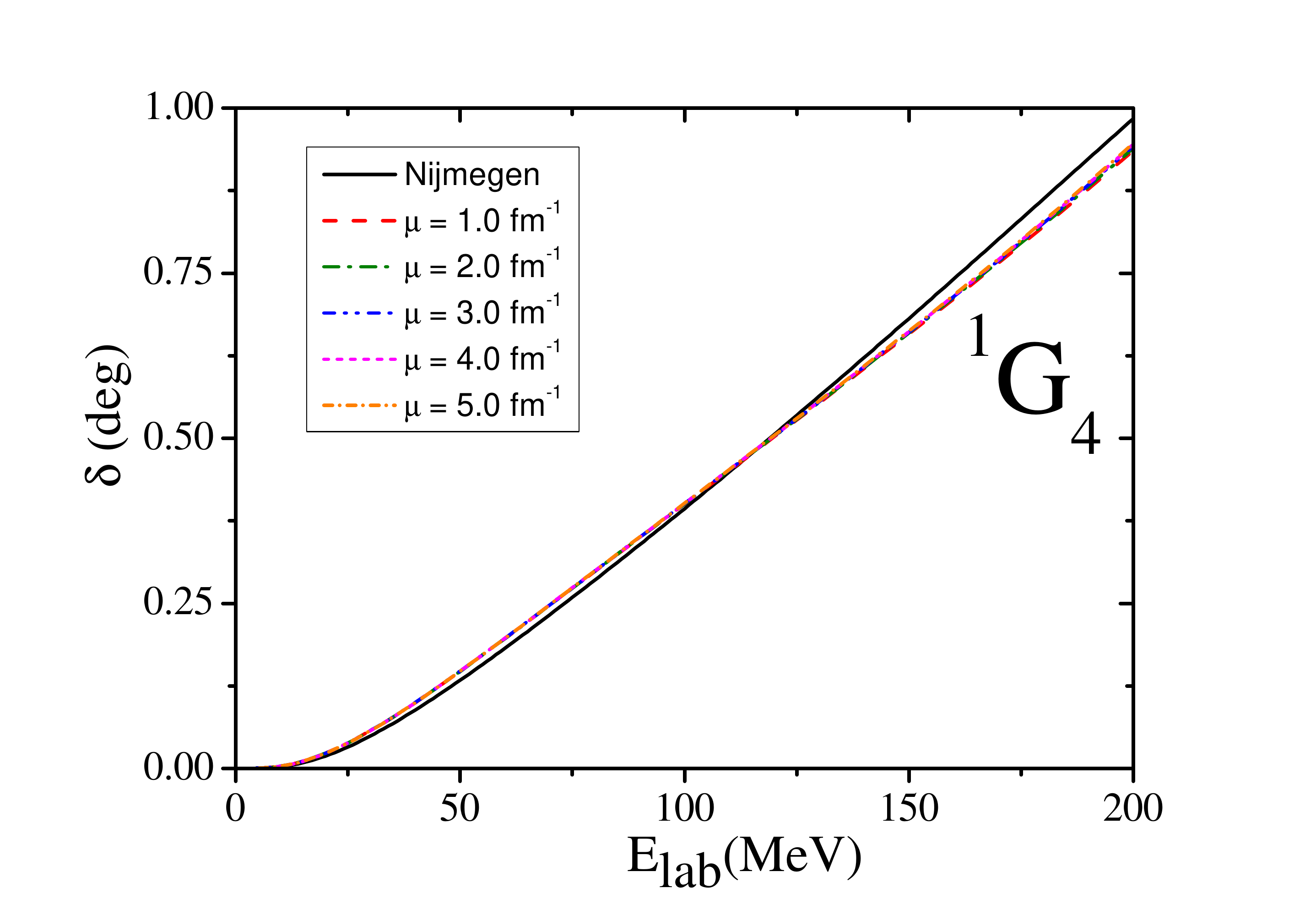}\hspace*{0.1cm}\includegraphics[scale=0.2]{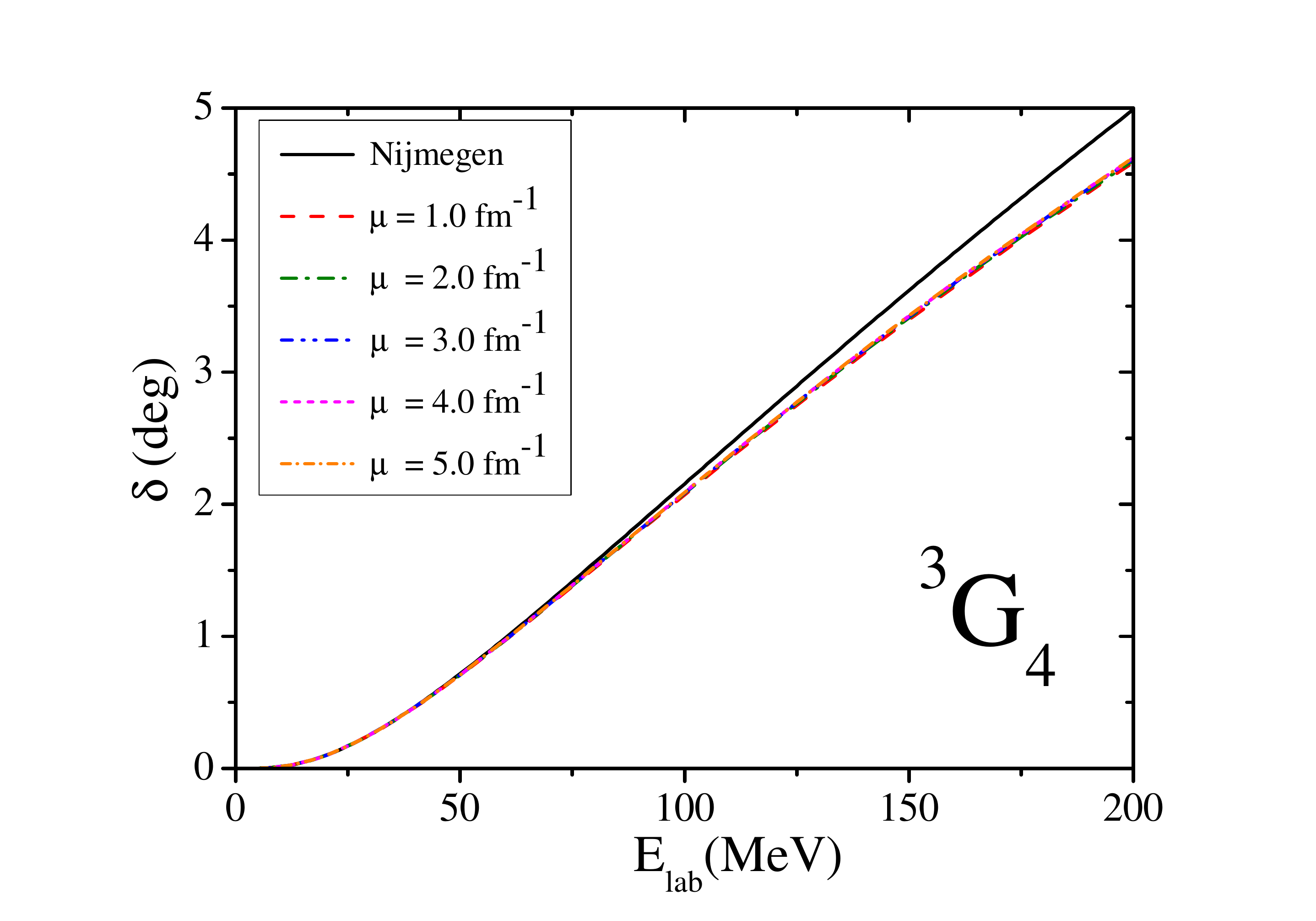} \\
\includegraphics[scale=0.2]{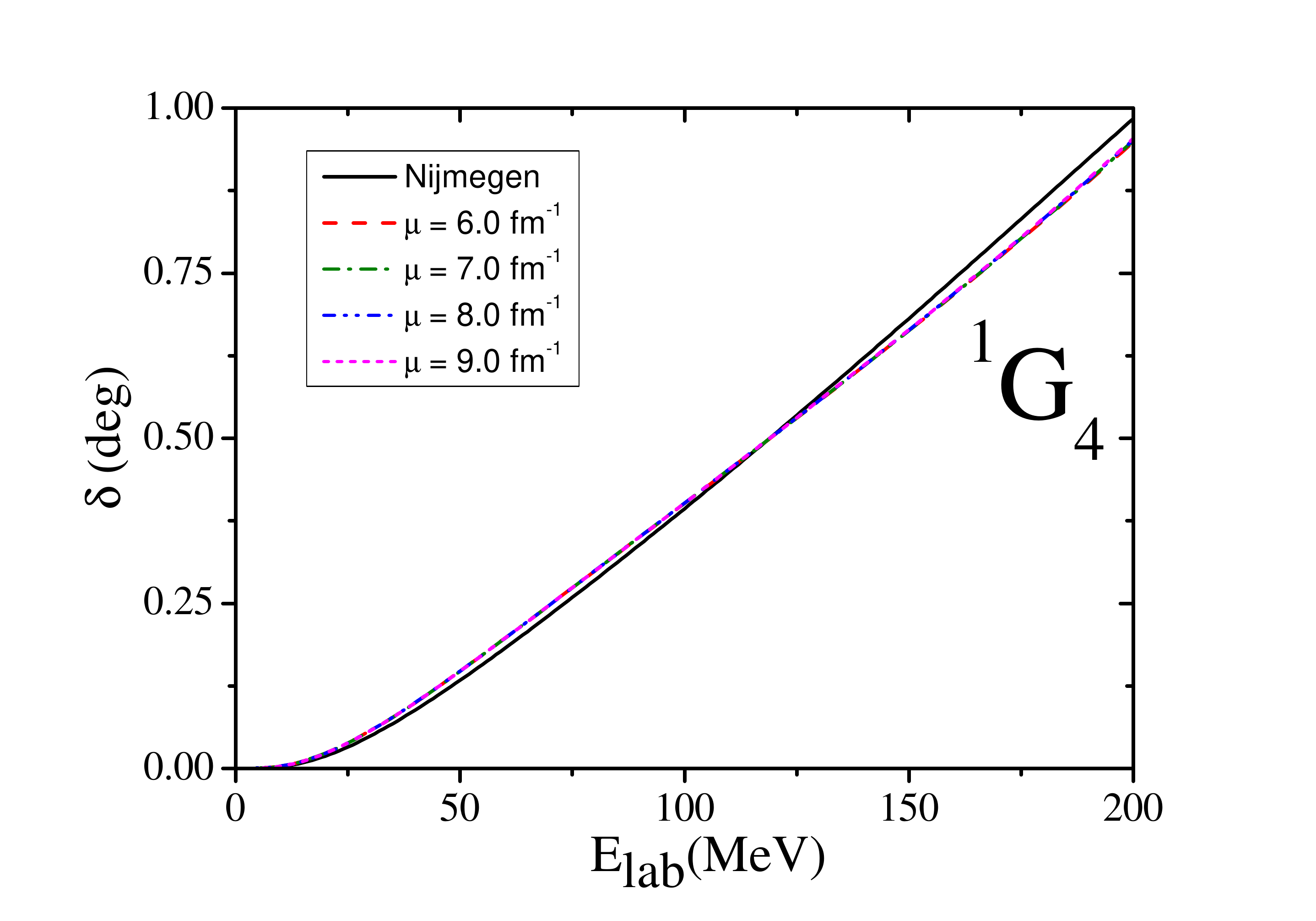}\hspace*{0.1cm}\includegraphics[scale=0.2]{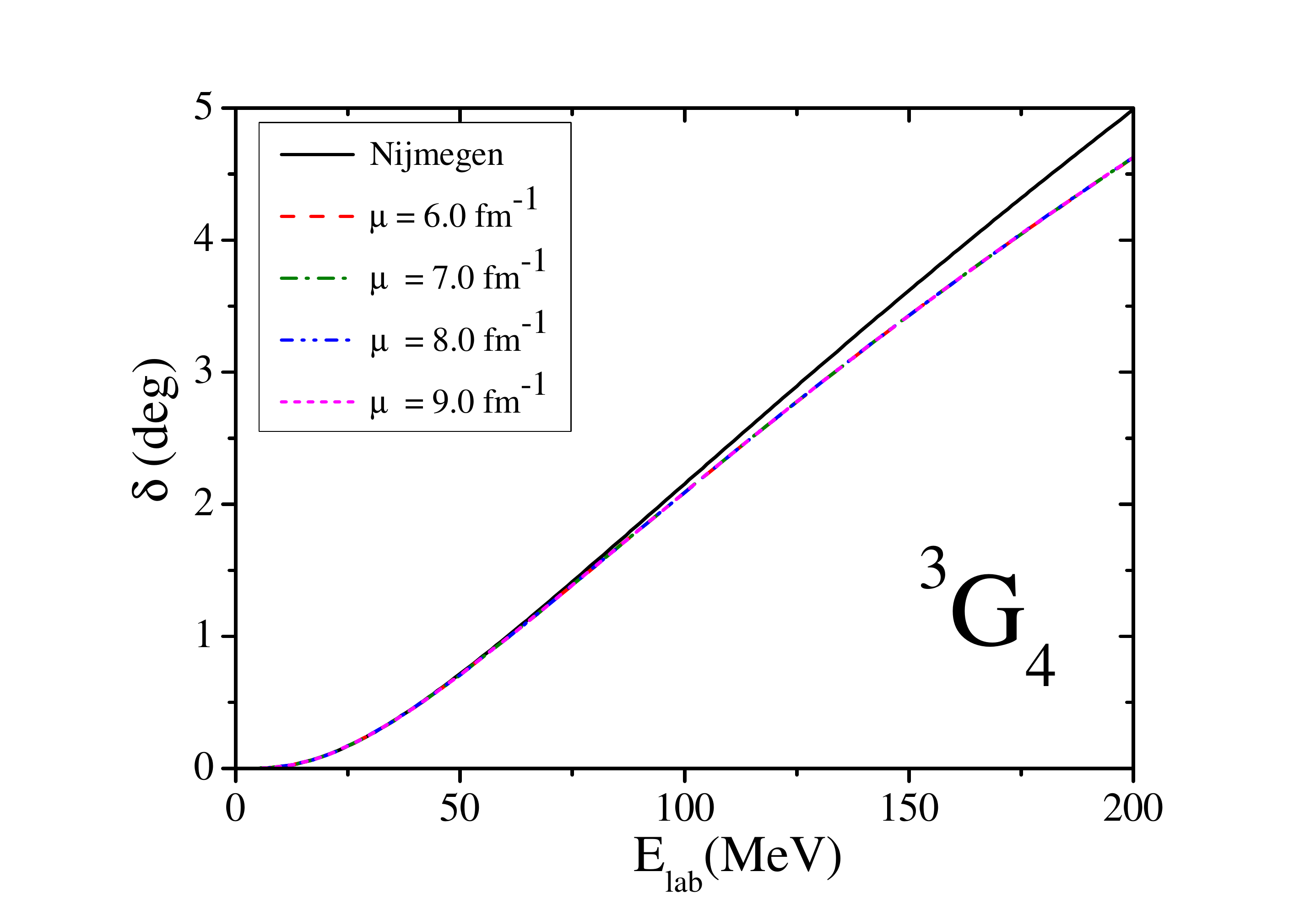} 
\end{center}
\caption{(Color on-line) Phase-shifts in the $^1G_4$ and $^3G_4$ uncoupled channels calculated from the solution of the subtracted LS equation for the $K$-matrix with five subtractions for the N3LO-EM potential for several values of the renormalization scale compared to the Nijmegen partial wave analysis.}
\label{fig2}
\end{figure*}

\begin{figure*}[t]
\begin{center}
\includegraphics[scale=0.2]{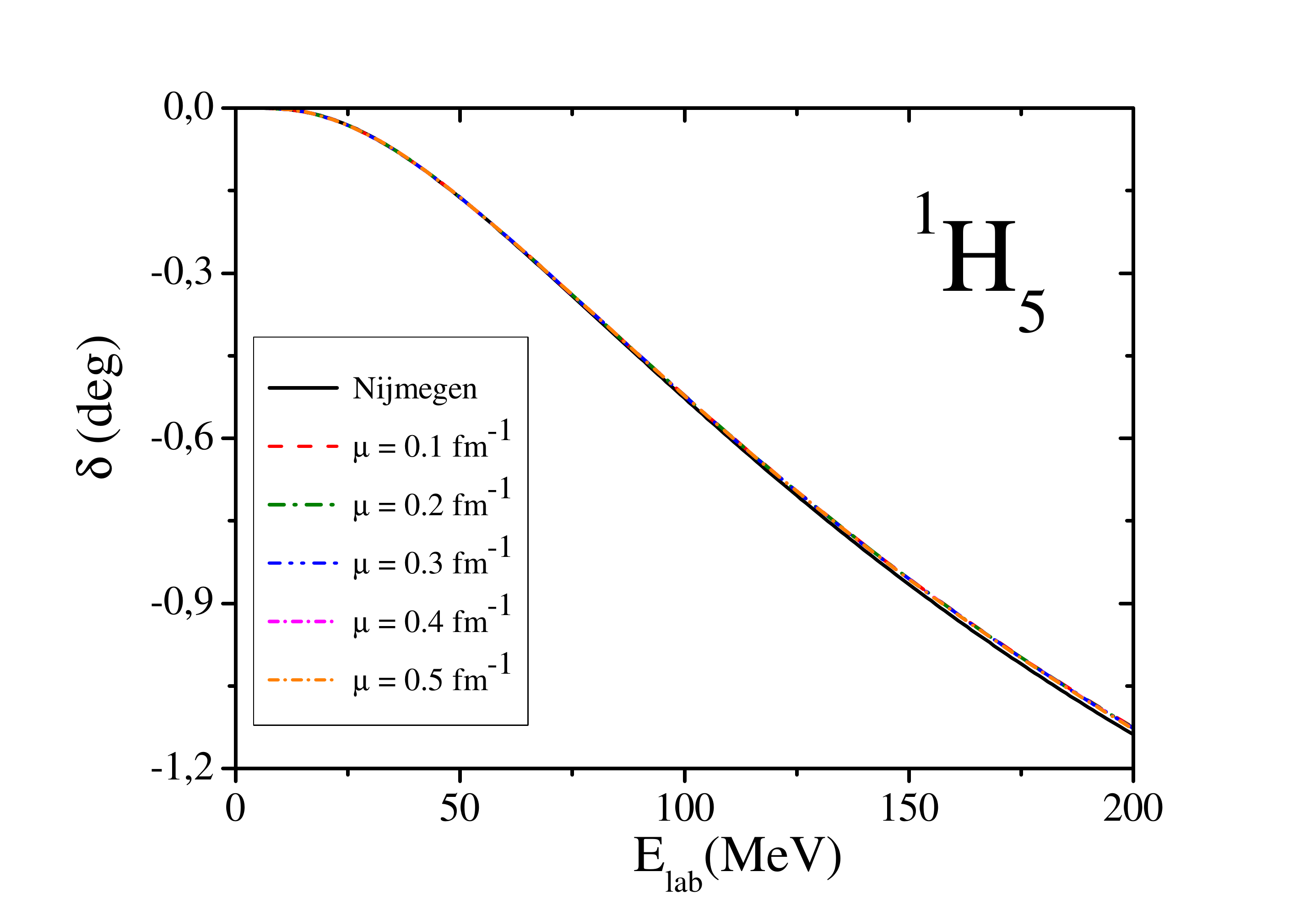}\hspace*{0.1cm}\includegraphics[scale=0.2]{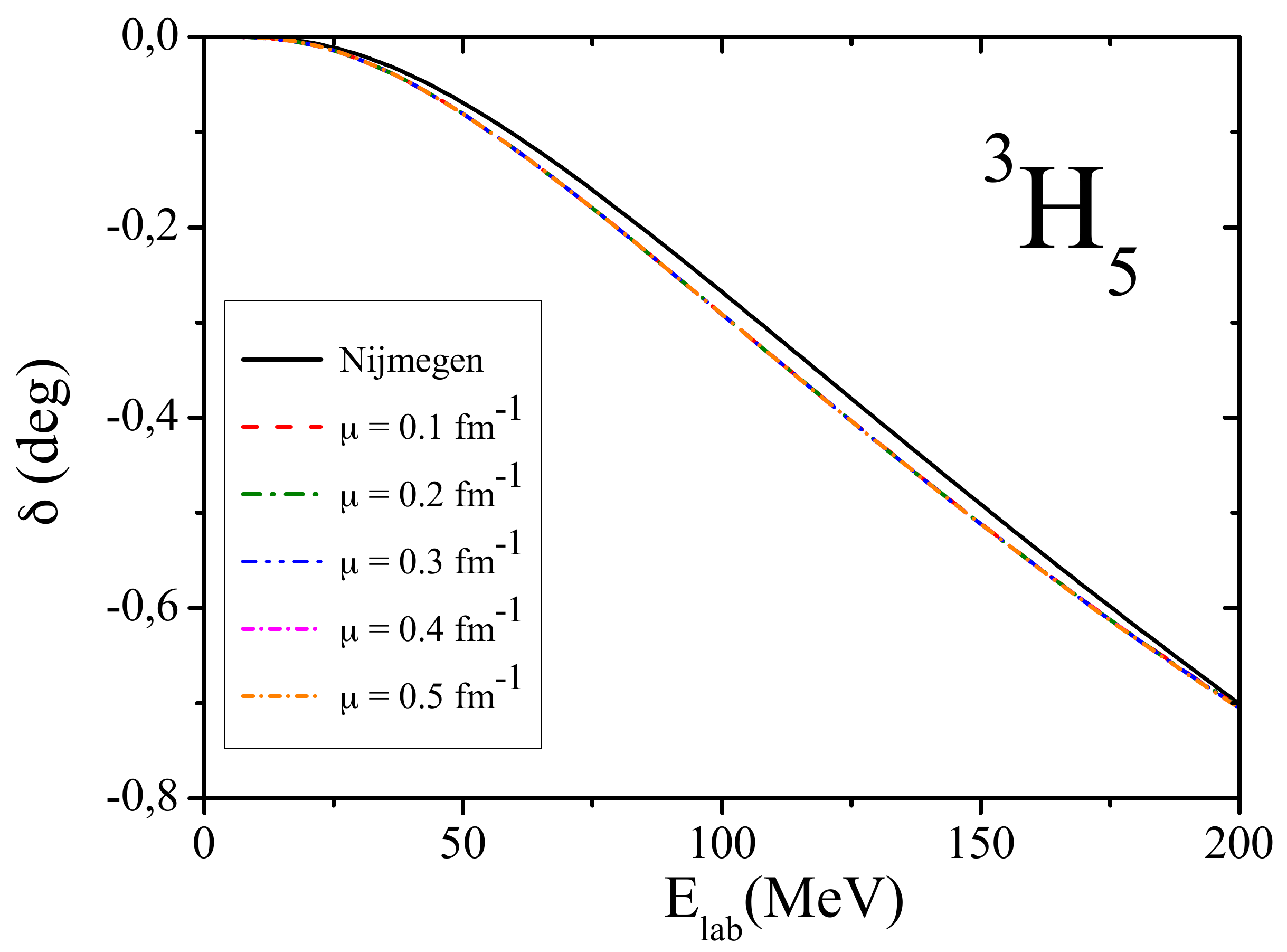} \\
\includegraphics[scale=0.2]{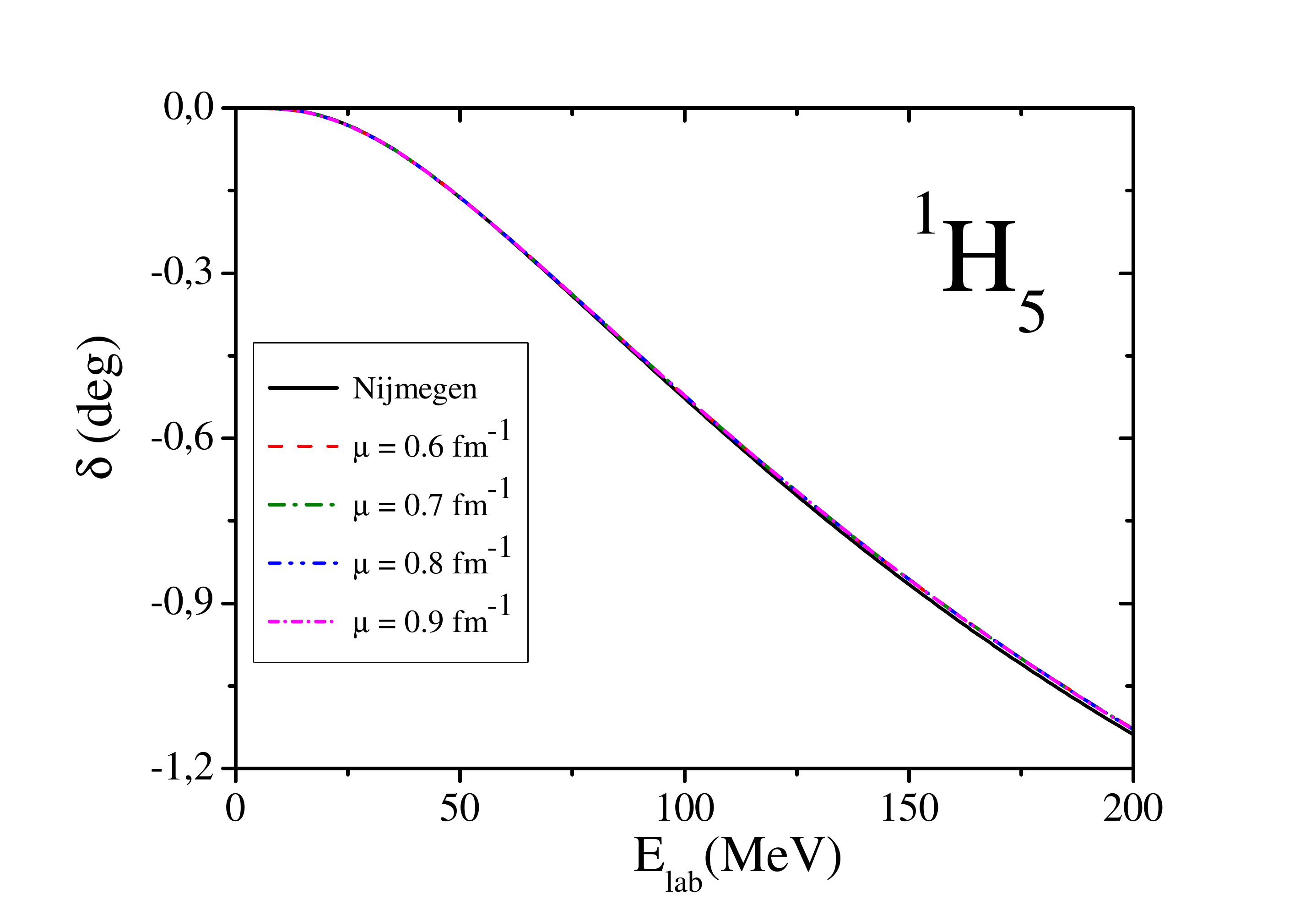}\hspace*{0.1cm}\includegraphics[scale=0.2]{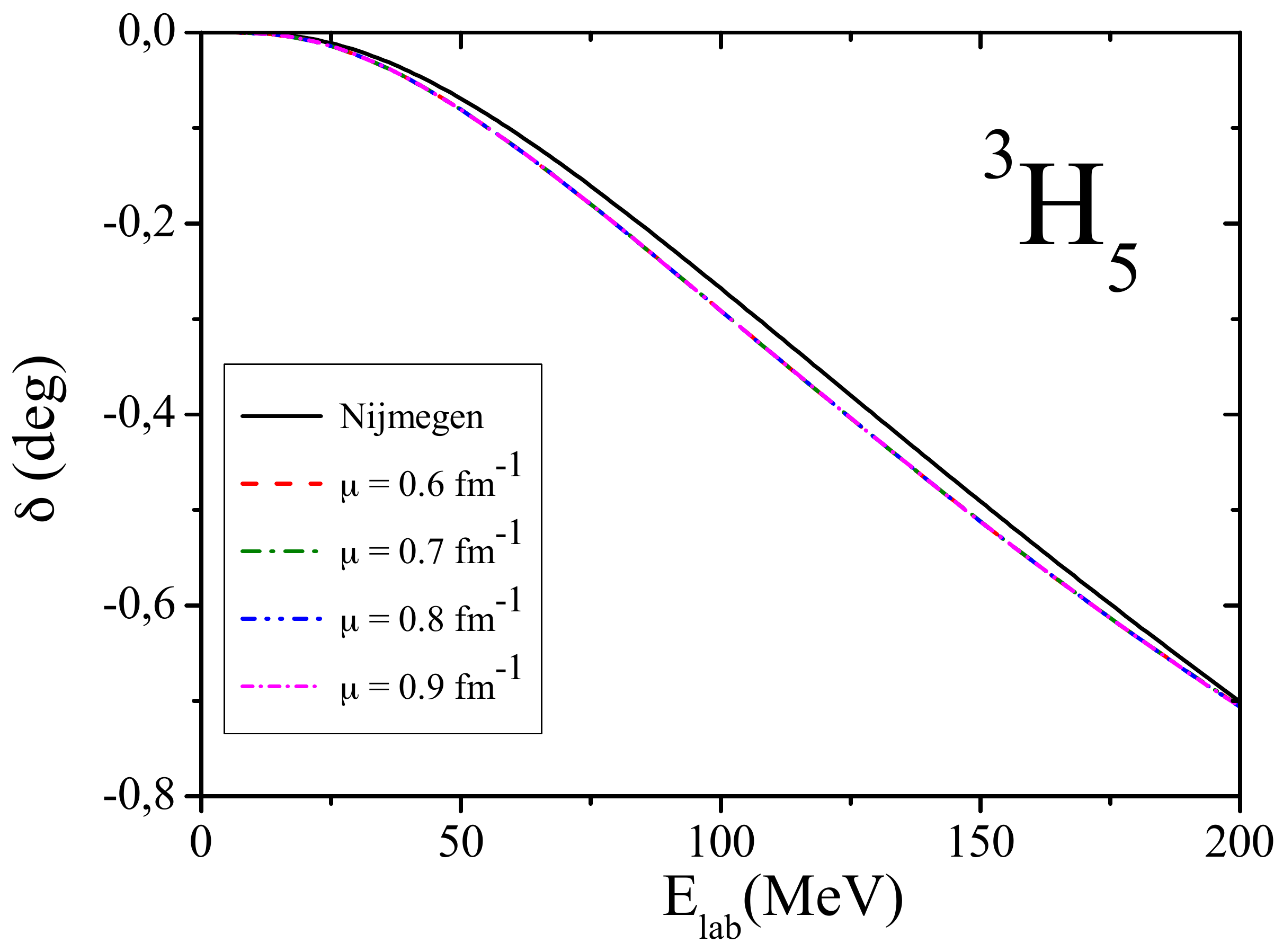} \\
\includegraphics[scale=0.2]{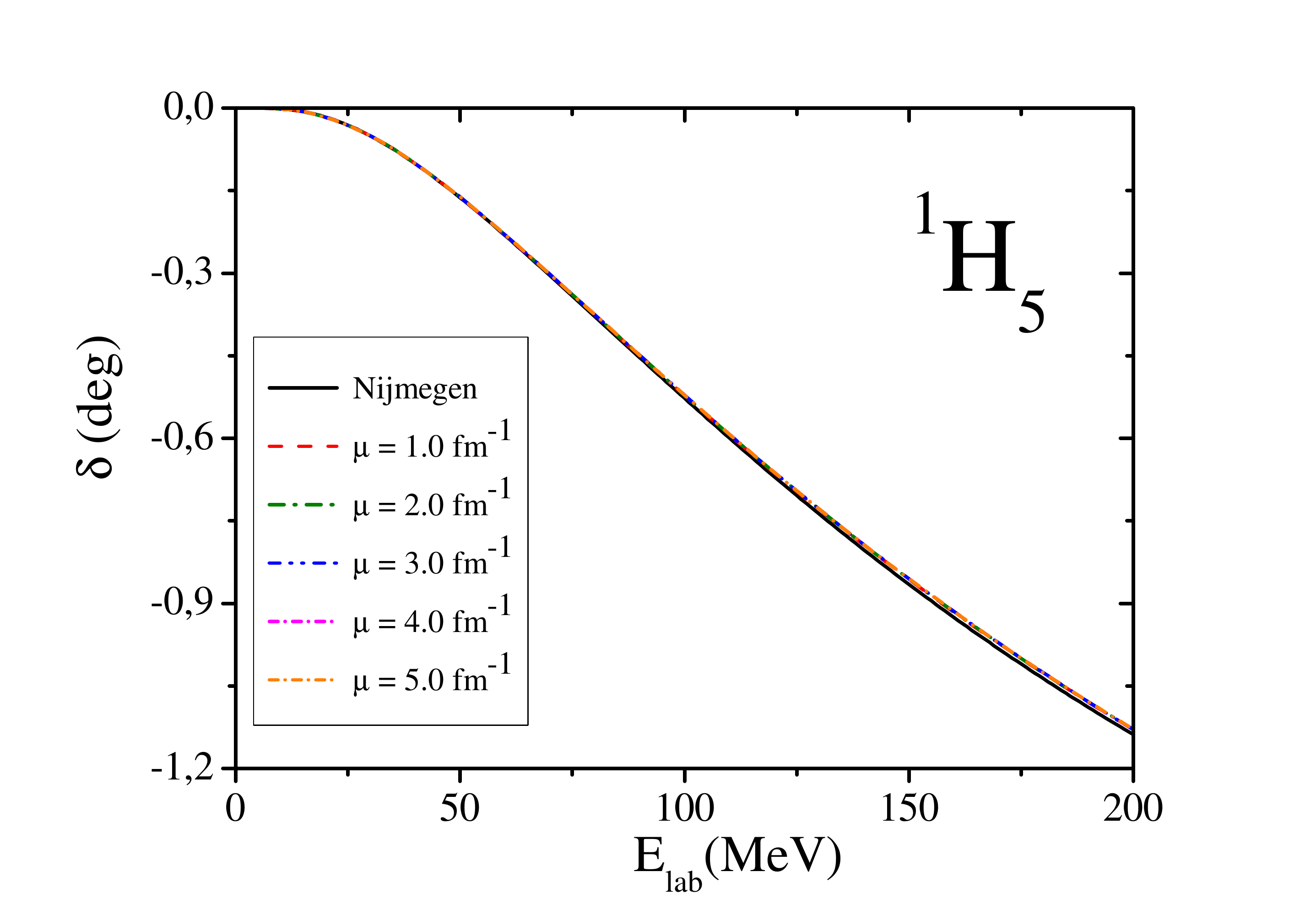}\hspace*{0.1cm}\includegraphics[scale=0.2]{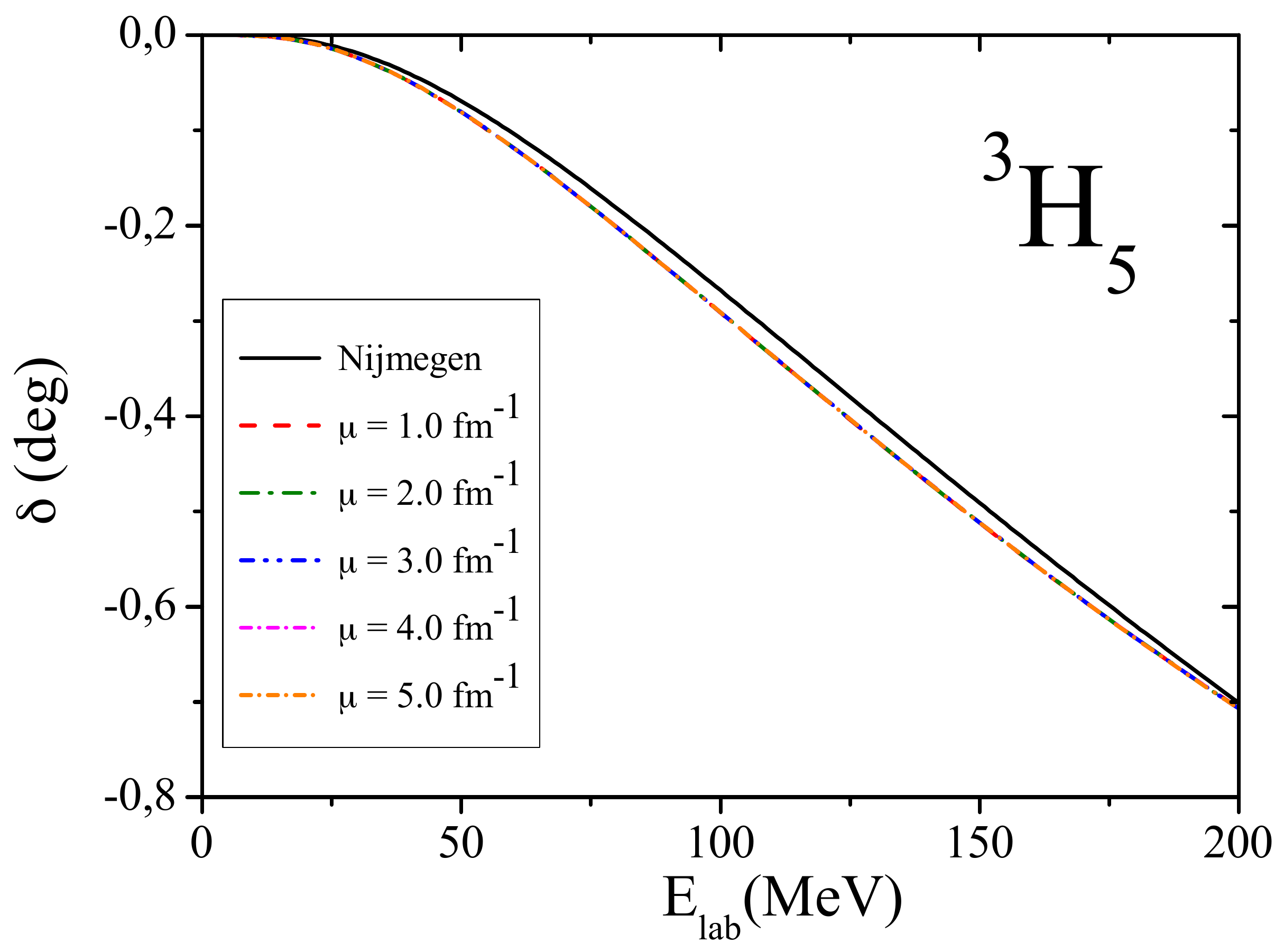} \\
\includegraphics[scale=0.2]{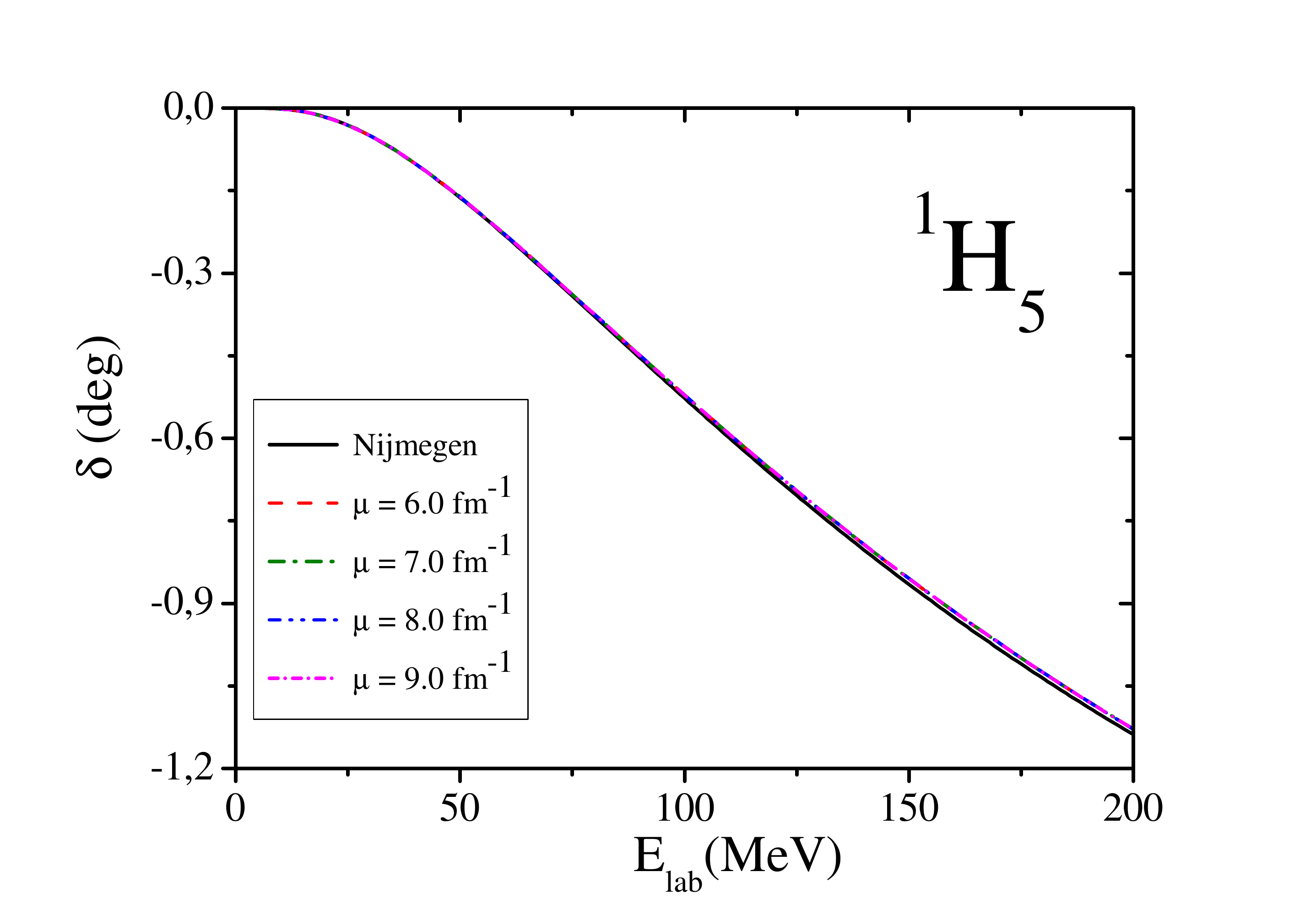}\hspace*{0.1cm}\includegraphics[scale=0.2]{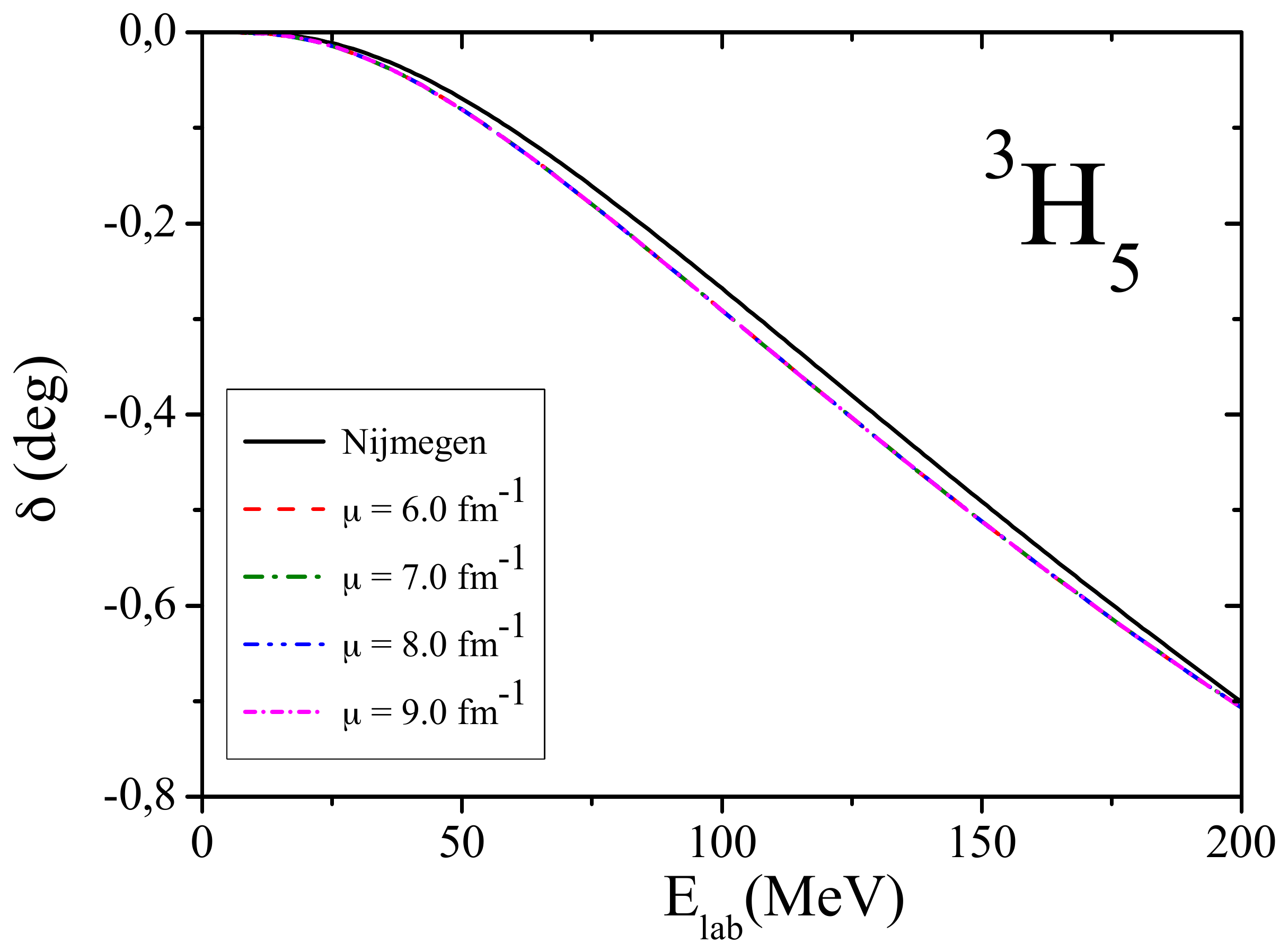} 
\end{center}
\caption{(Color on-line) Phase-shifts in the $^1H_5$ and $^3H_5$ uncoupled channels calculated from the solution of the subtracted LS equation for the $K$-matrix with five subtractions for the N3LO-EM potential for several values of the renormalization scale compared to the Nijmegen partial wave analysis.}
\label{fig3}
\end{figure*}

\begin{figure}[t]
\begin{center}
\includegraphics[scale=0.2]{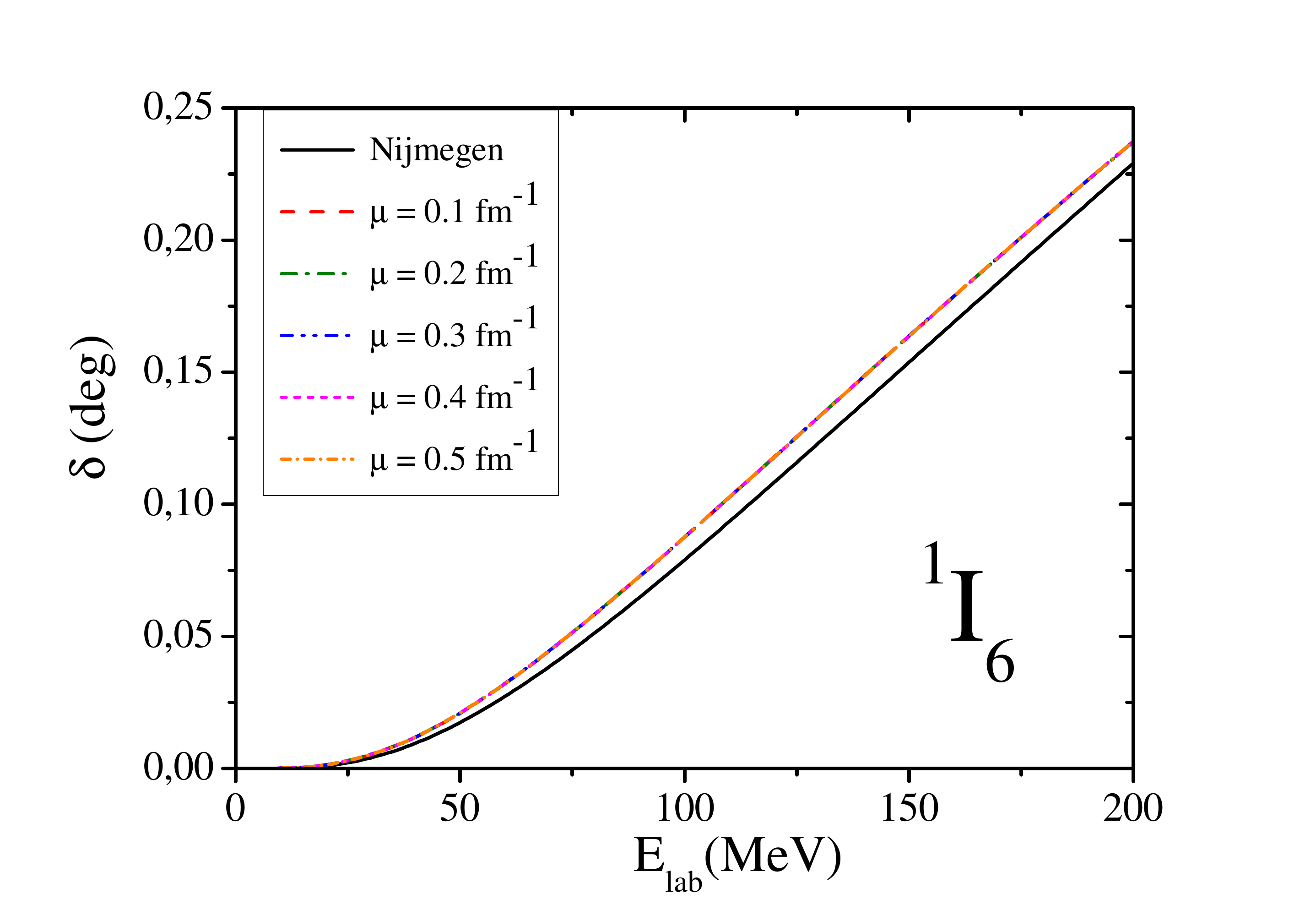}\hspace*{0.1cm}\includegraphics[scale=0.2]{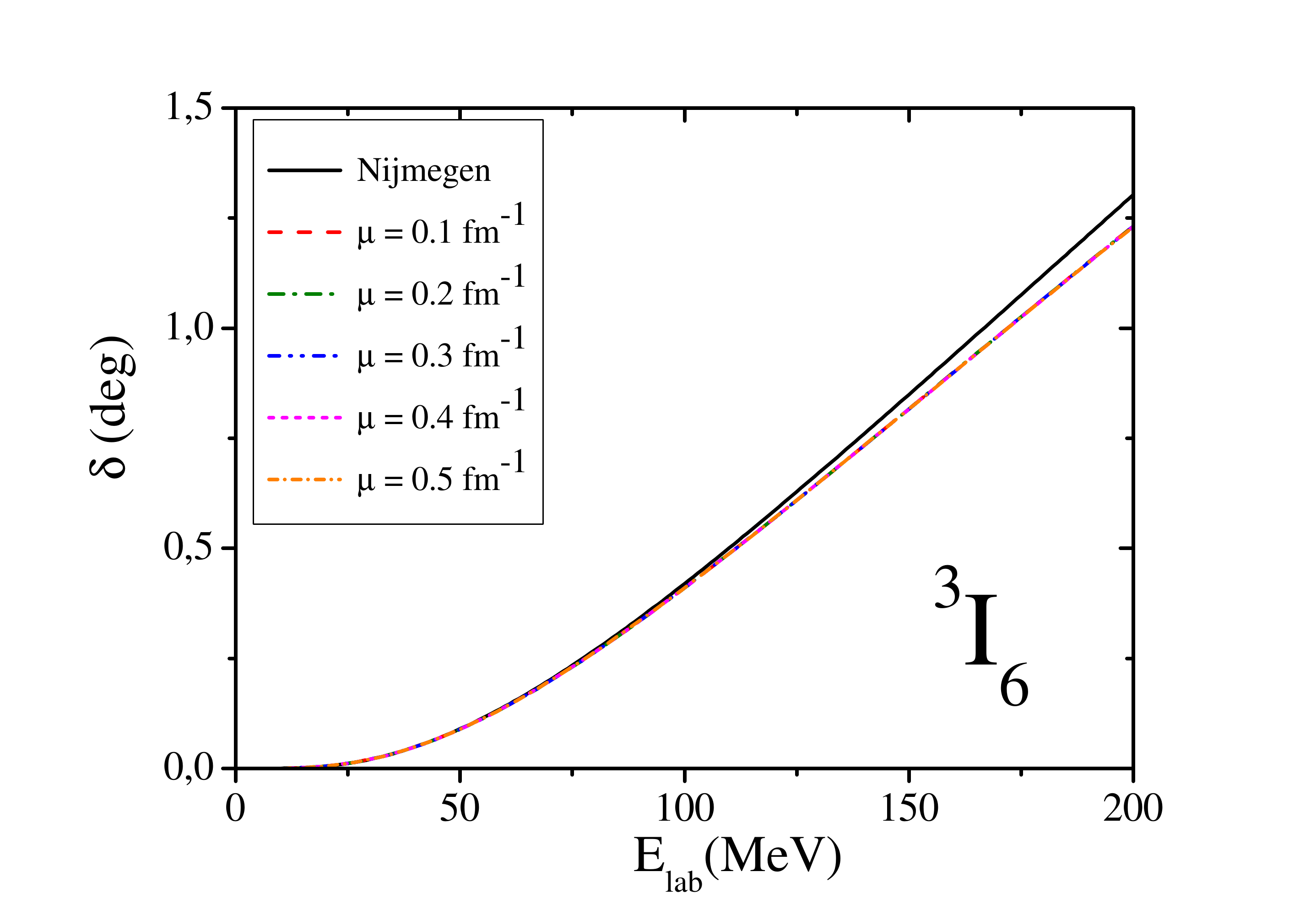} \\
\includegraphics[scale=0.2]{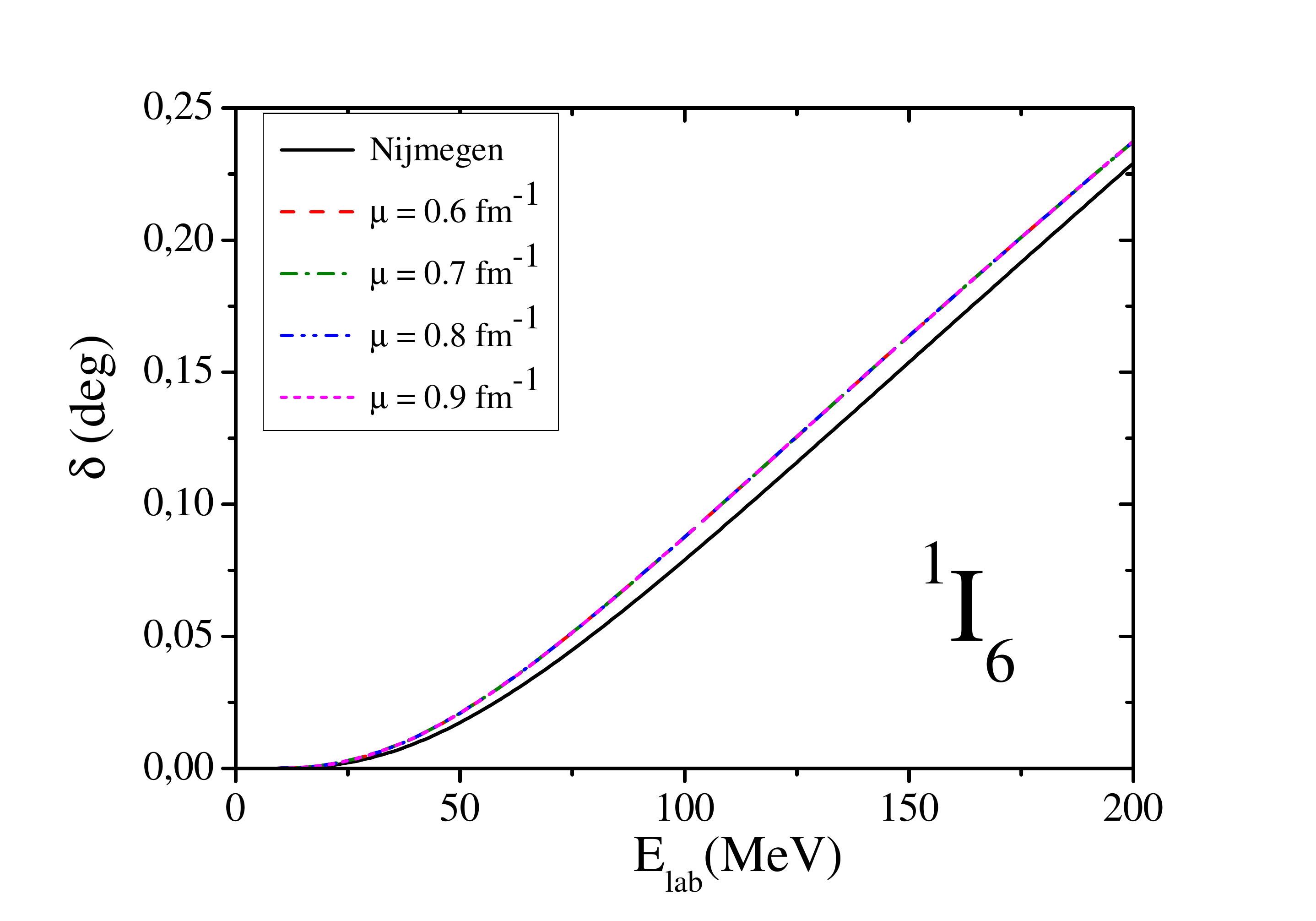}\hspace*{0.1cm}\includegraphics[scale=0.2]{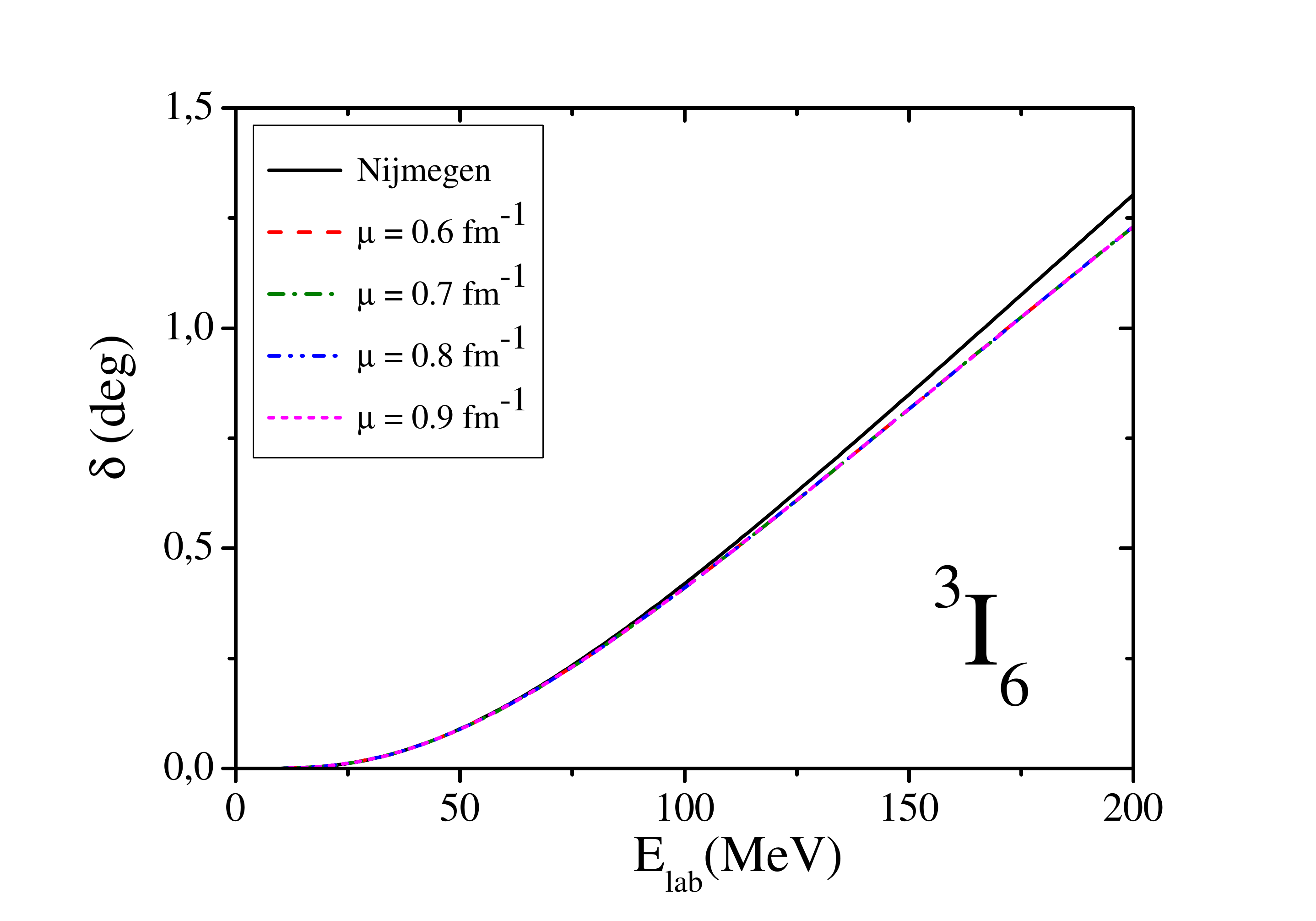} \\
\includegraphics[scale=0.2]{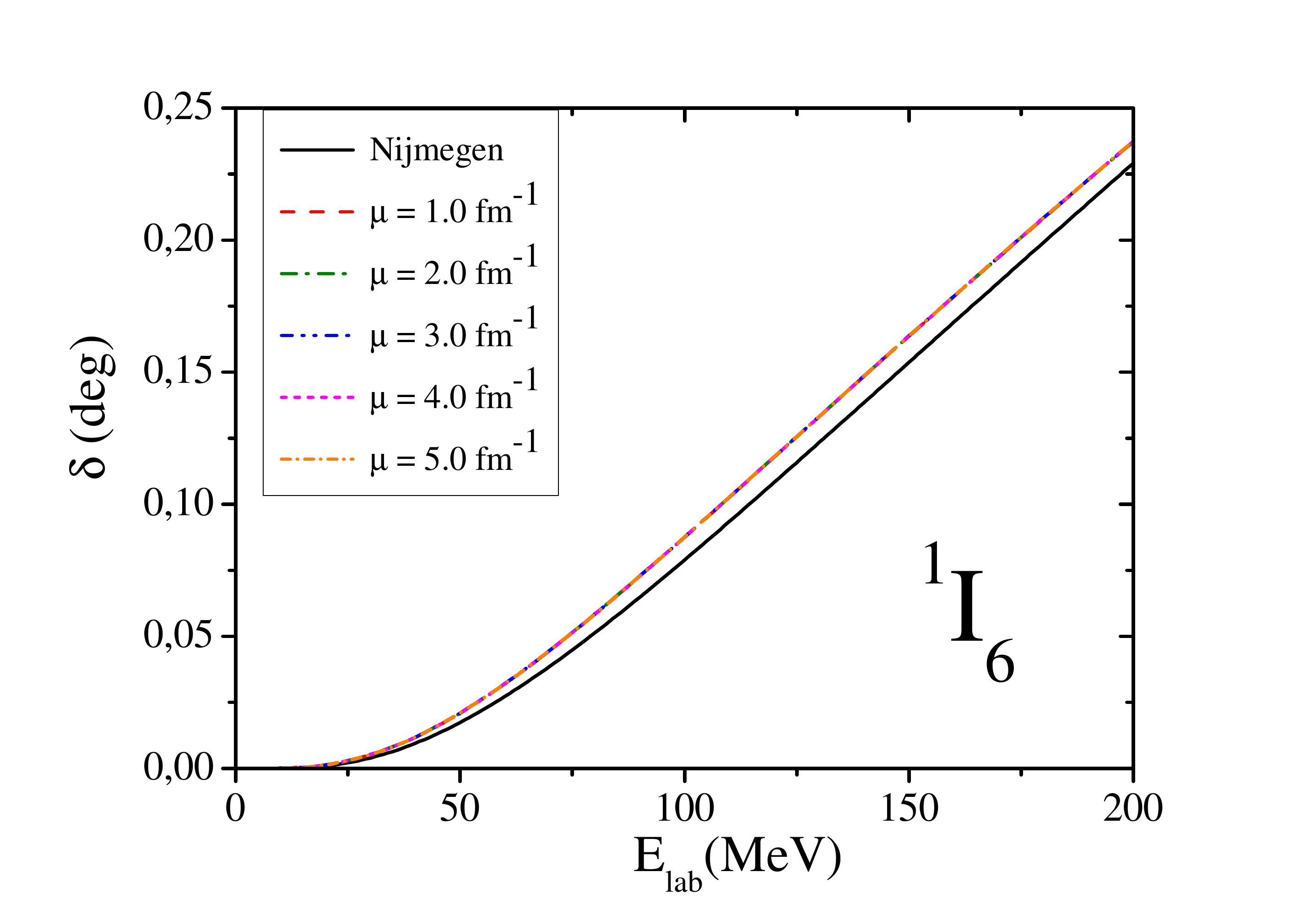}\hspace*{0.1cm}\includegraphics[scale=0.2]{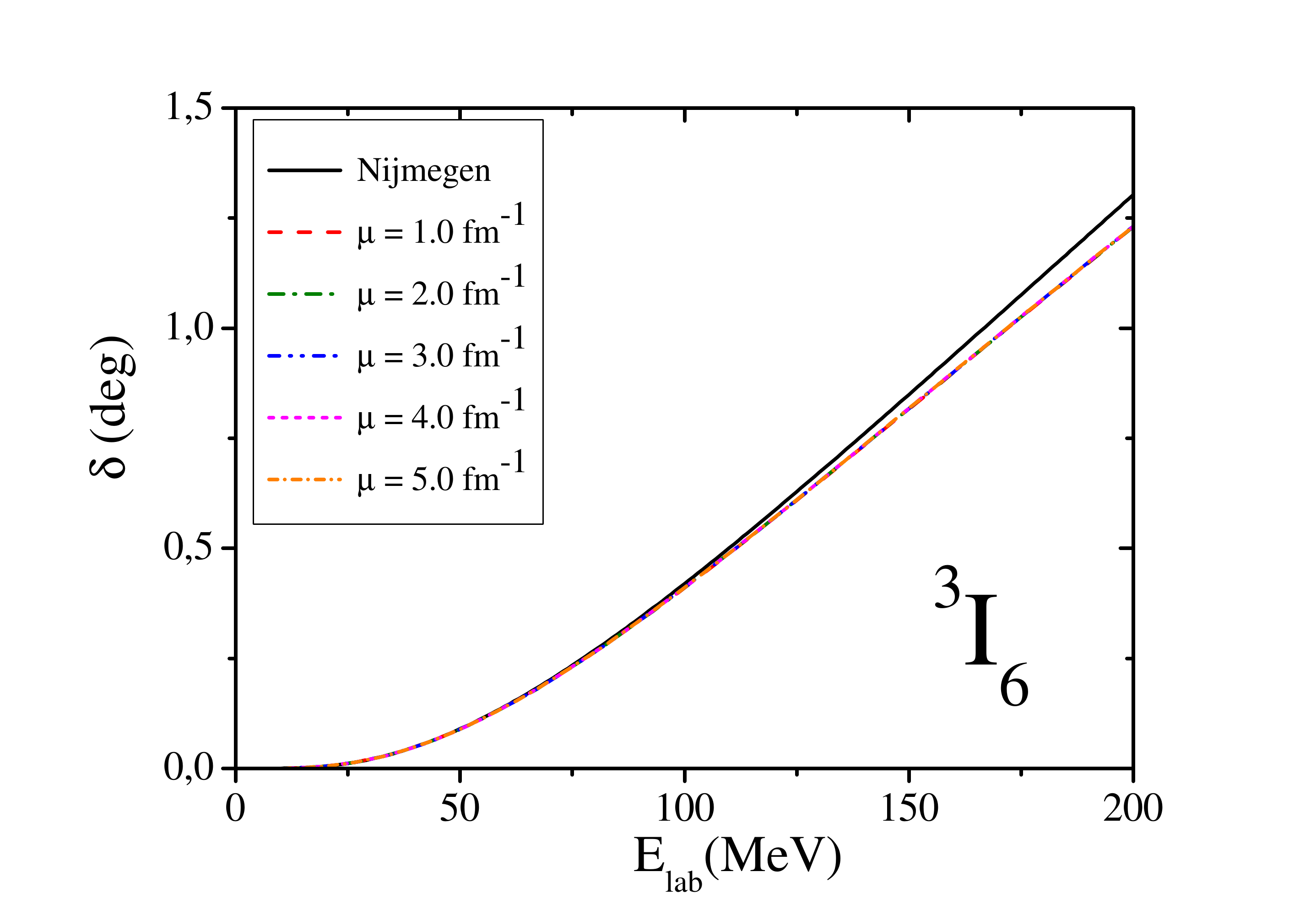} \\
\includegraphics[scale=0.2]{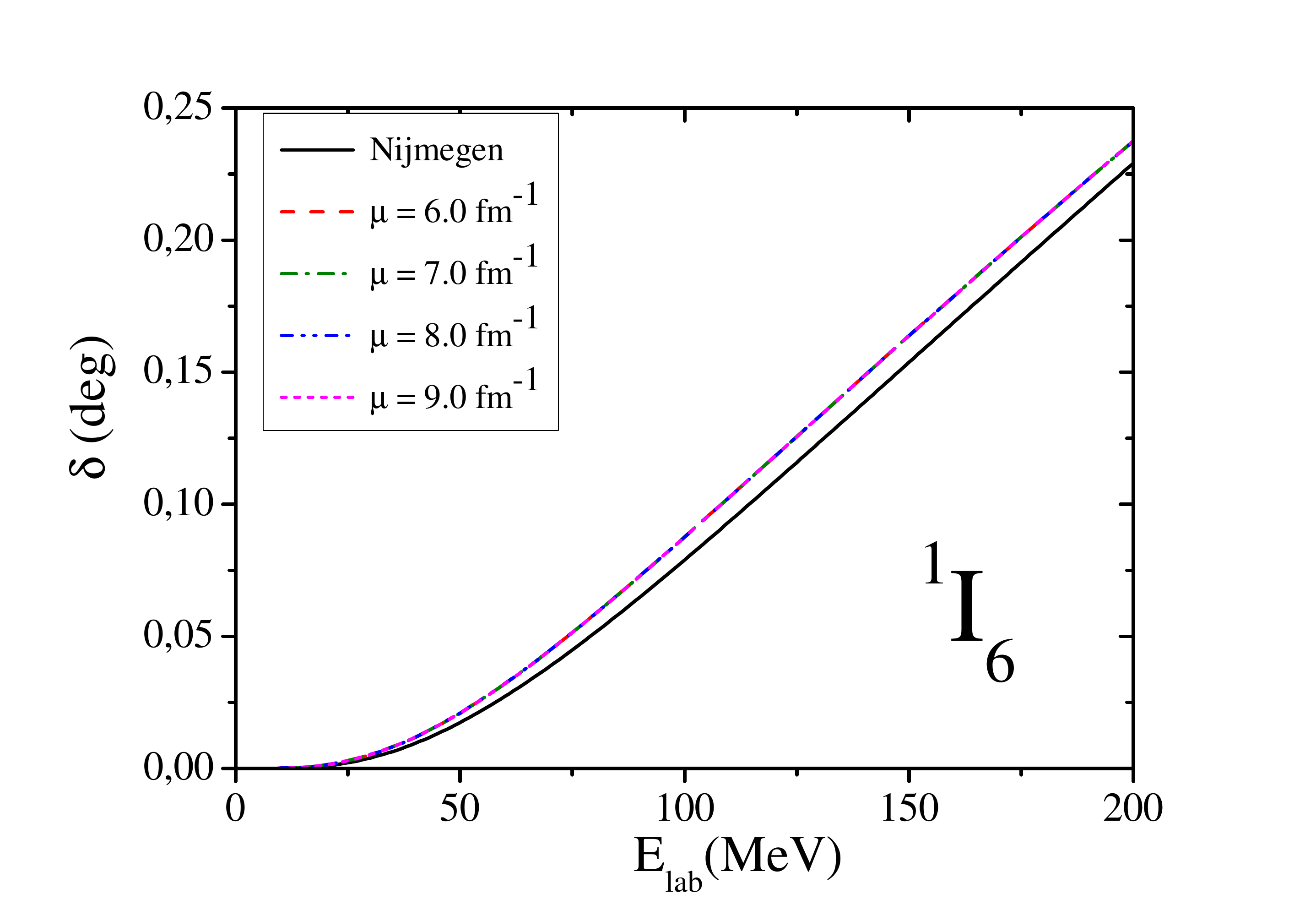}\hspace*{0.1cm}\includegraphics[scale=0.2]{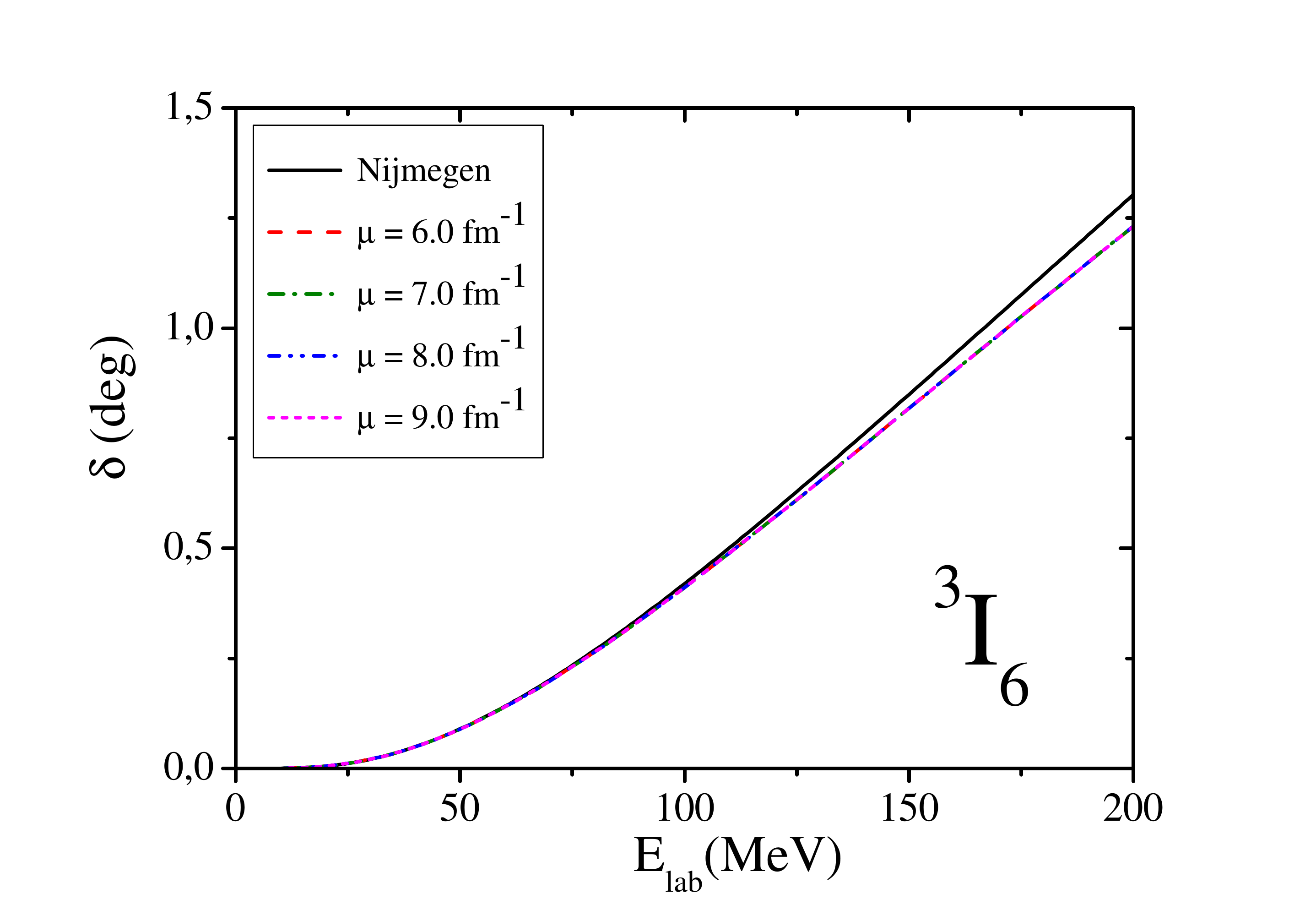}  
\end{center}
\caption{(Color on-line) Phase-shifts in the $^1I_6$ and $^3I_6$ uncoupled channels calculated from the solution of the subtracted LS equation for the $K$-matrix with five subtractions for the N3LO-EM potential for several values of the renormalization scale compared to the Nijmegen partial wave analysis.}
\label{fig4}
\end{figure}

\begin{figure}[t]
\begin{center}
\includegraphics[scale=0.17]{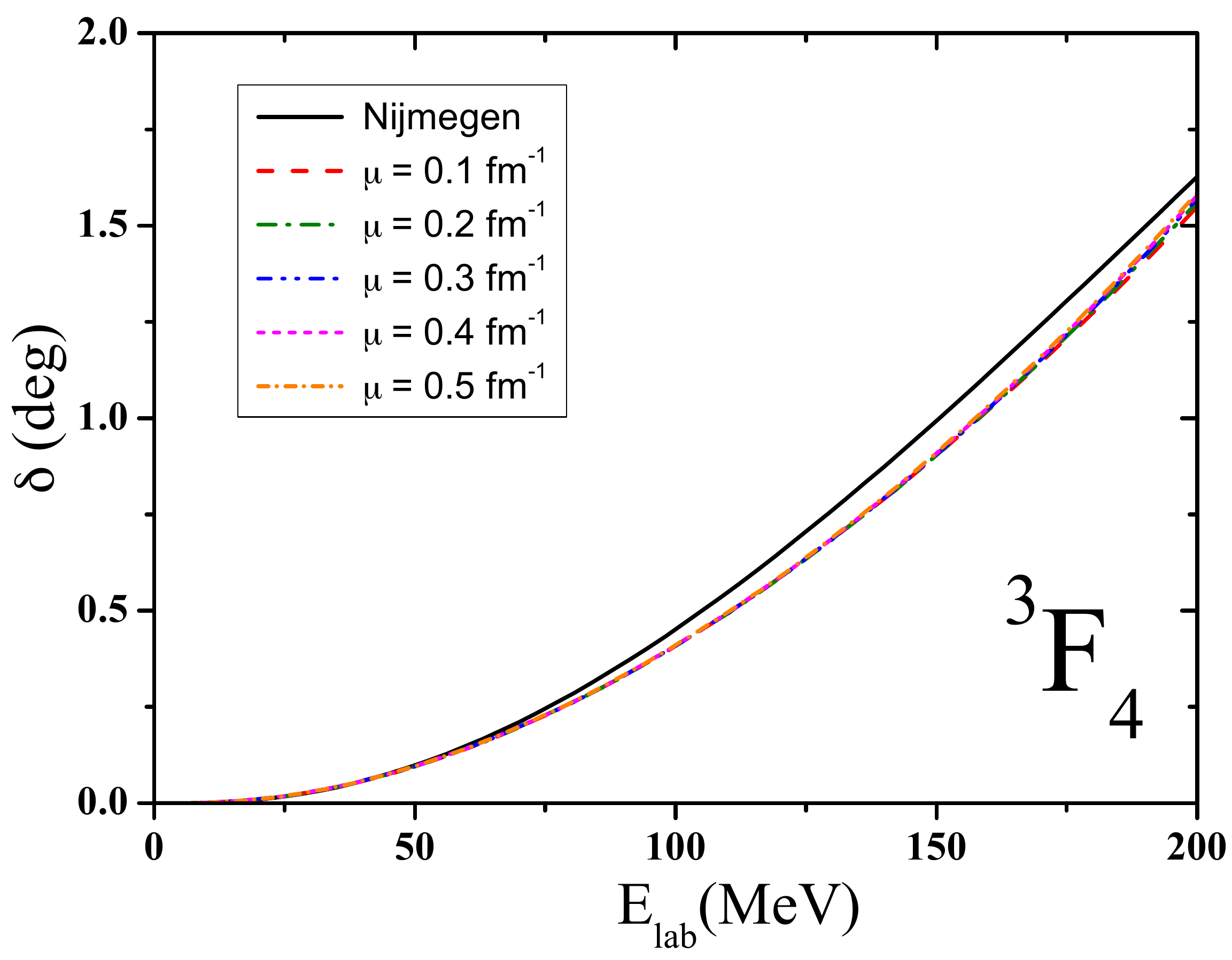}\includegraphics[scale=0.17]{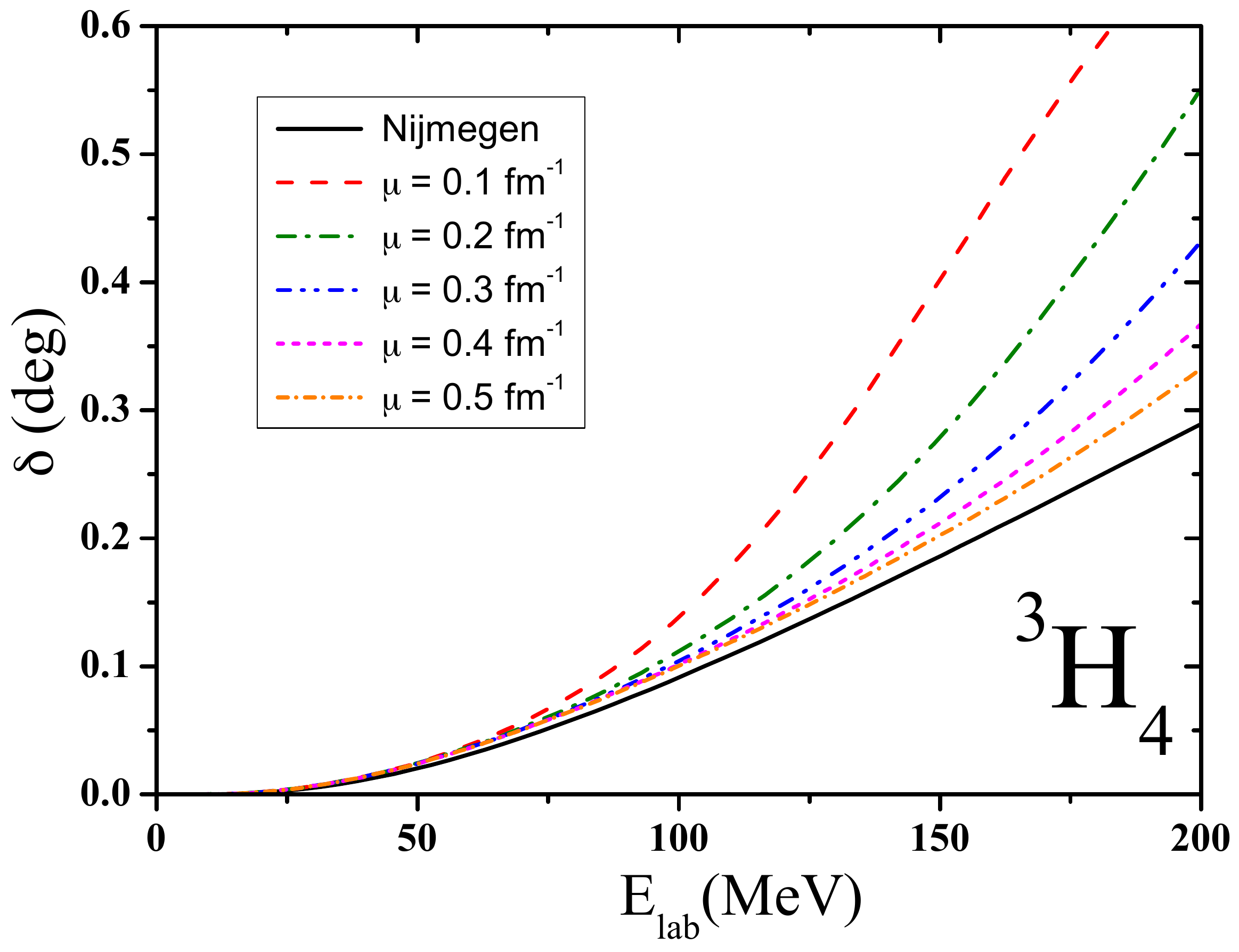}\includegraphics[scale=0.17]{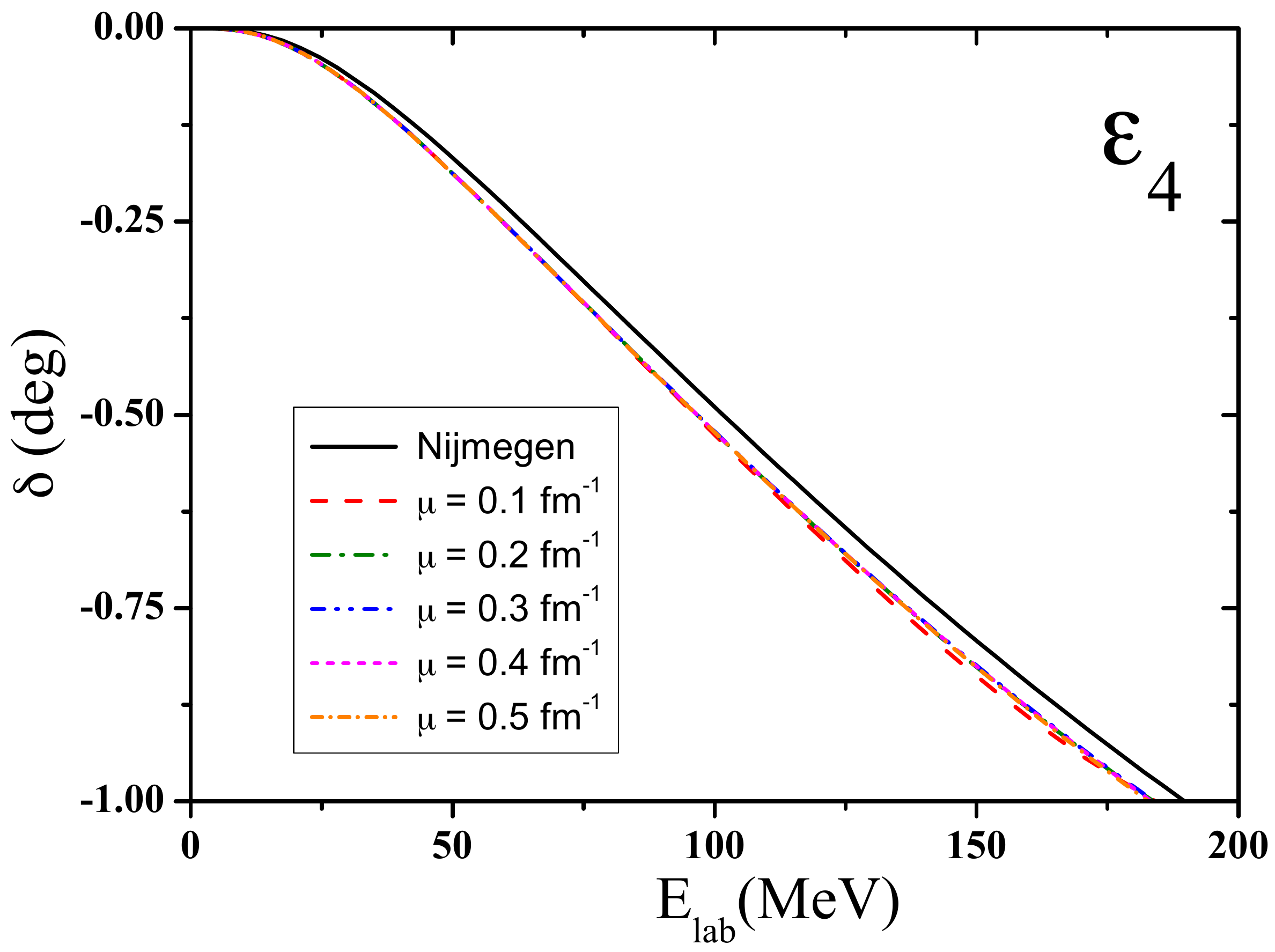} \\ \vspace*{1cm}
\includegraphics[scale=0.17]{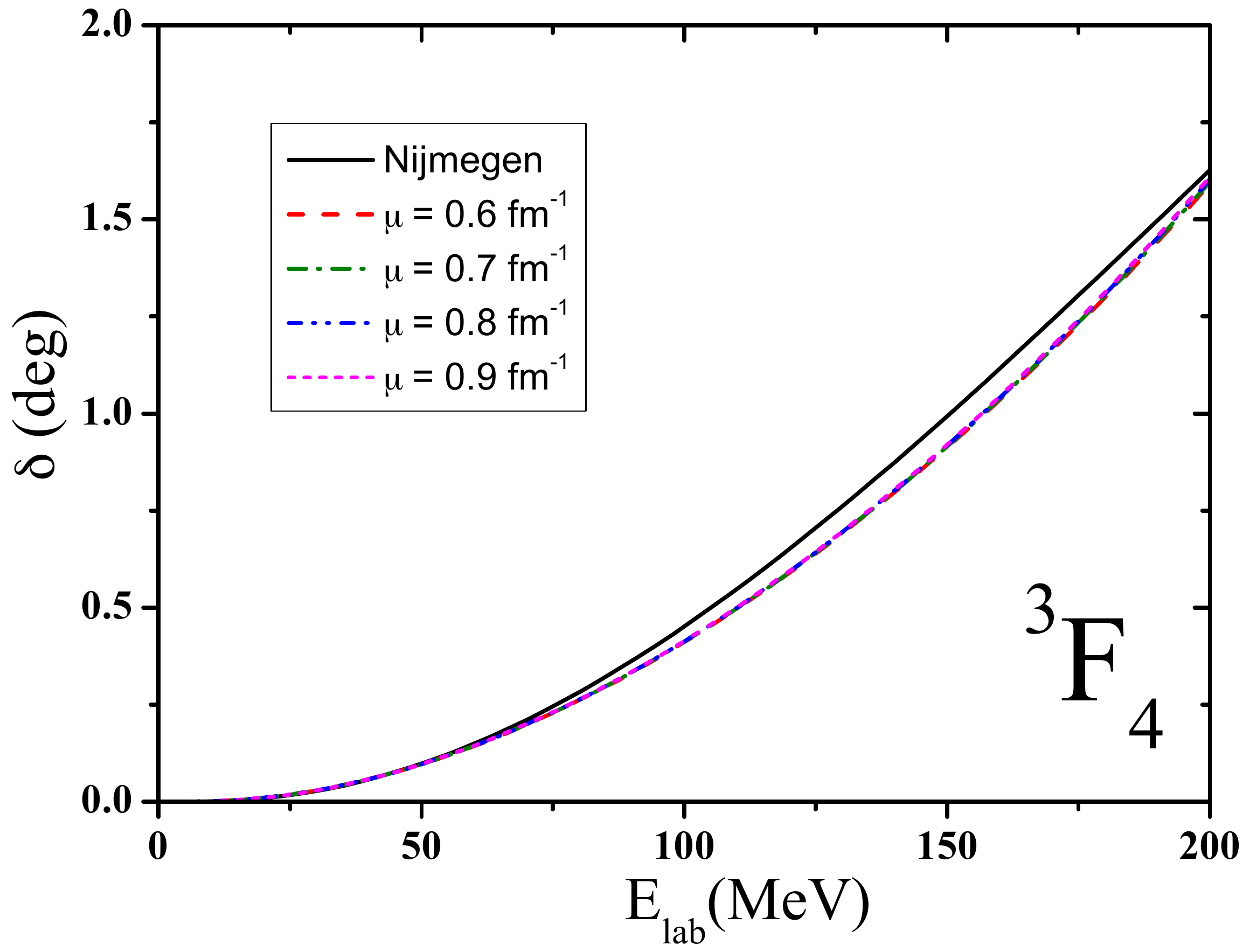}\includegraphics[scale=0.17]{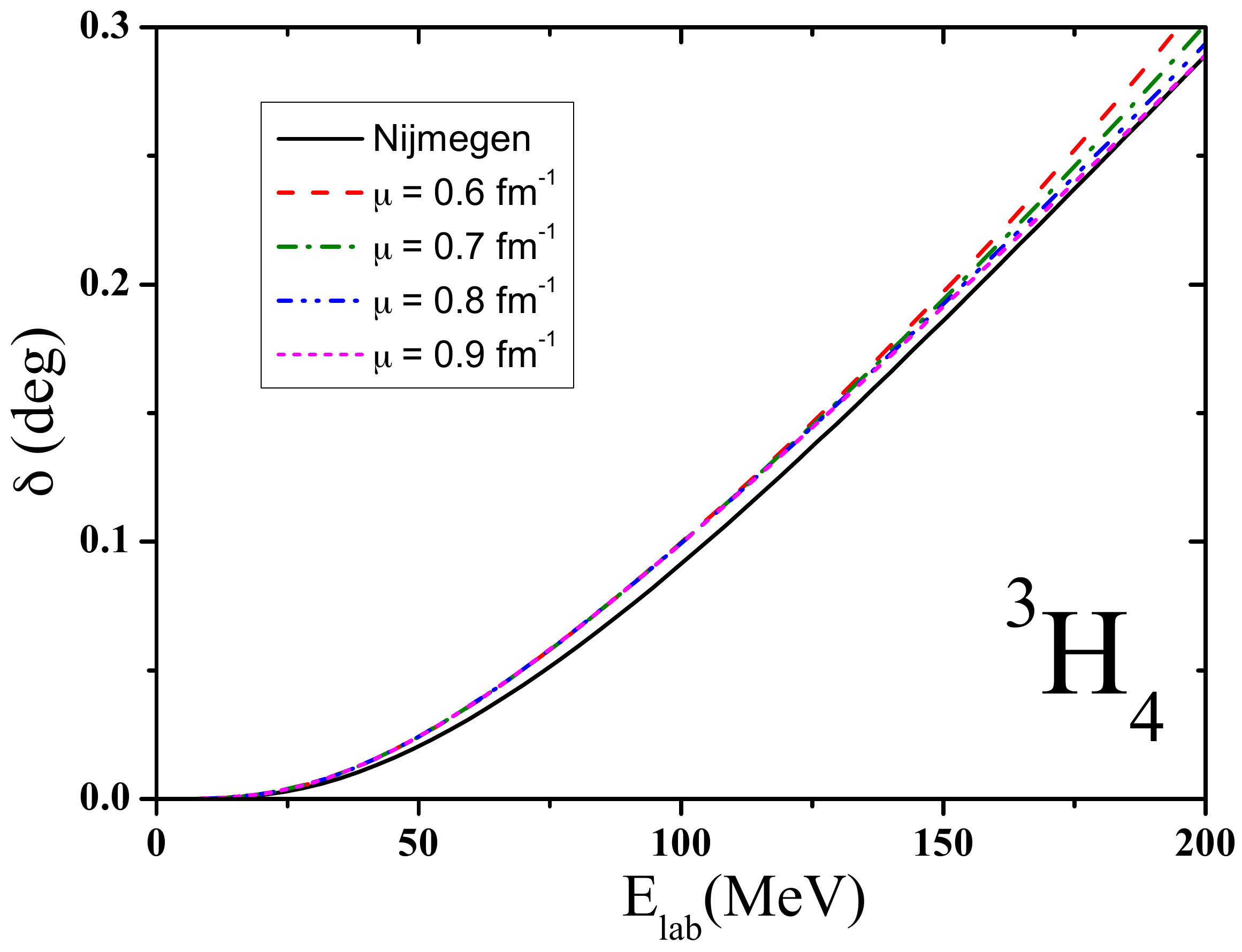}\includegraphics[scale=0.17]{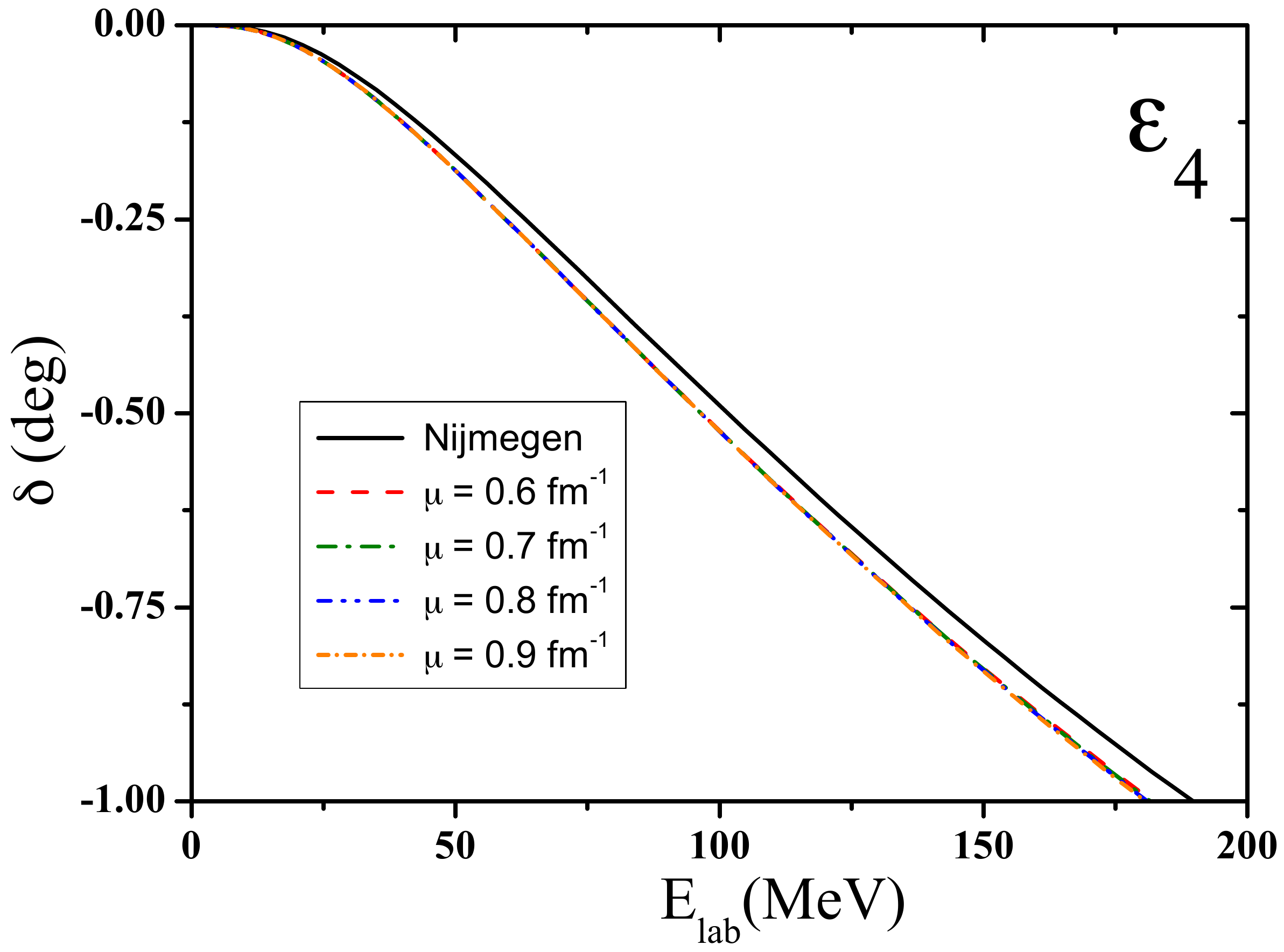}\\ \vspace*{1cm}
\includegraphics[scale=0.17]{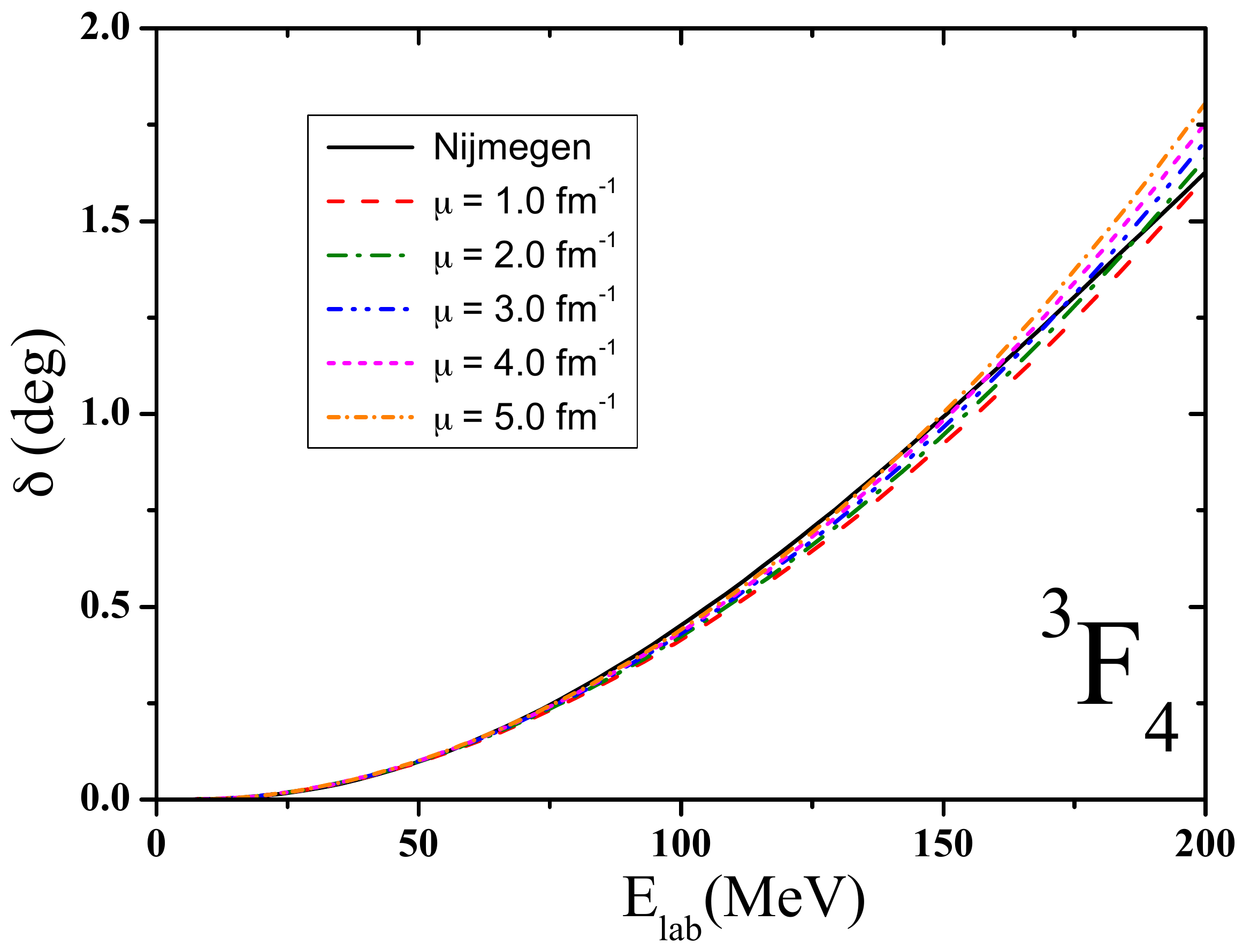}\includegraphics[scale=0.17]{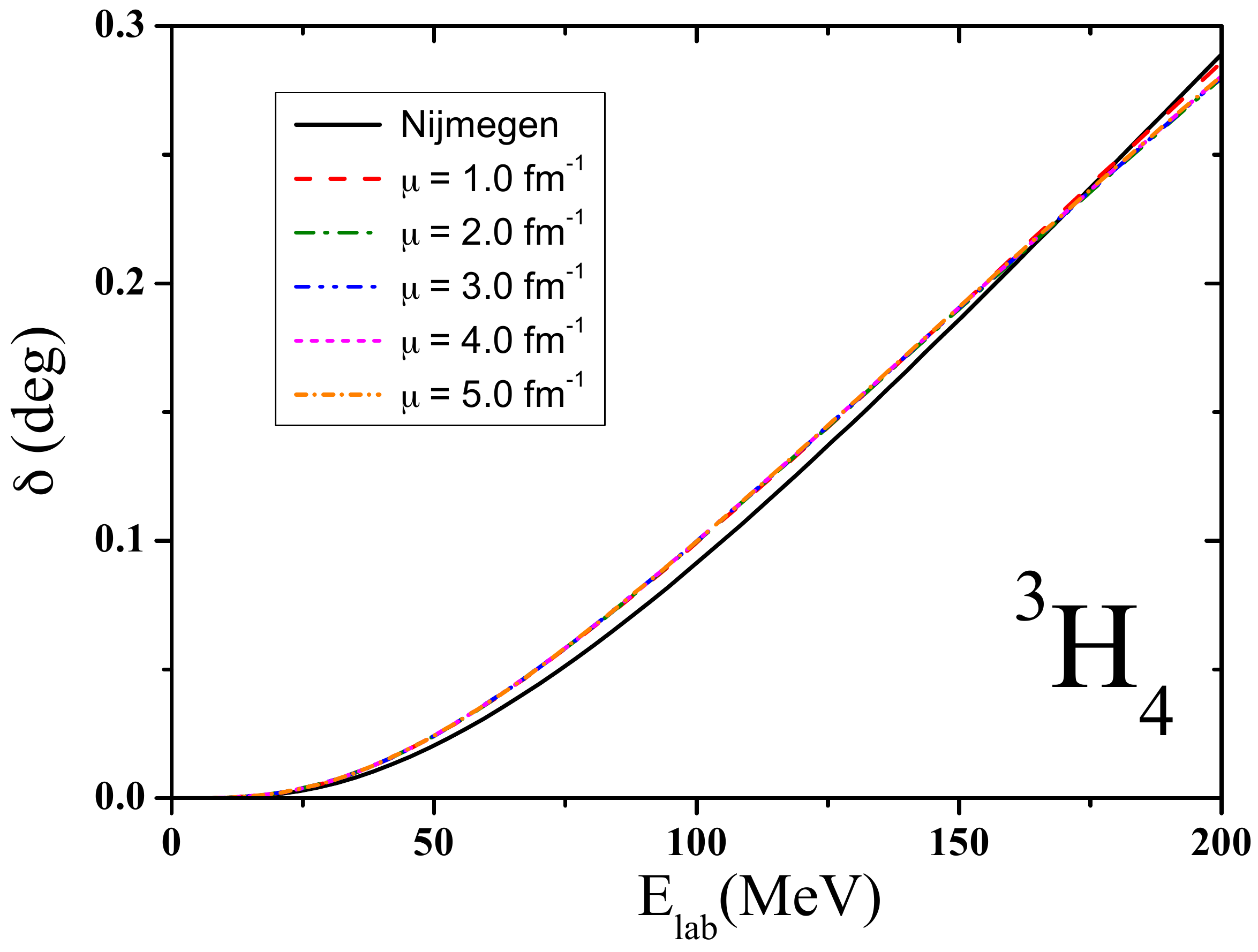}\includegraphics[scale=0.17]{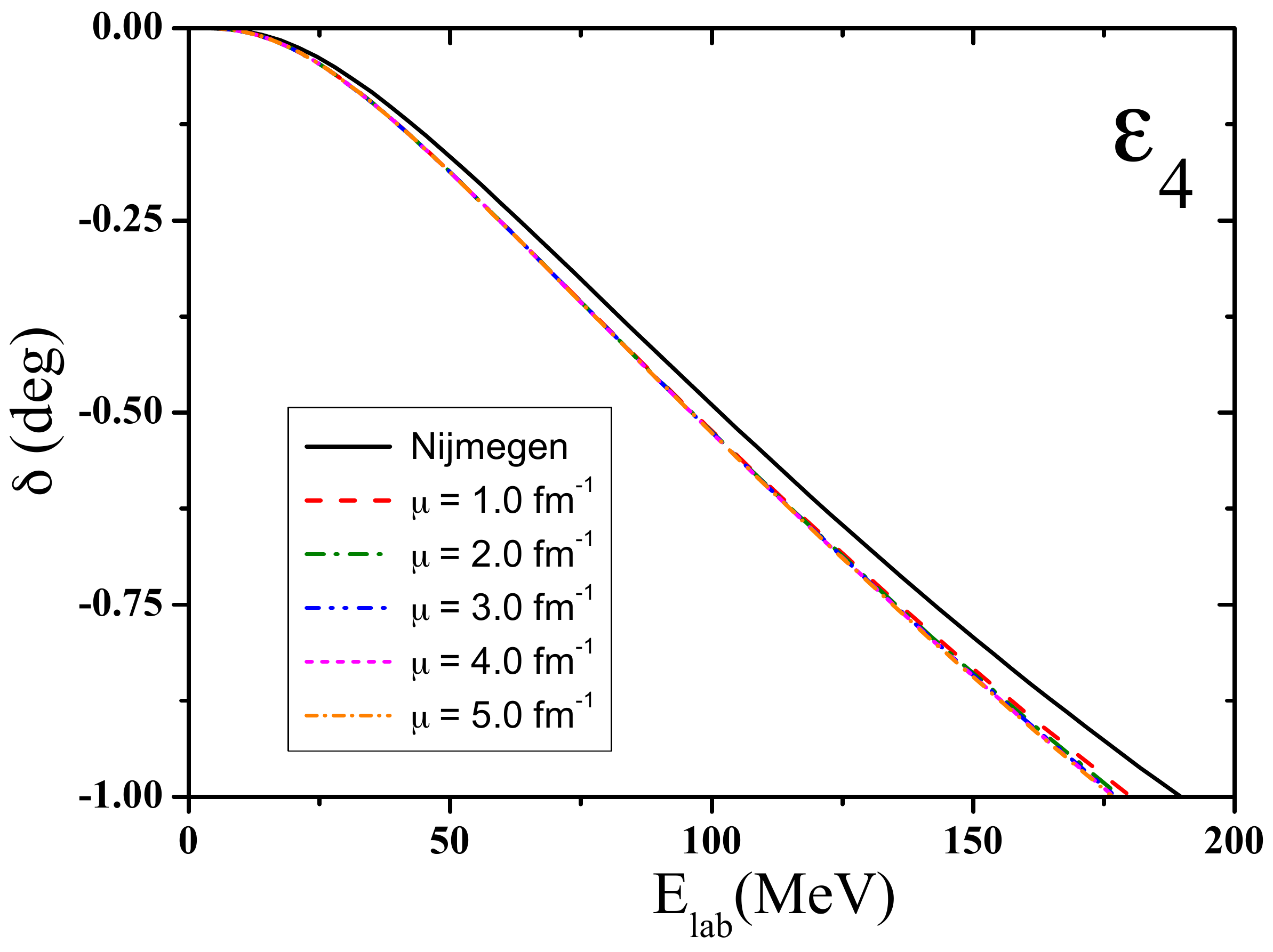}\\ \vspace*{1cm}
\includegraphics[scale=0.17]{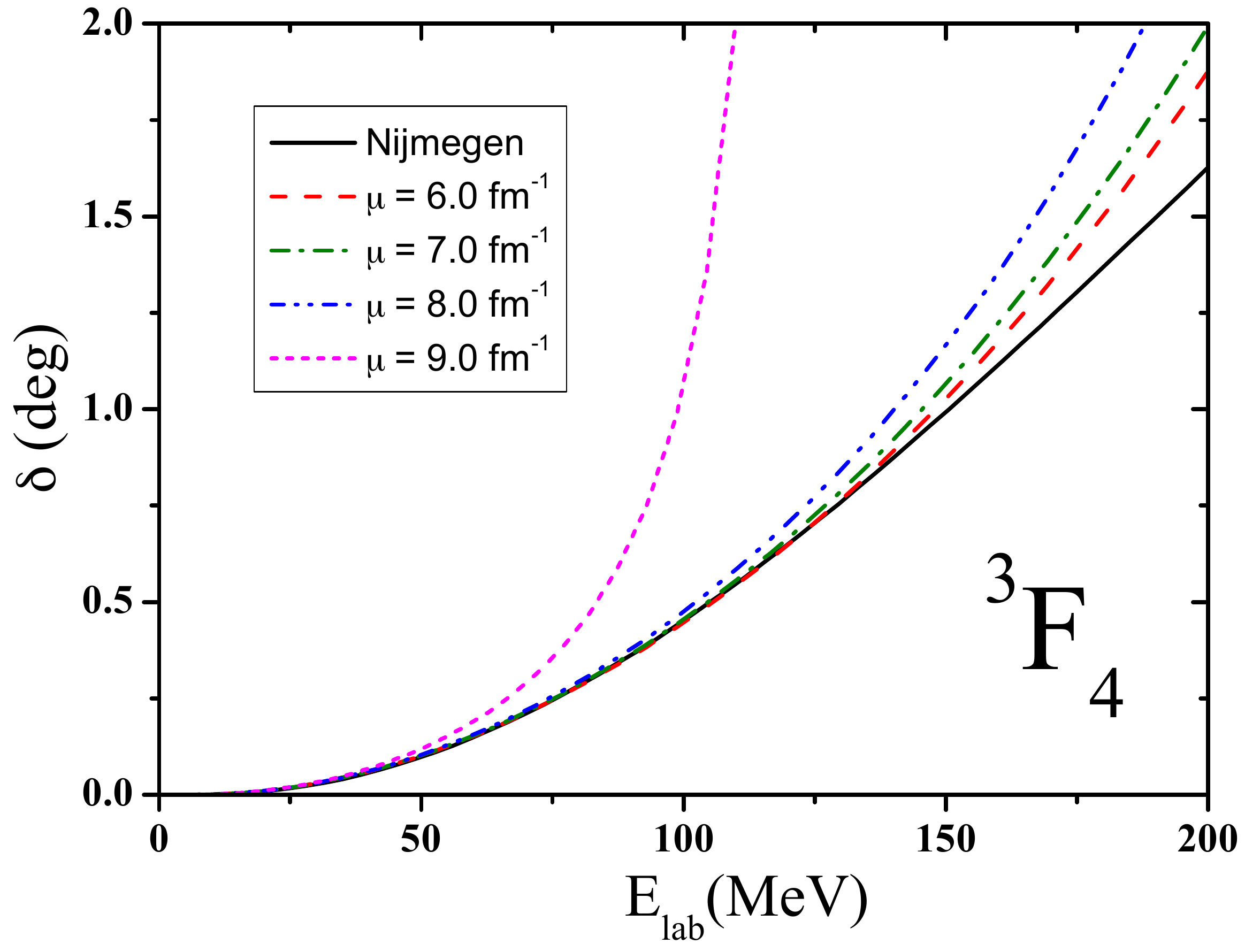}\includegraphics[scale=0.17]{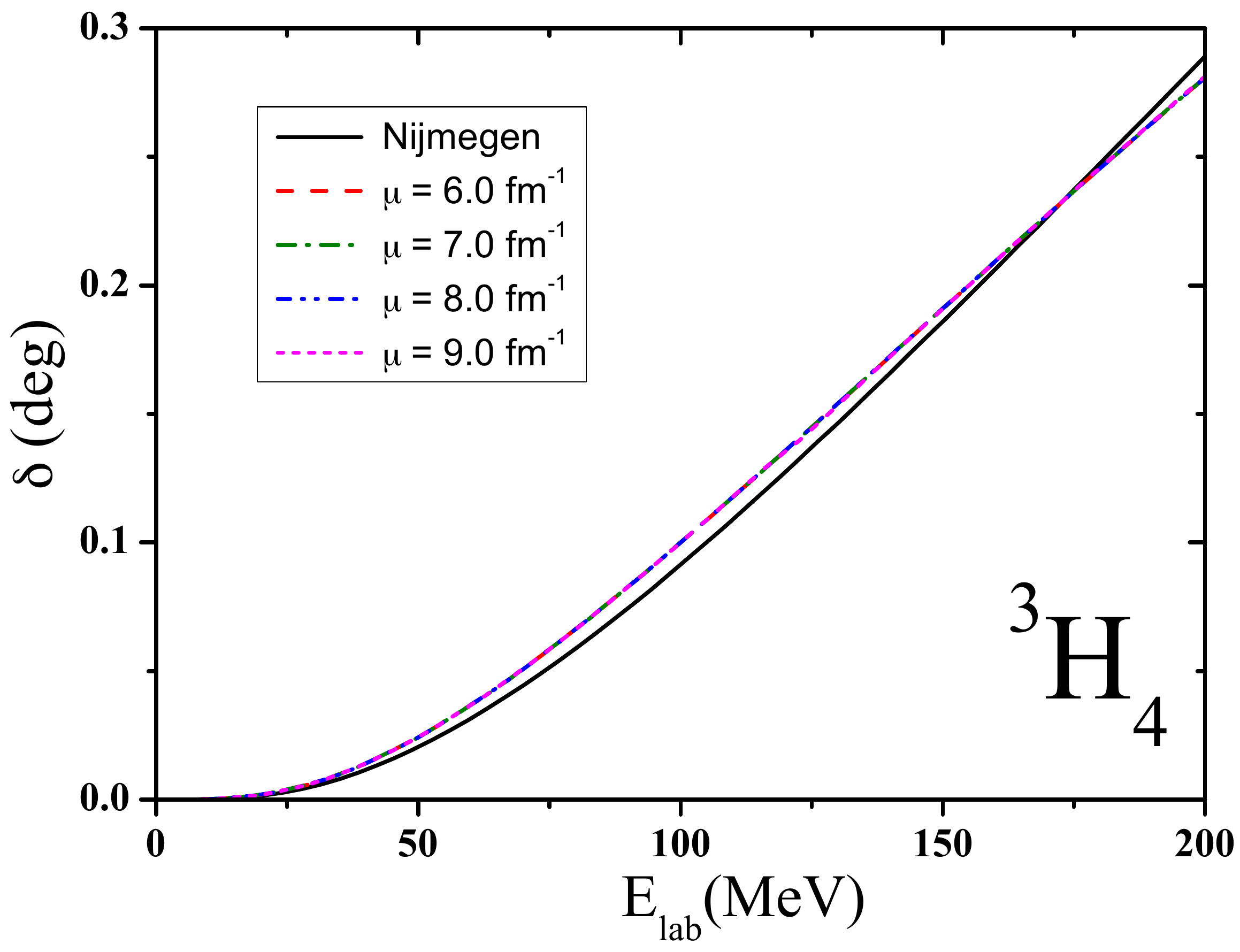}\includegraphics[scale=0.17]{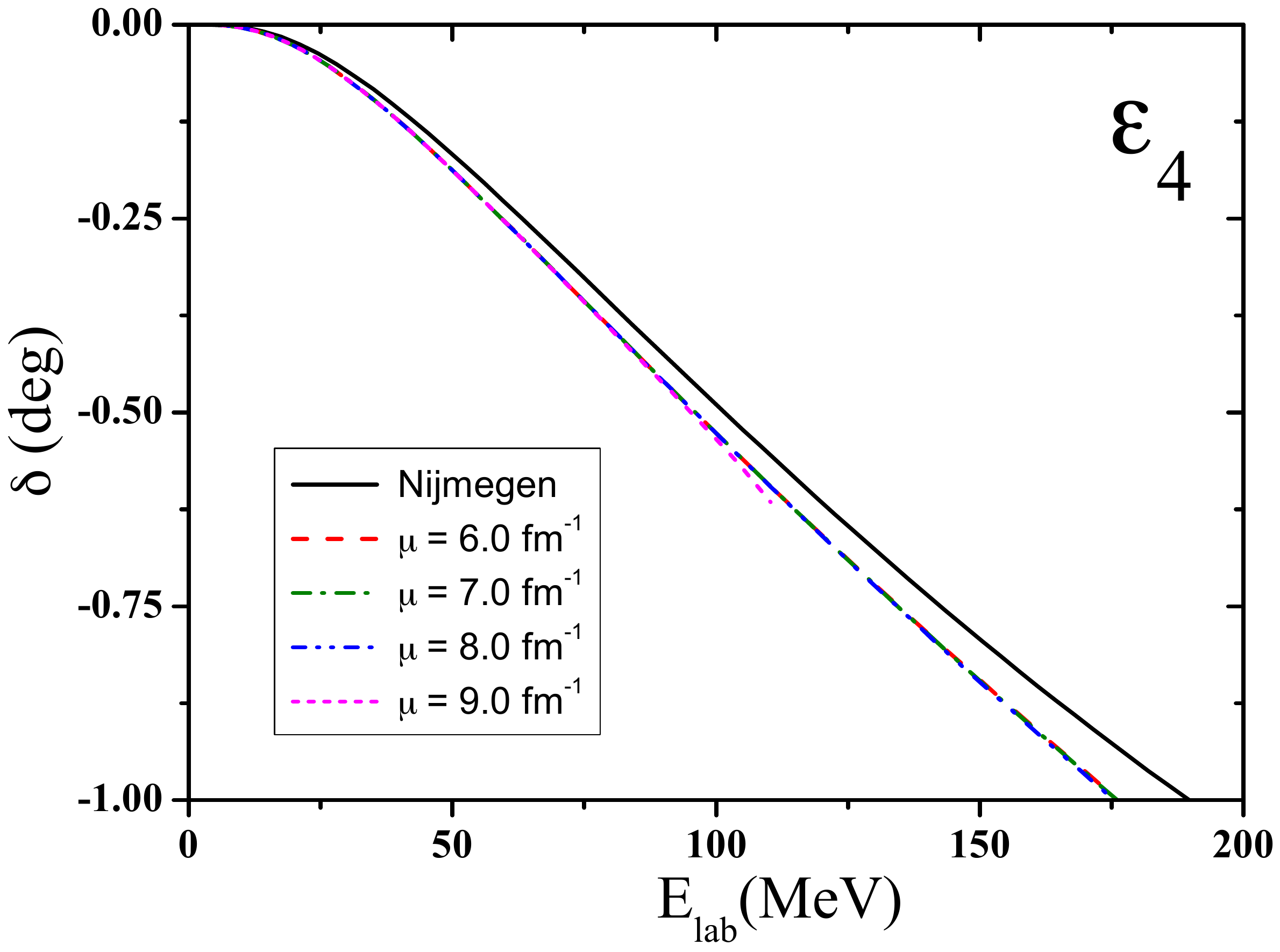}
\end{center}
\caption{(Color on-line) Phase-shifts in the $^3F_4 - ^3H_4$ coupled channels calculated from the solution of the subtracted LS equation for the $K$-matrix with five subtractions for the N3LO-EM potential for several values of the renormalization scale compared to the Nijmegen partial wave analysis.}
\label{fig5}
\end{figure}

\begin{figure}[t]
\begin{center}
\includegraphics[scale=0.17]{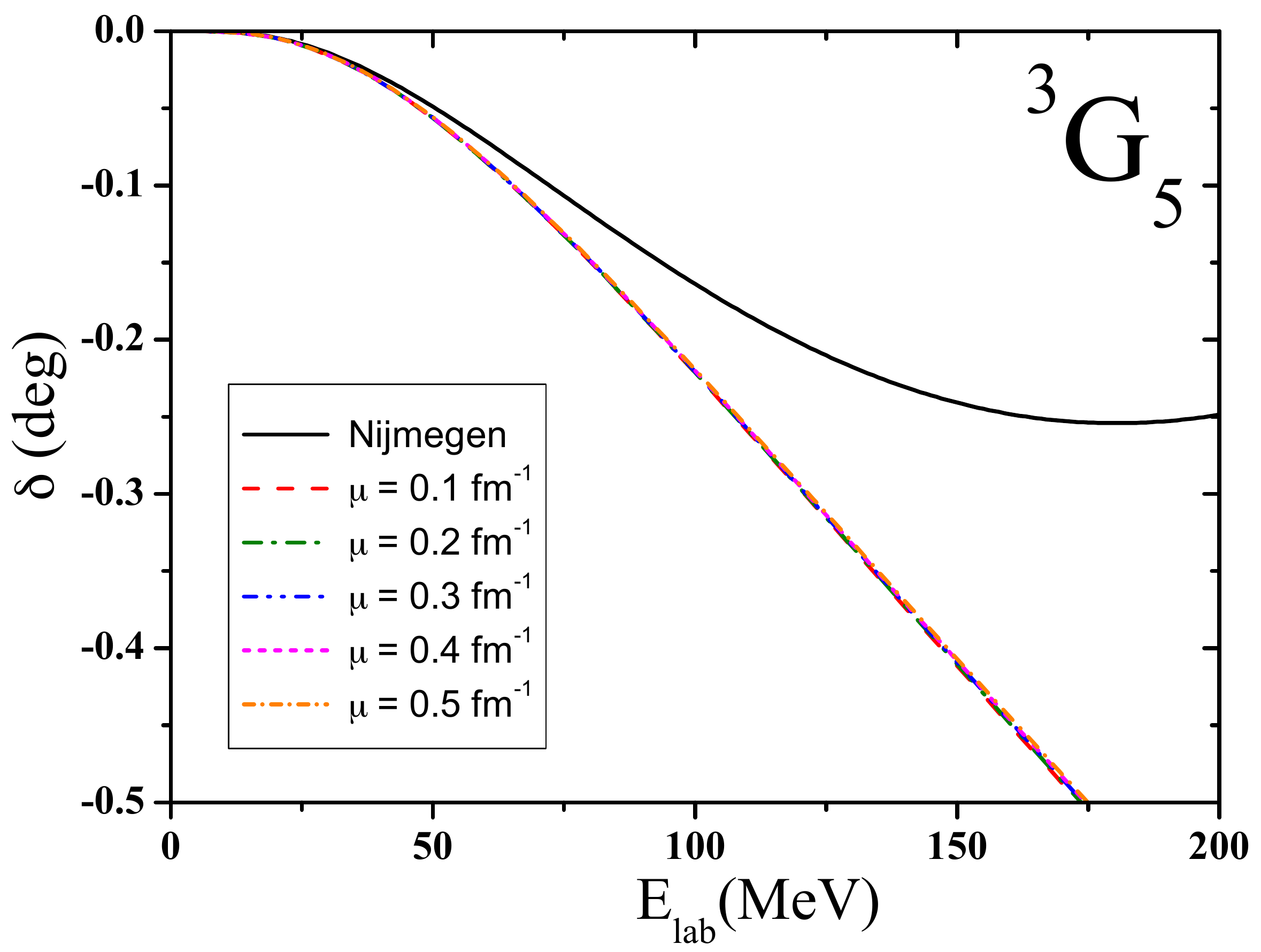}\includegraphics[scale=0.17]{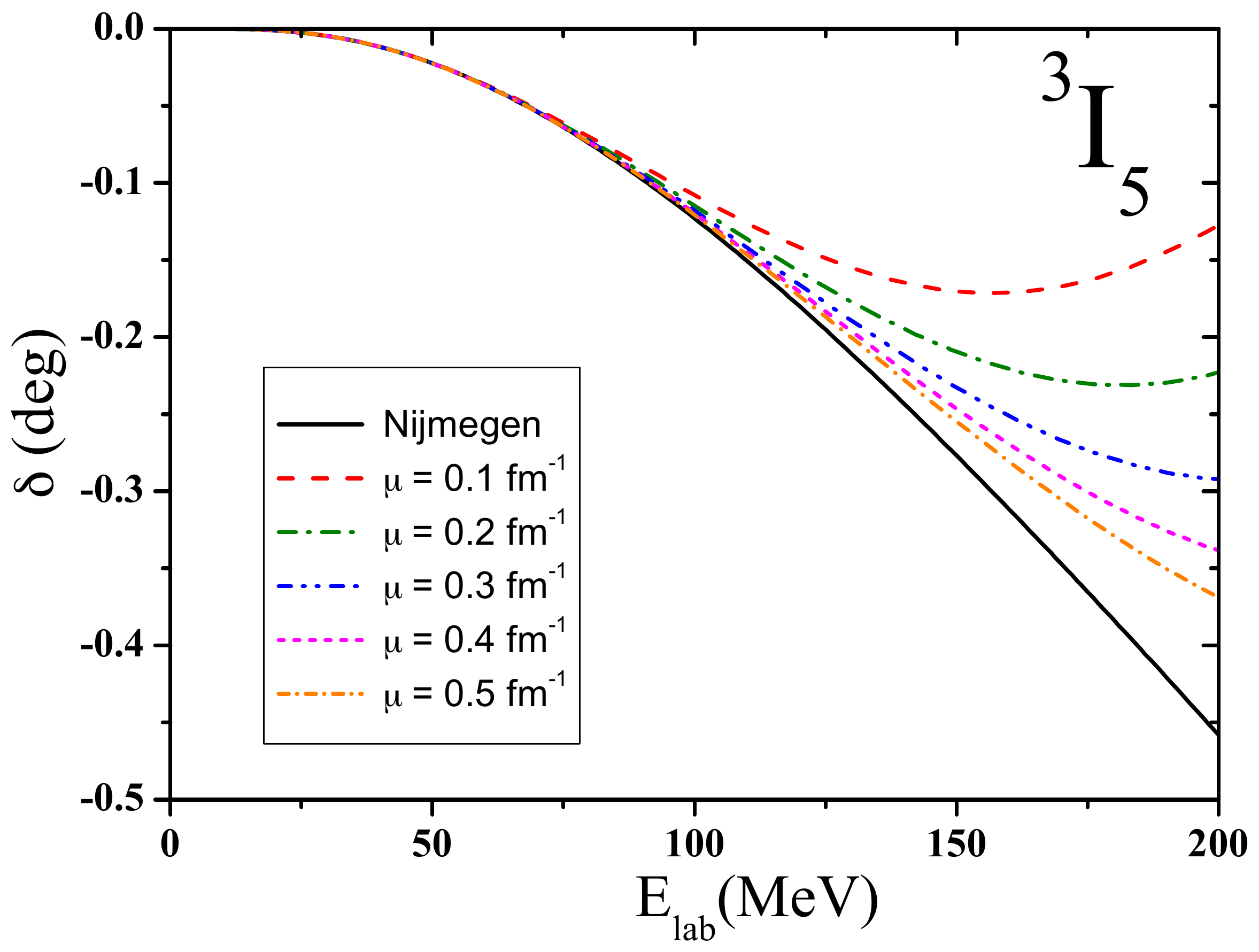}\includegraphics[scale=0.17]{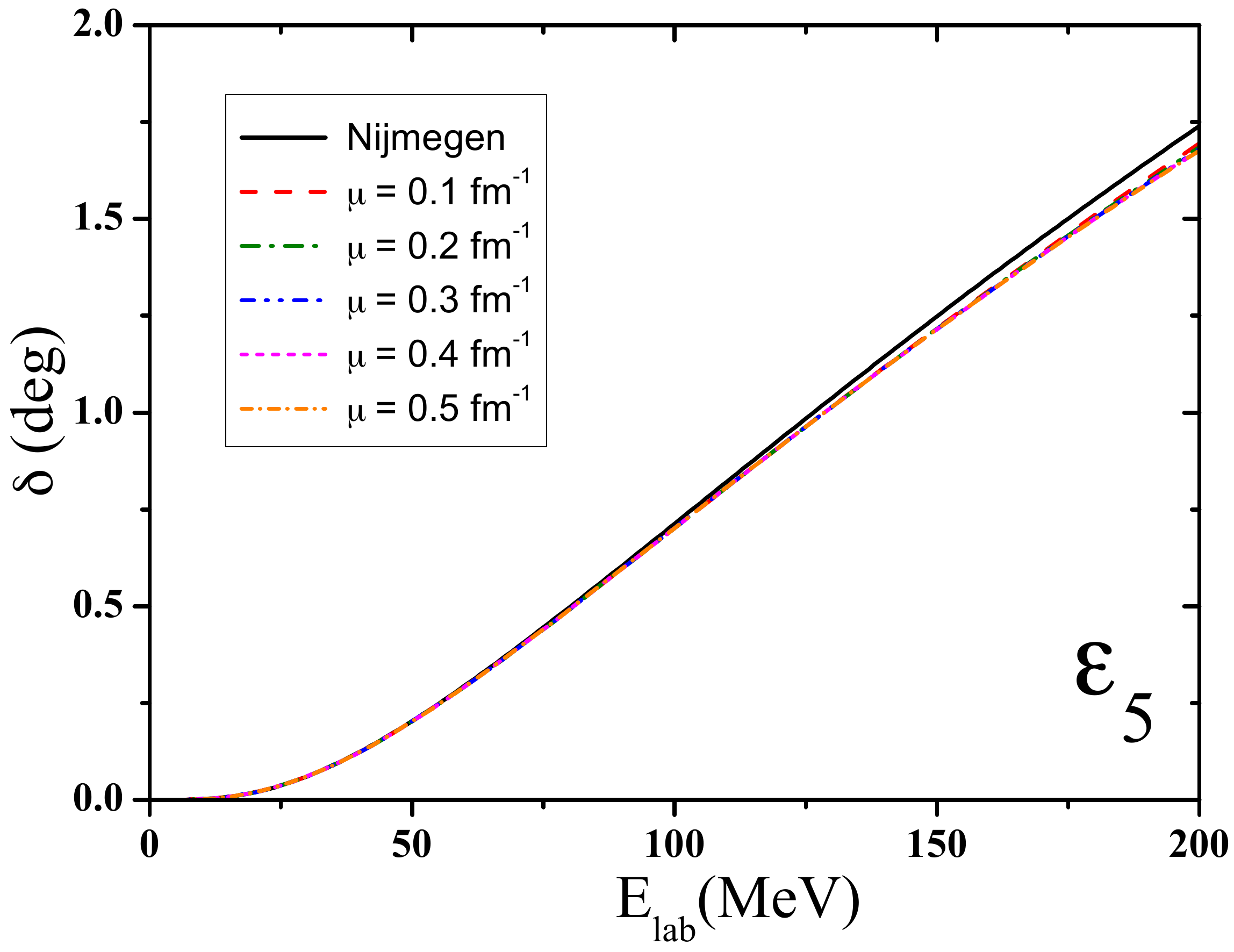} \\ \vspace*{1cm}
\includegraphics[scale=0.17]{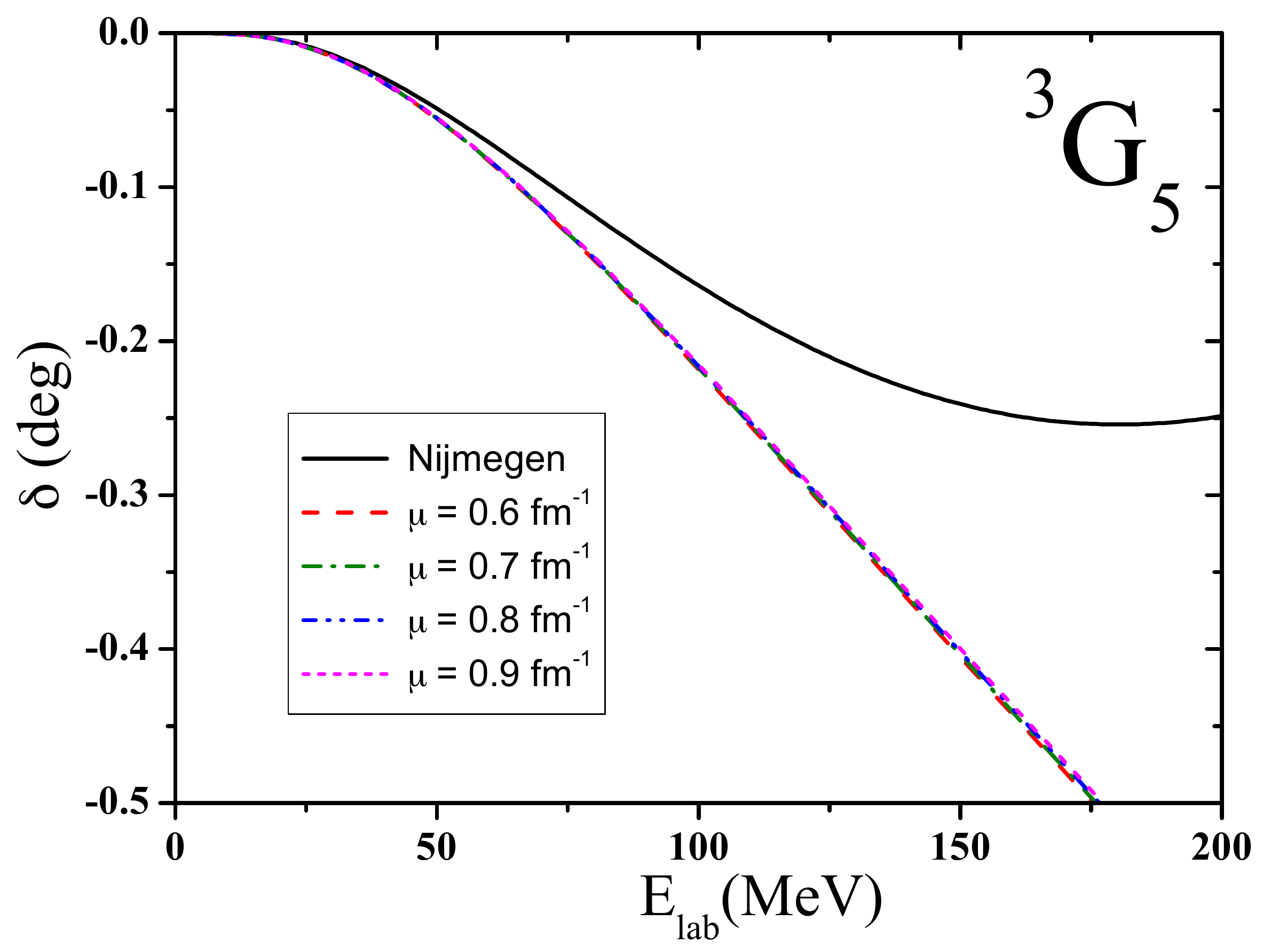}\includegraphics[scale=0.17]{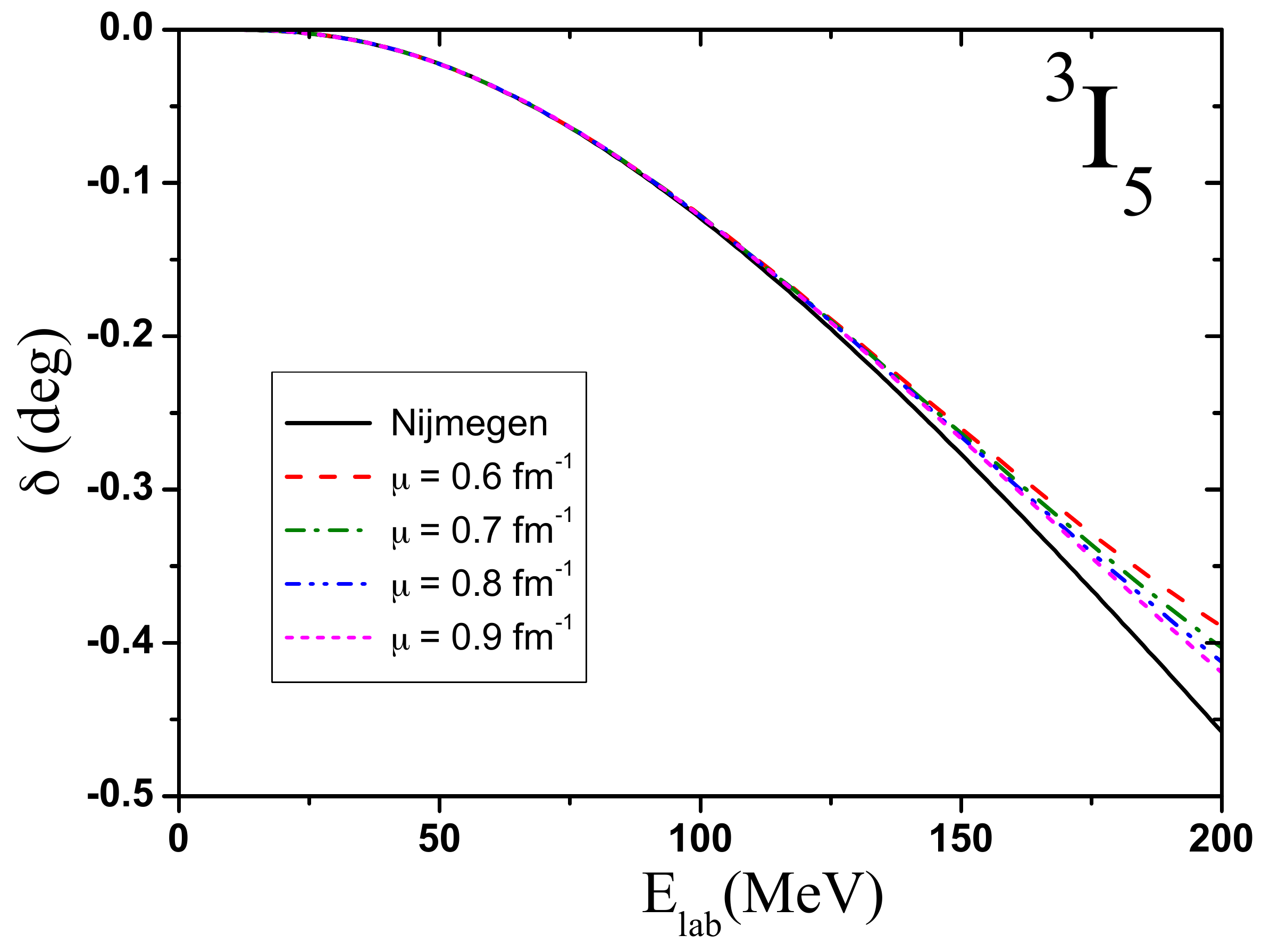}\includegraphics[scale=0.17]{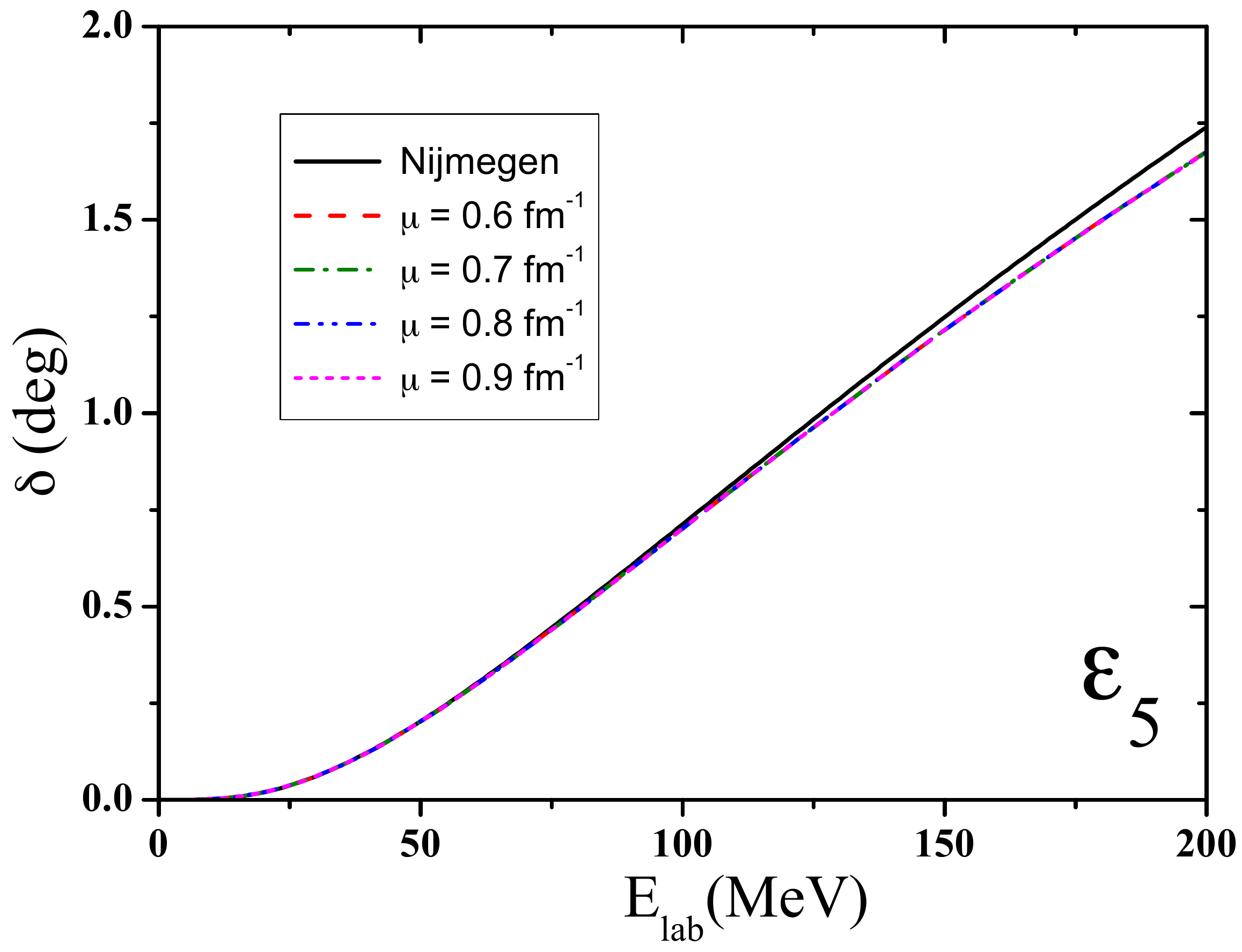}\\ \vspace*{1cm}
\includegraphics[scale=0.17]{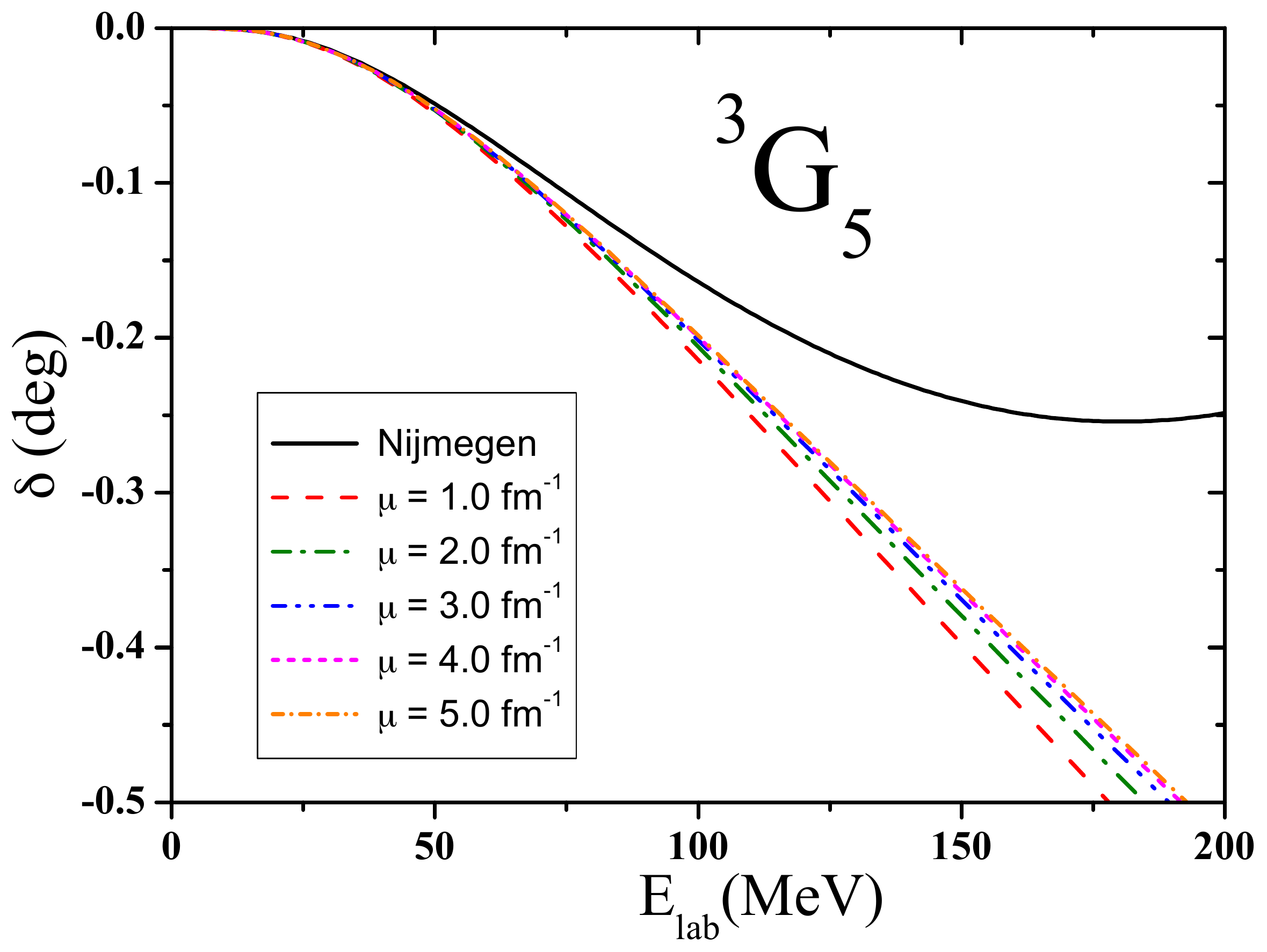}\includegraphics[scale=0.17]{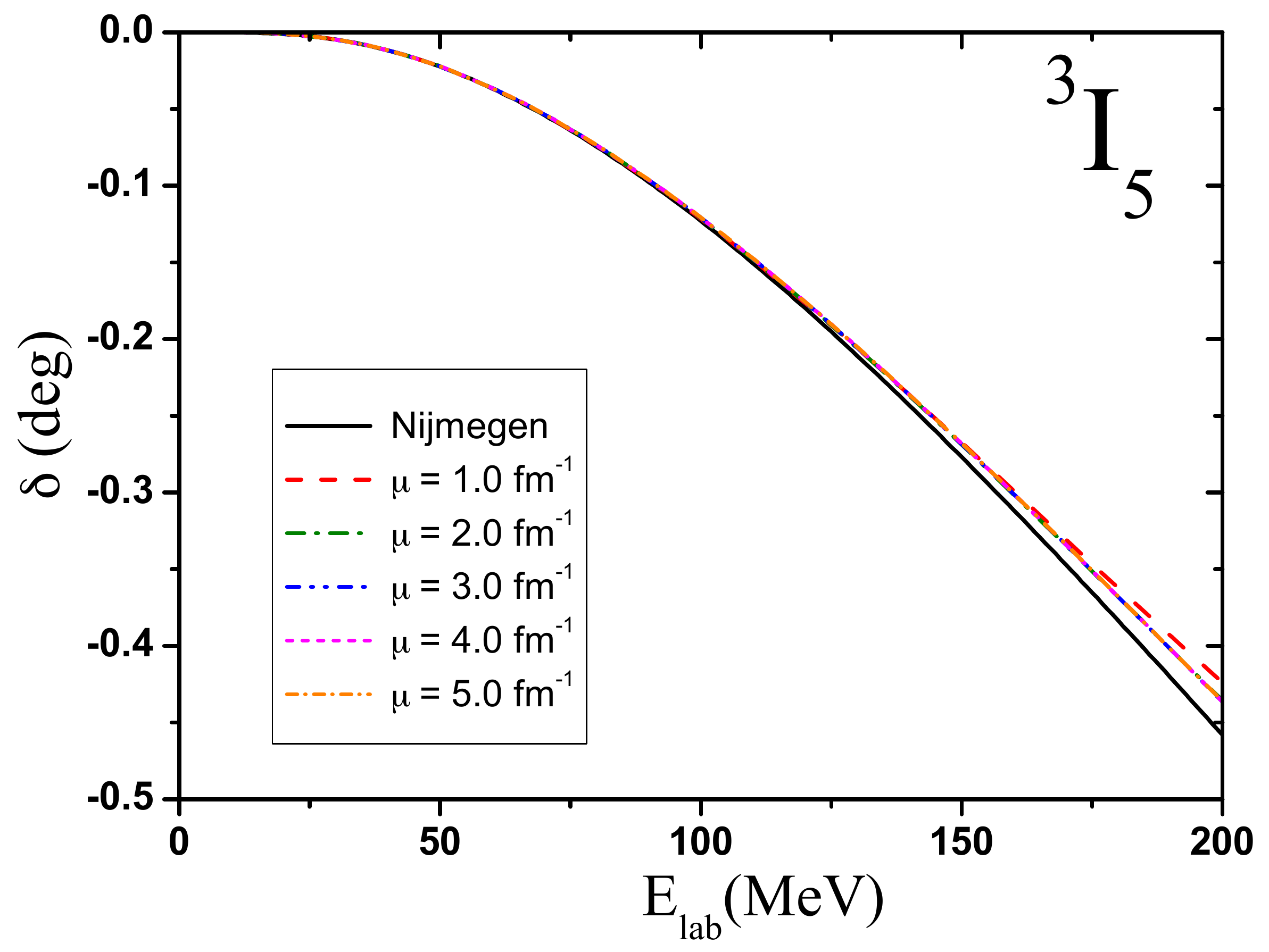}\includegraphics[scale=0.17]{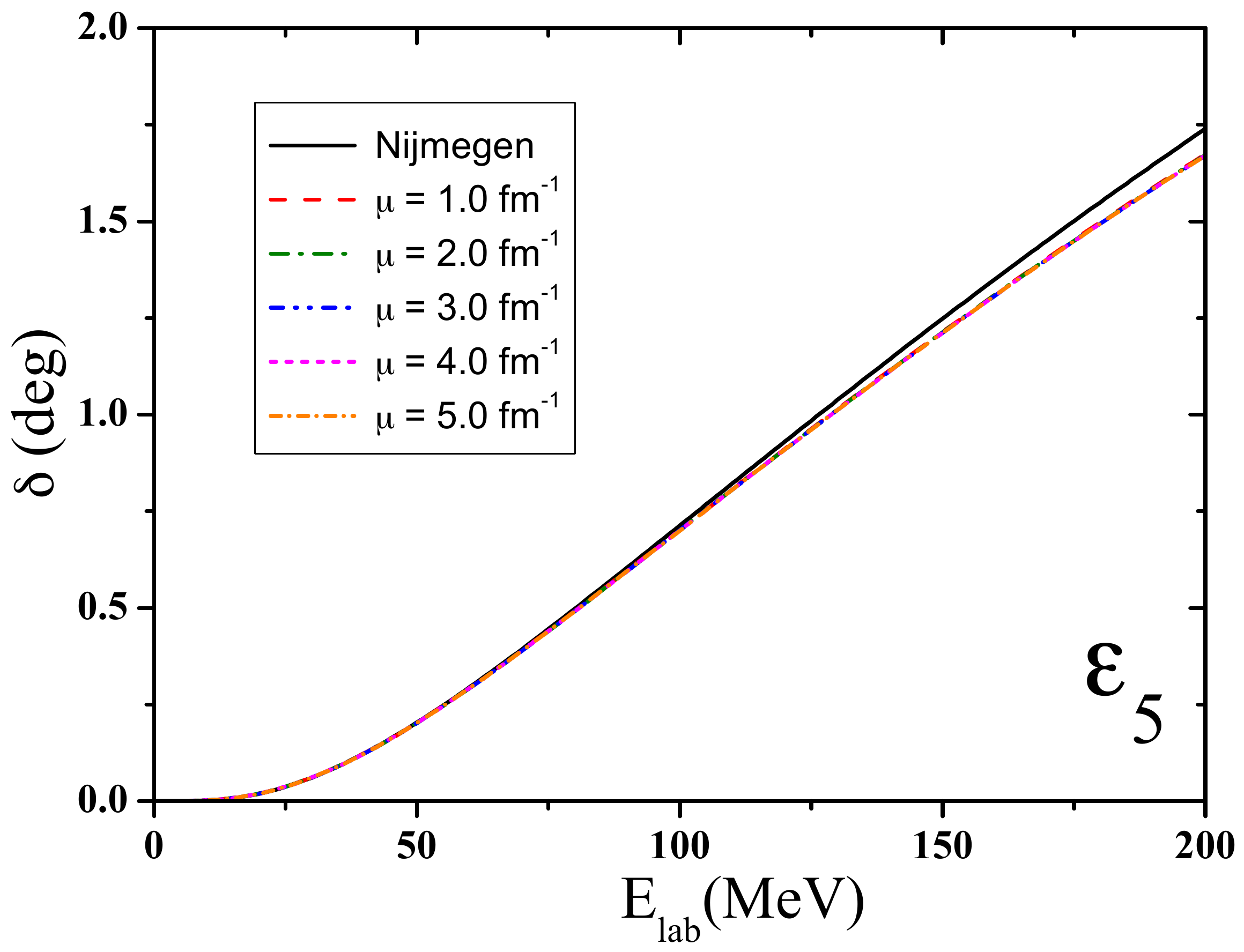}\\ \vspace*{1cm}
\includegraphics[scale=0.17]{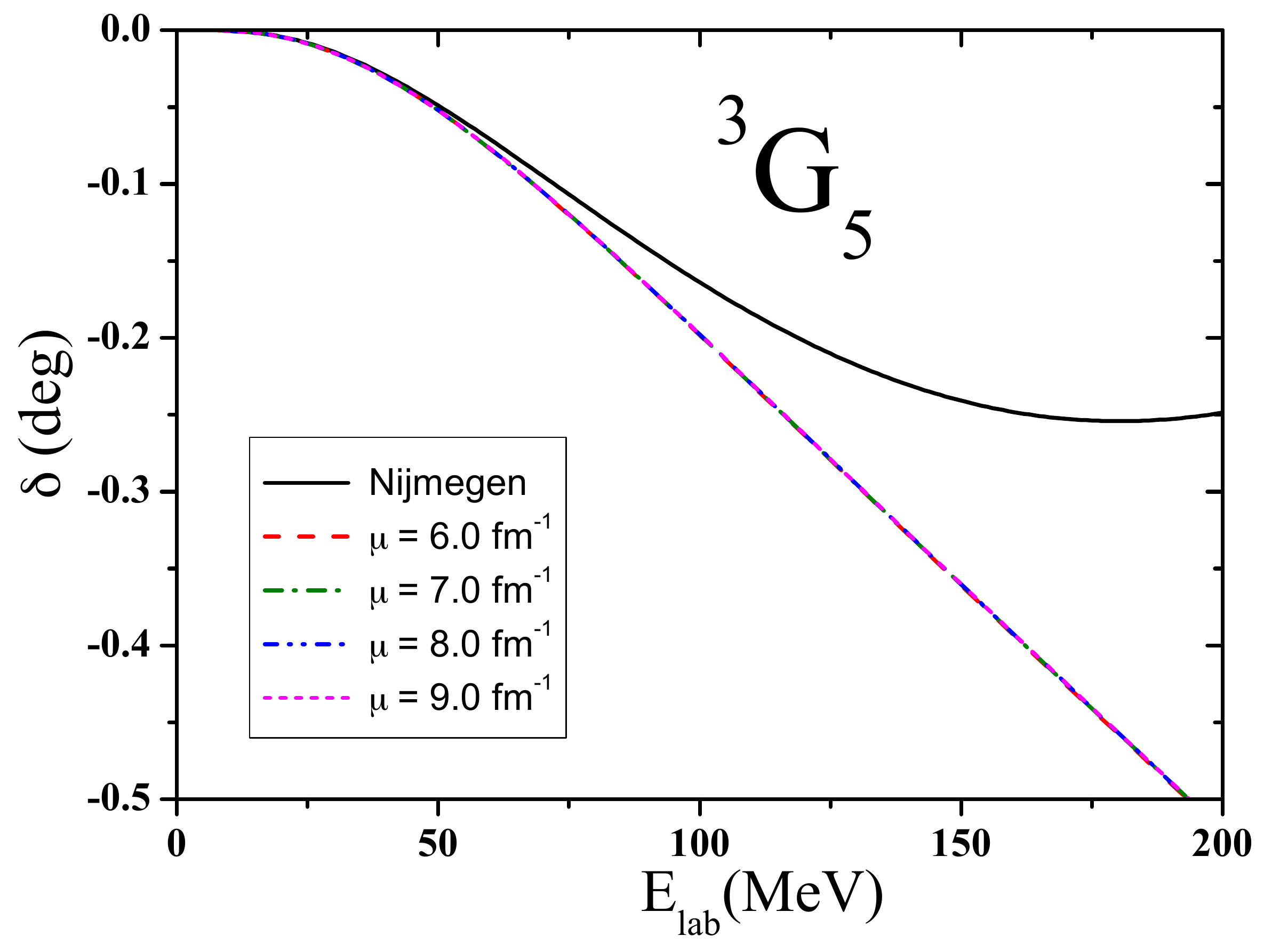}\includegraphics[scale=0.17]{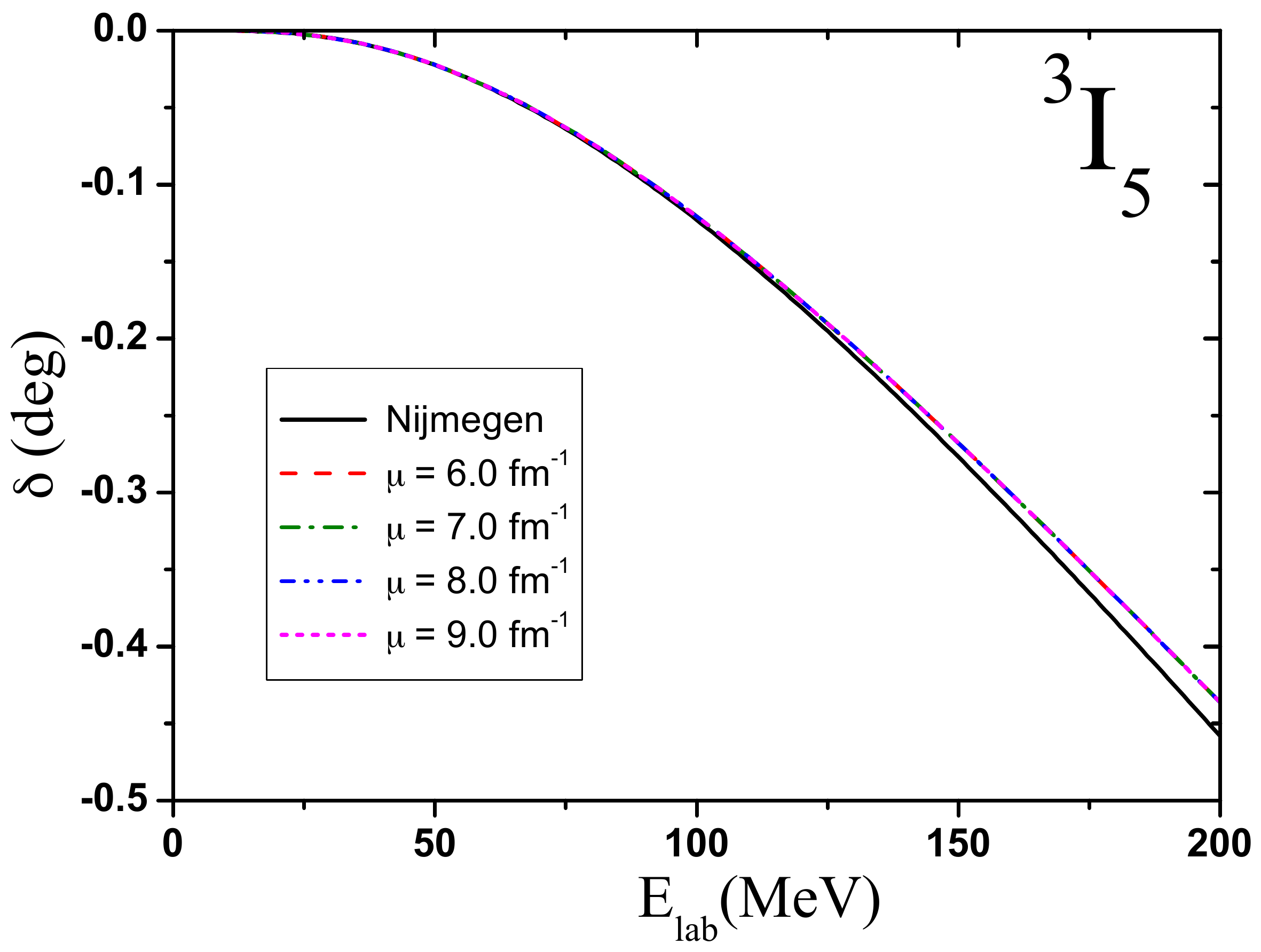}\includegraphics[scale=0.17]{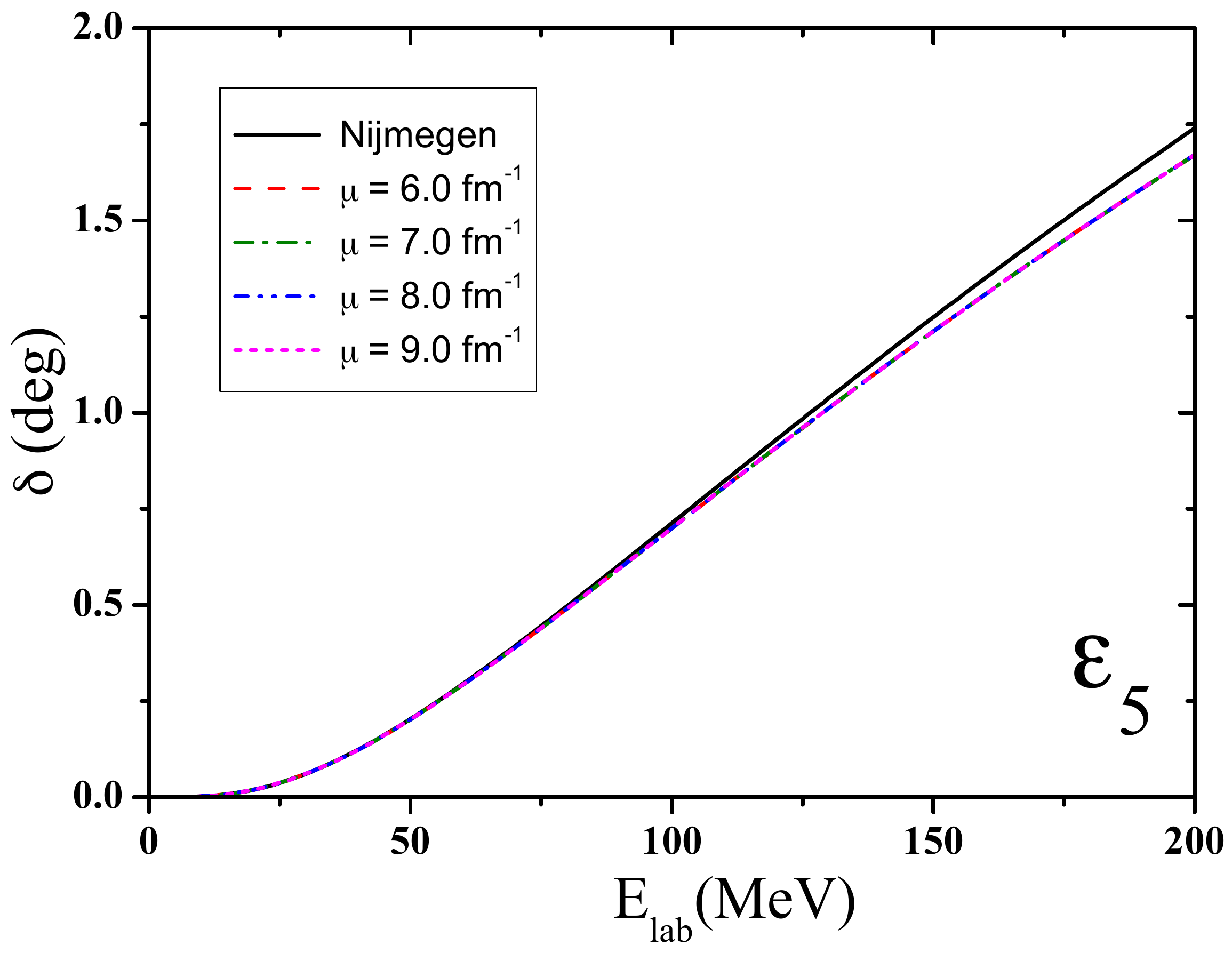}
\end{center}
\caption{(Color on-line) Phase-shifts in the $^3G_5 - ^3I_5$ coupled channels calculated from the solution of the subtracted LS equation for the $K$-matrix with five subtractions for the N3LO-EM potential for several values of the renormalization scale compared to the Nijmegen partial wave analysis.}
\label{fig6}
\end{figure}

\begin{figure*}[t]
\begin{center}
\includegraphics[scale=0.2]{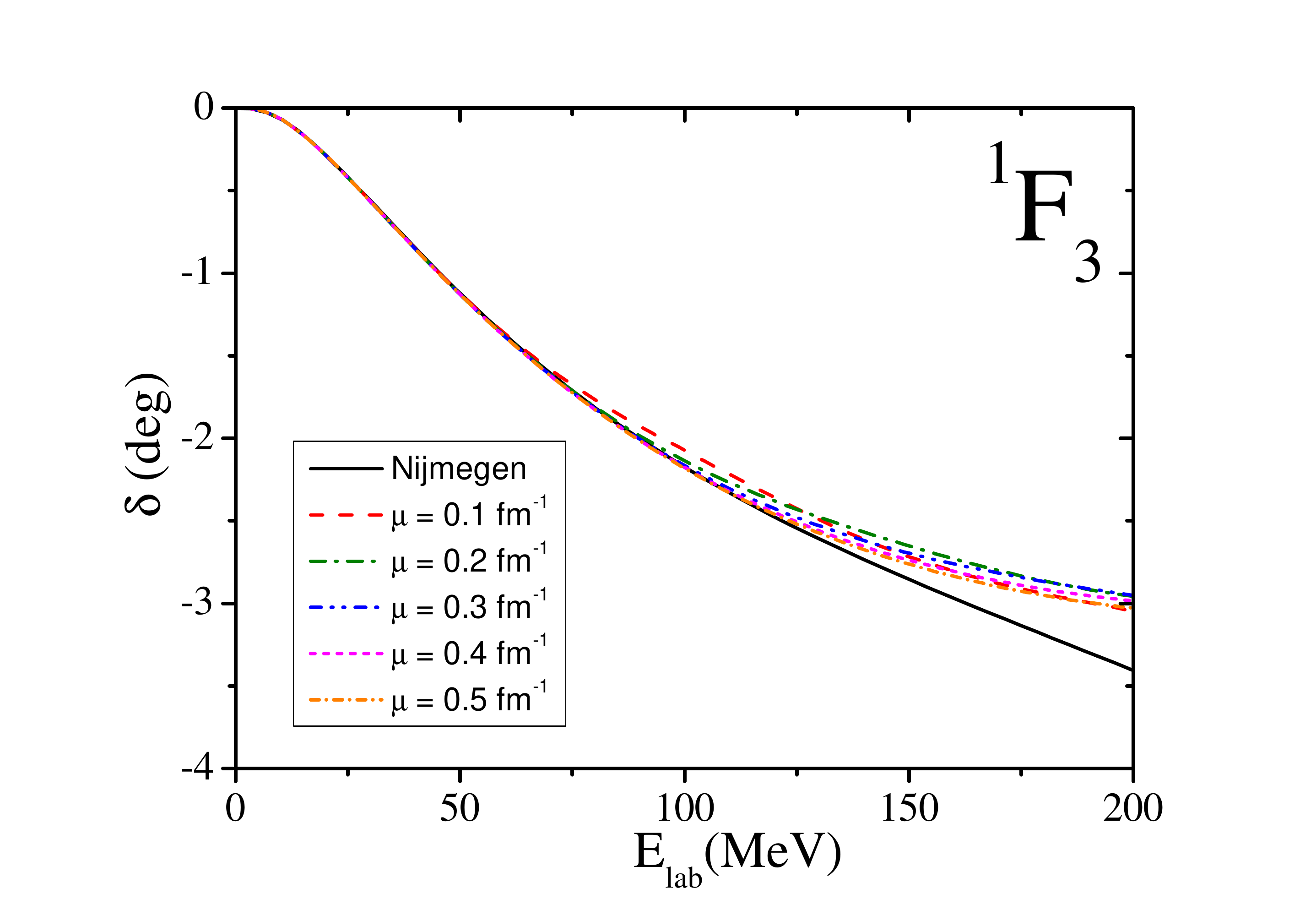}\hspace*{0.1cm}\includegraphics[scale=0.2]{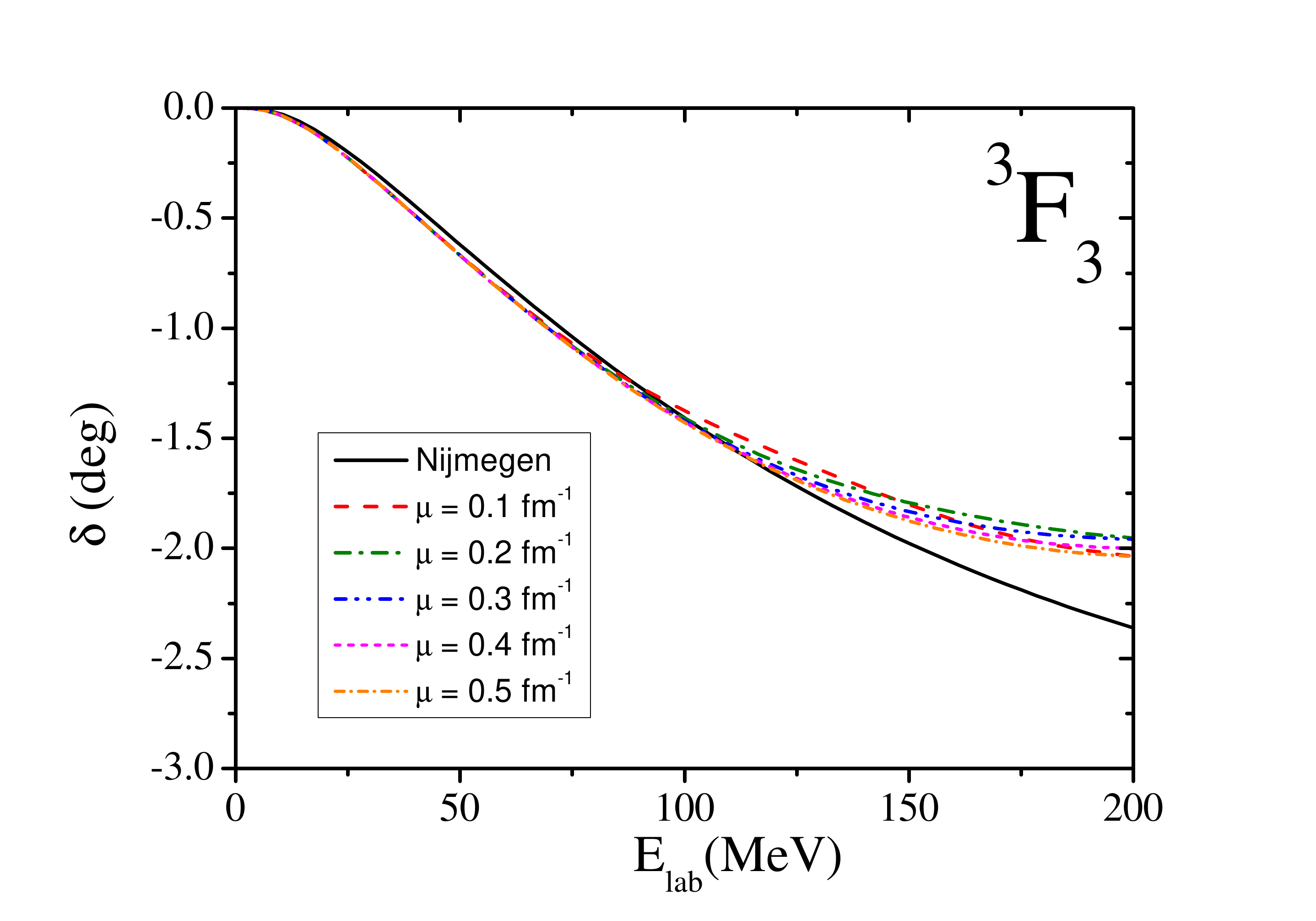} \\
\includegraphics[scale=0.2]{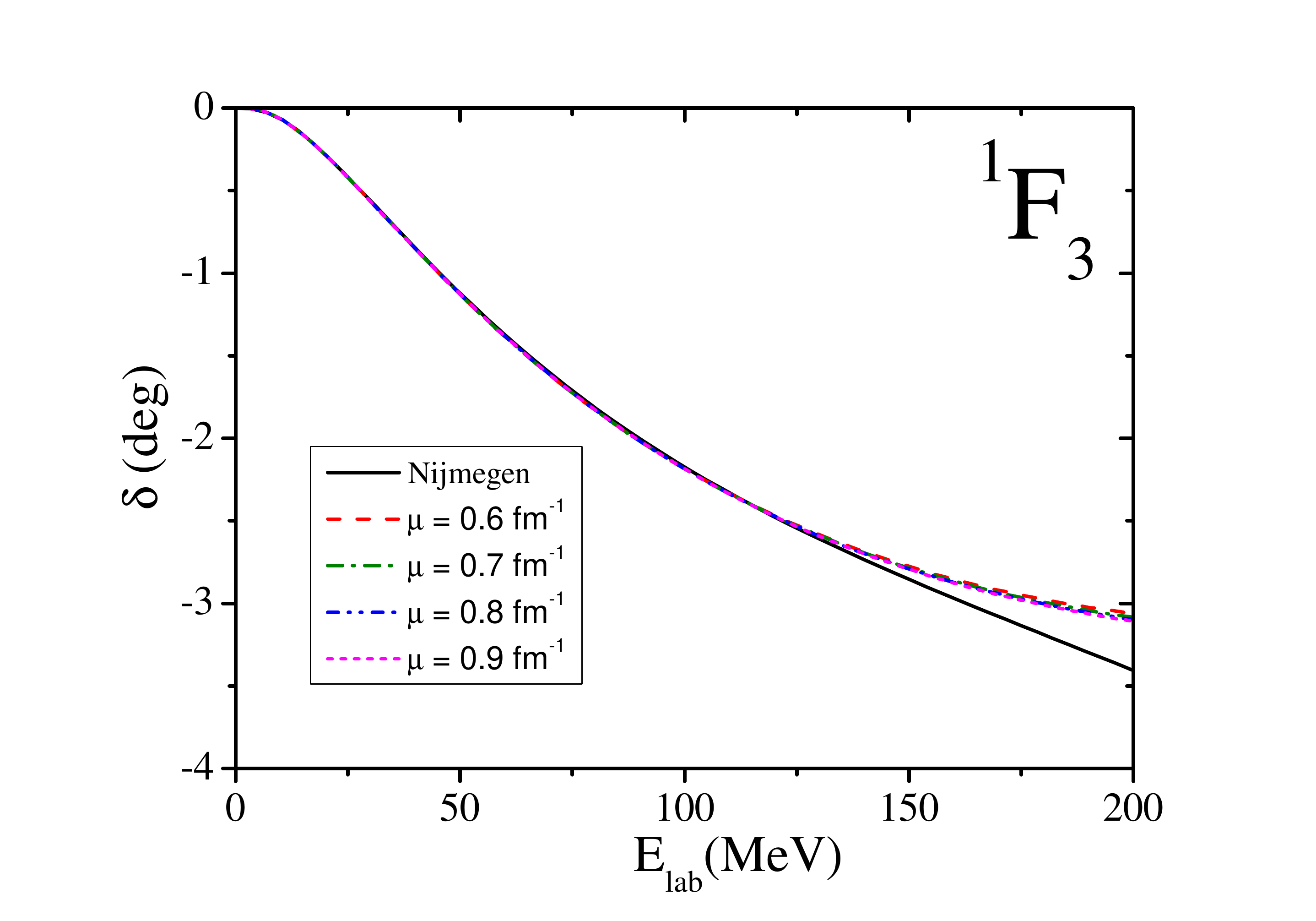}\hspace*{0.1cm}\includegraphics[scale=0.2]{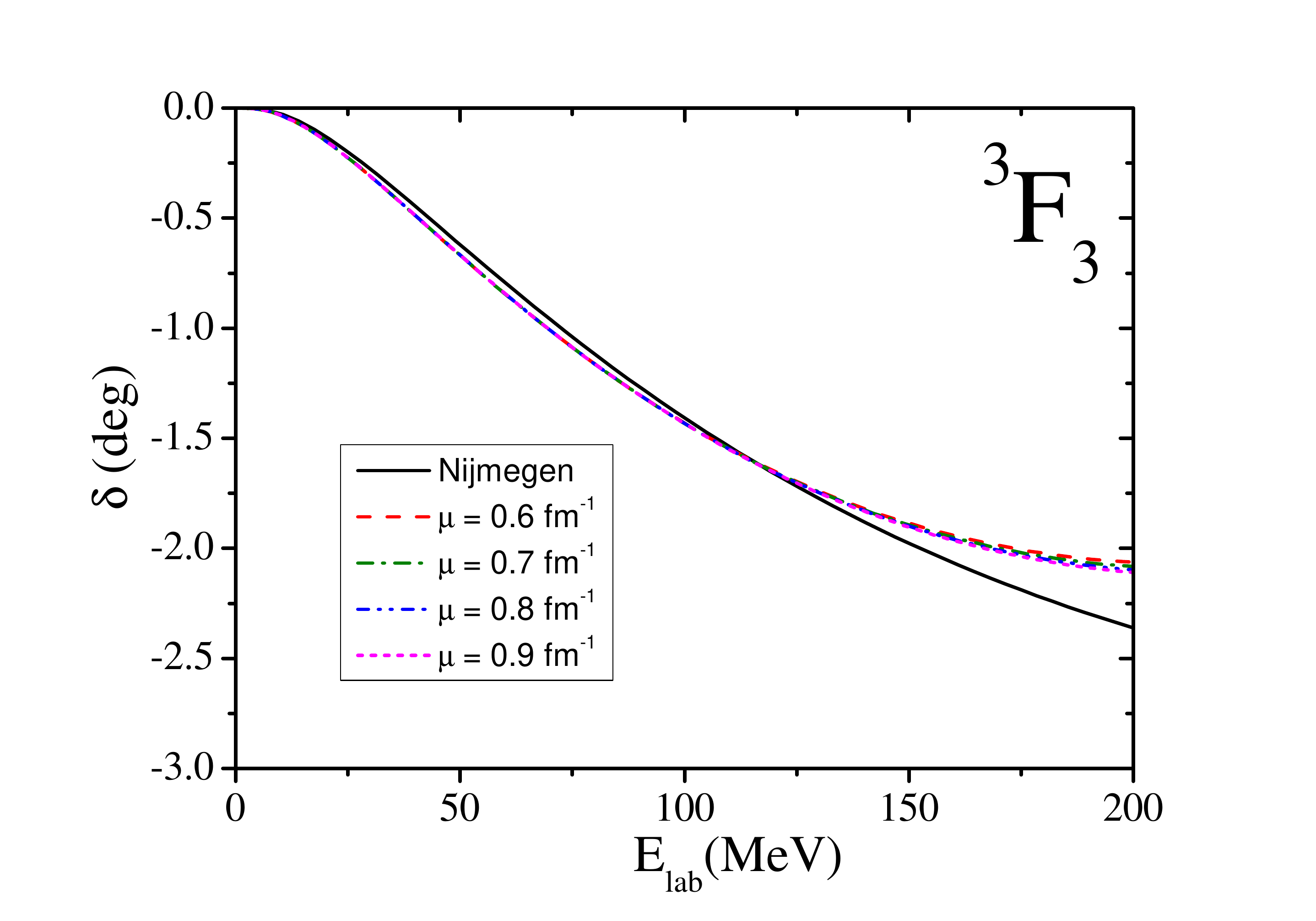} \\
\includegraphics[scale=0.2]{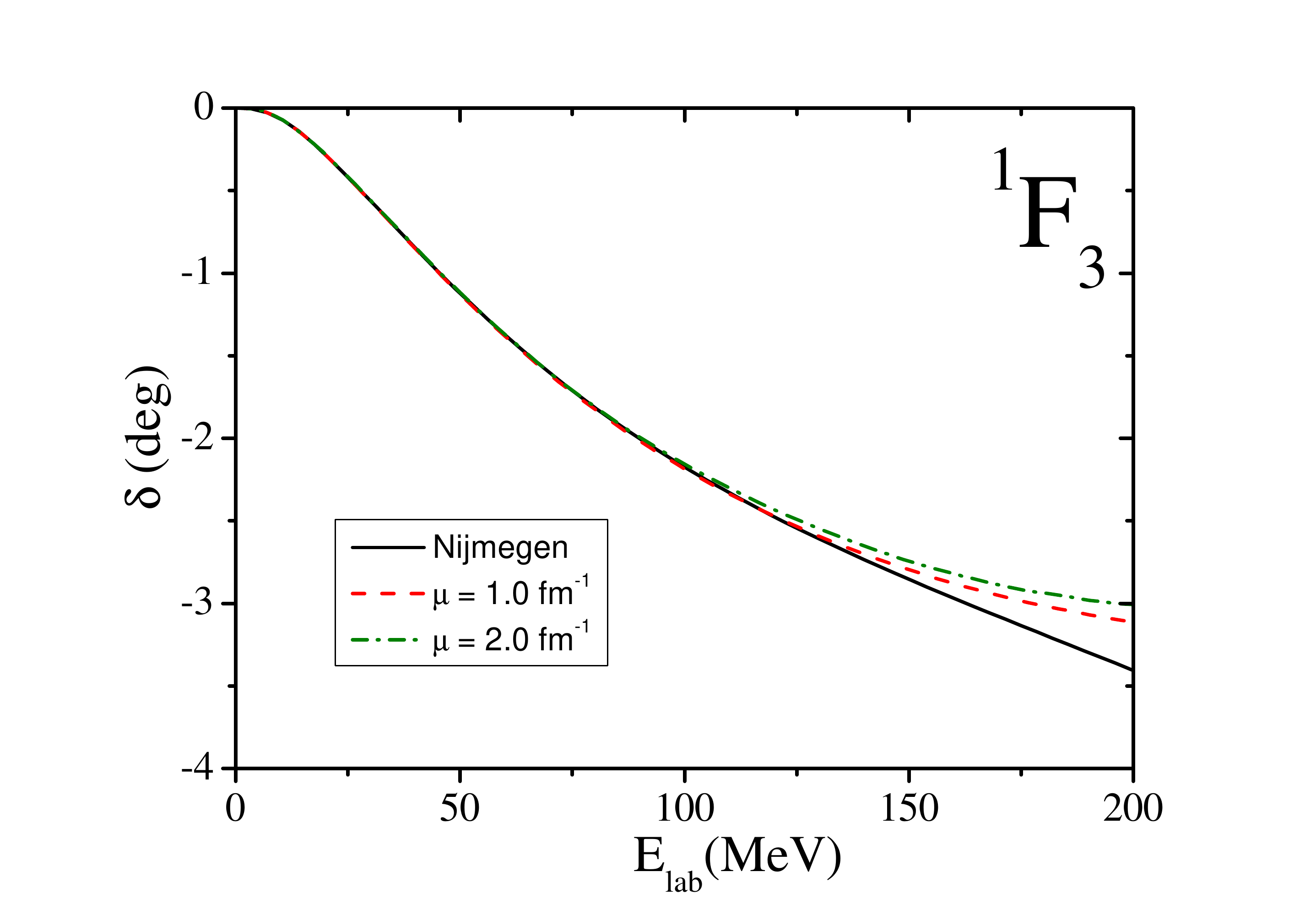}\hspace*{0.1cm}\includegraphics[scale=0.2]{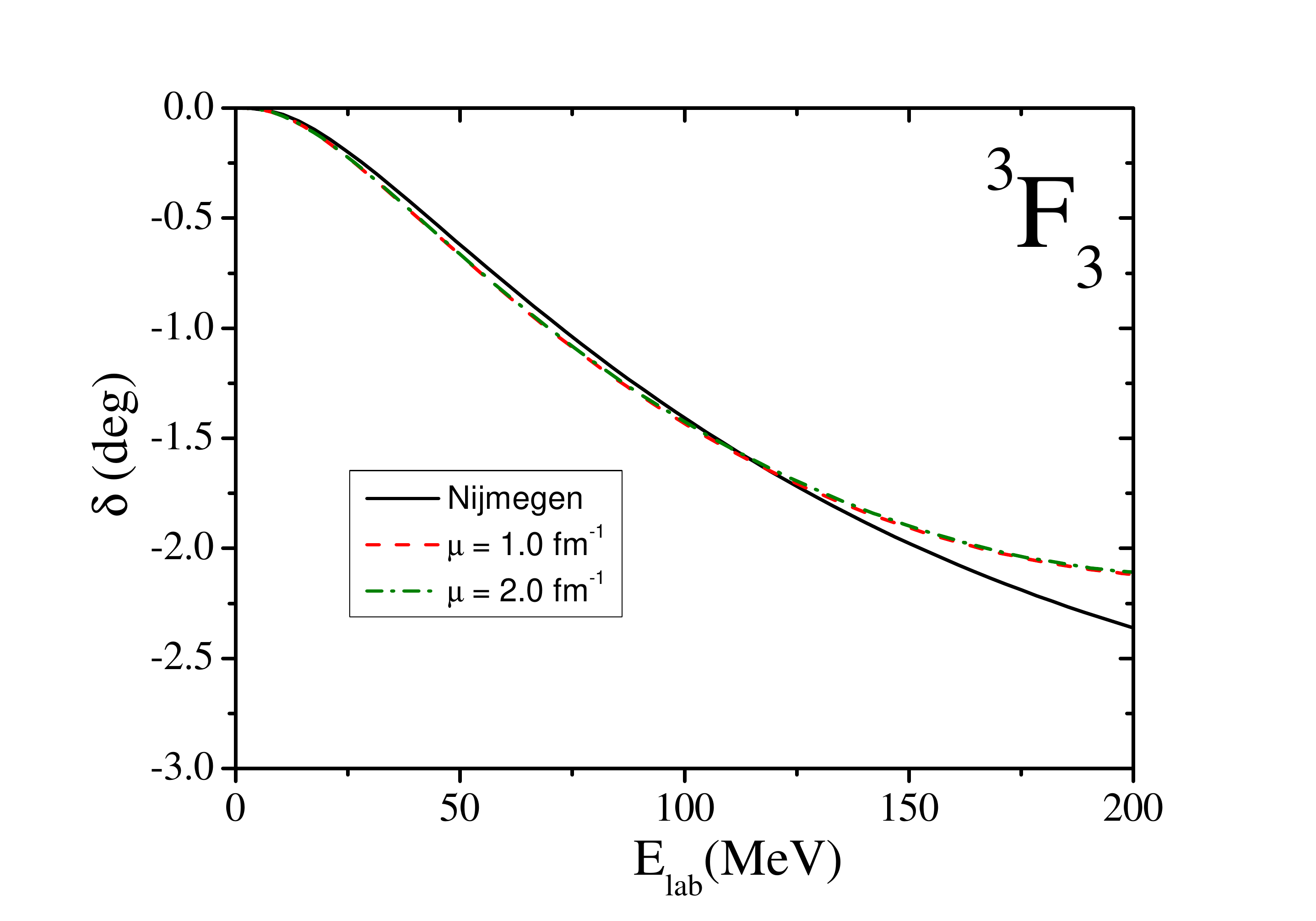} 
\end{center}
\caption{(Color on-line) Phase-shifts in the $^1F_3$ and $^3F_3$ channels calculated from the solution of the subtracted LS equation for the $K$-matrix with five subtractions for the N3LO-EGM potential for several values of the renormalization scale compared to the Nijmegen partial wave analysis.}
\label{fig7}
\end{figure*}

\begin{figure*}[t]
\begin{center}
\includegraphics[scale=0.2]{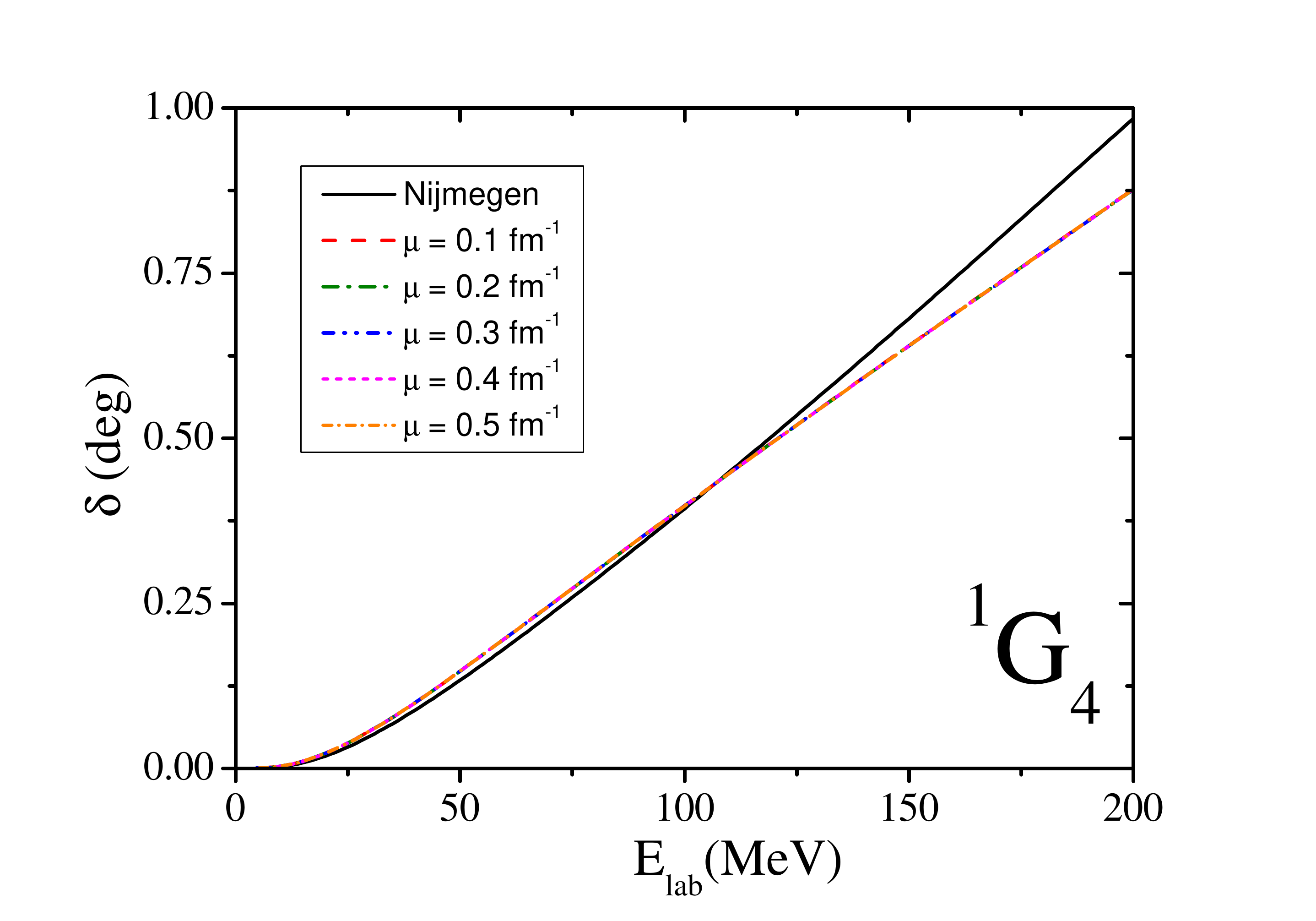}\hspace*{0.1cm}\includegraphics[scale=0.2]{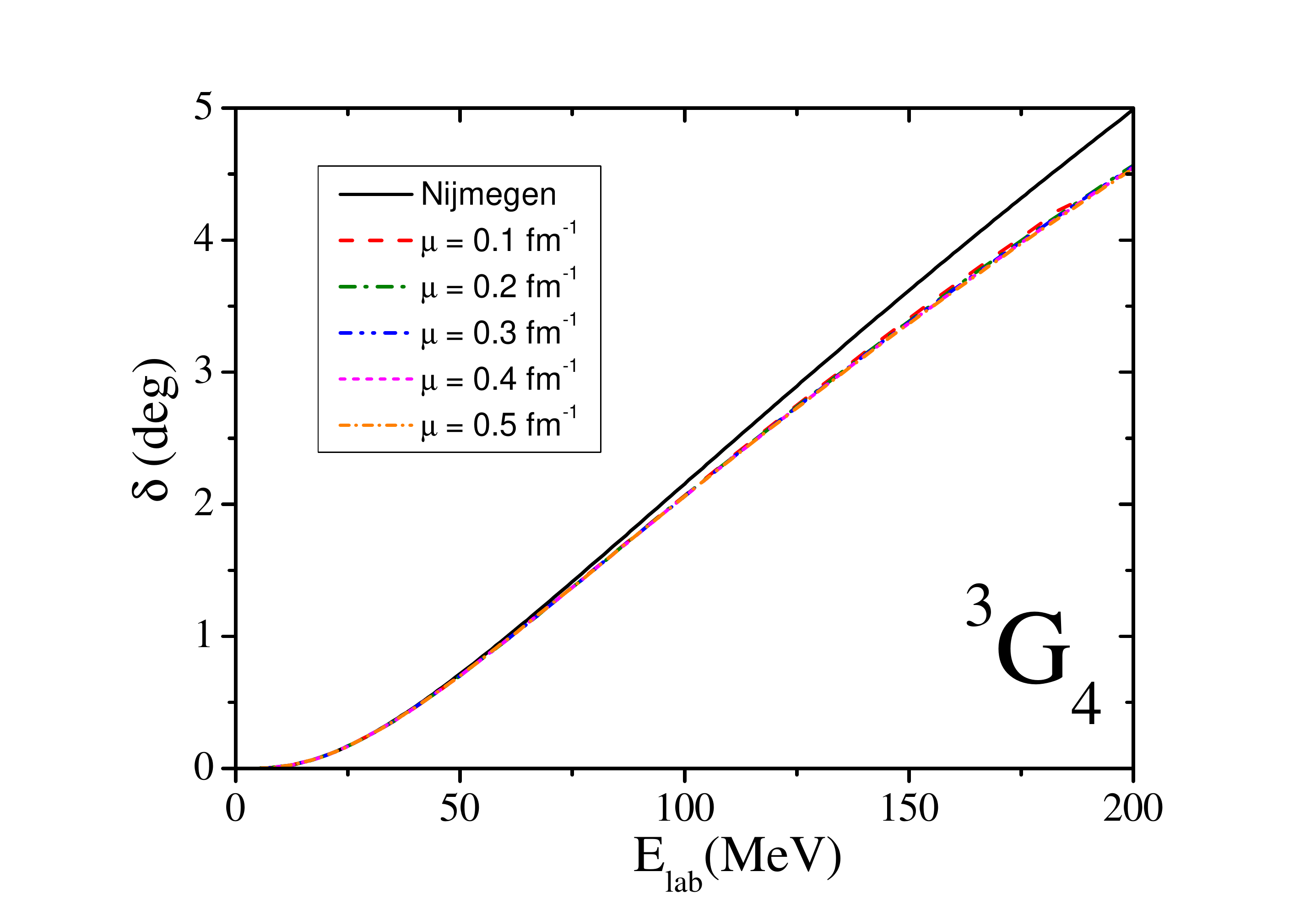} \\
\includegraphics[scale=0.2]{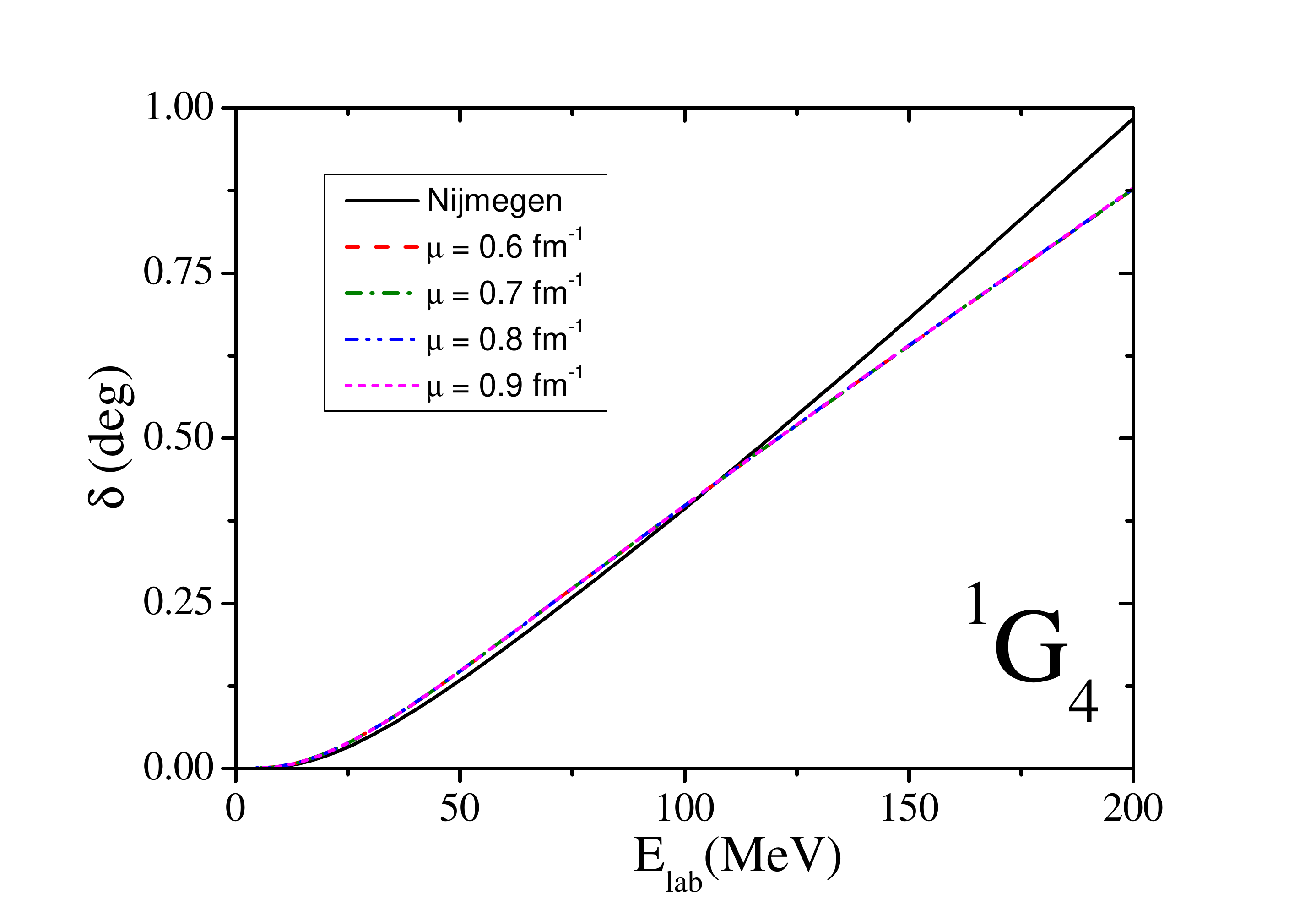}\hspace*{0.1cm}\includegraphics[scale=0.2]{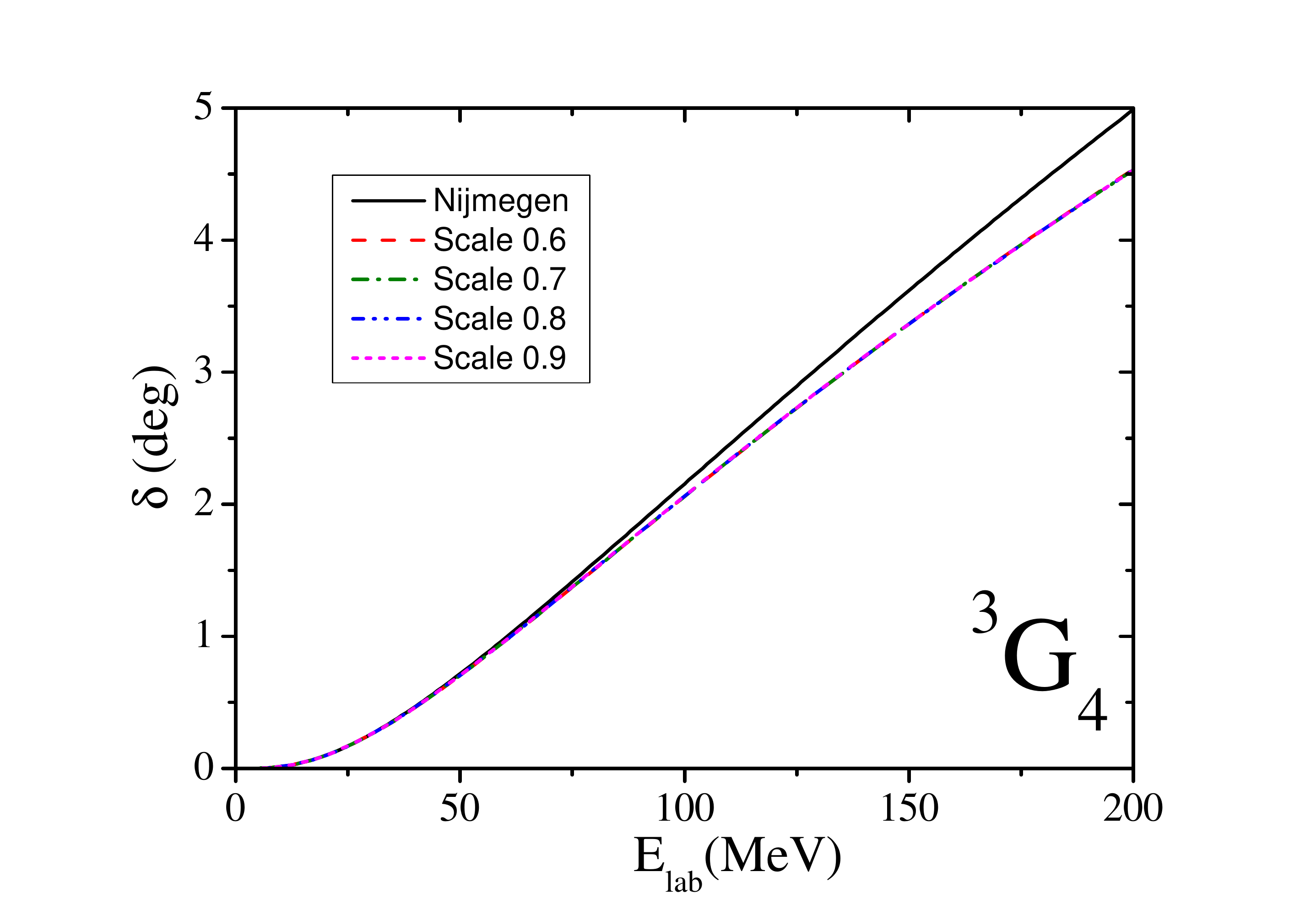} \\
\includegraphics[scale=0.2]{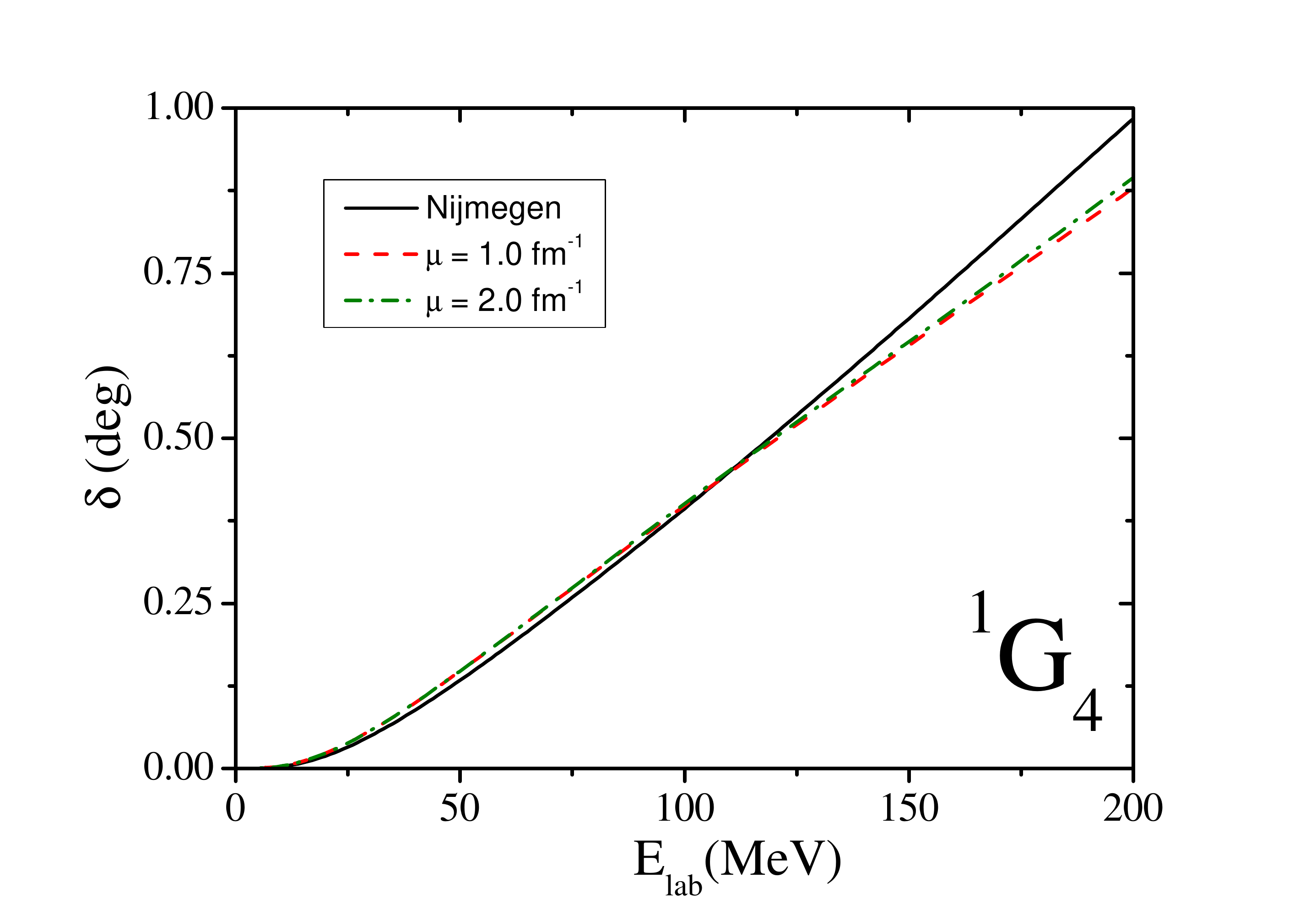}\hspace*{0.1cm}\includegraphics[scale=0.2]{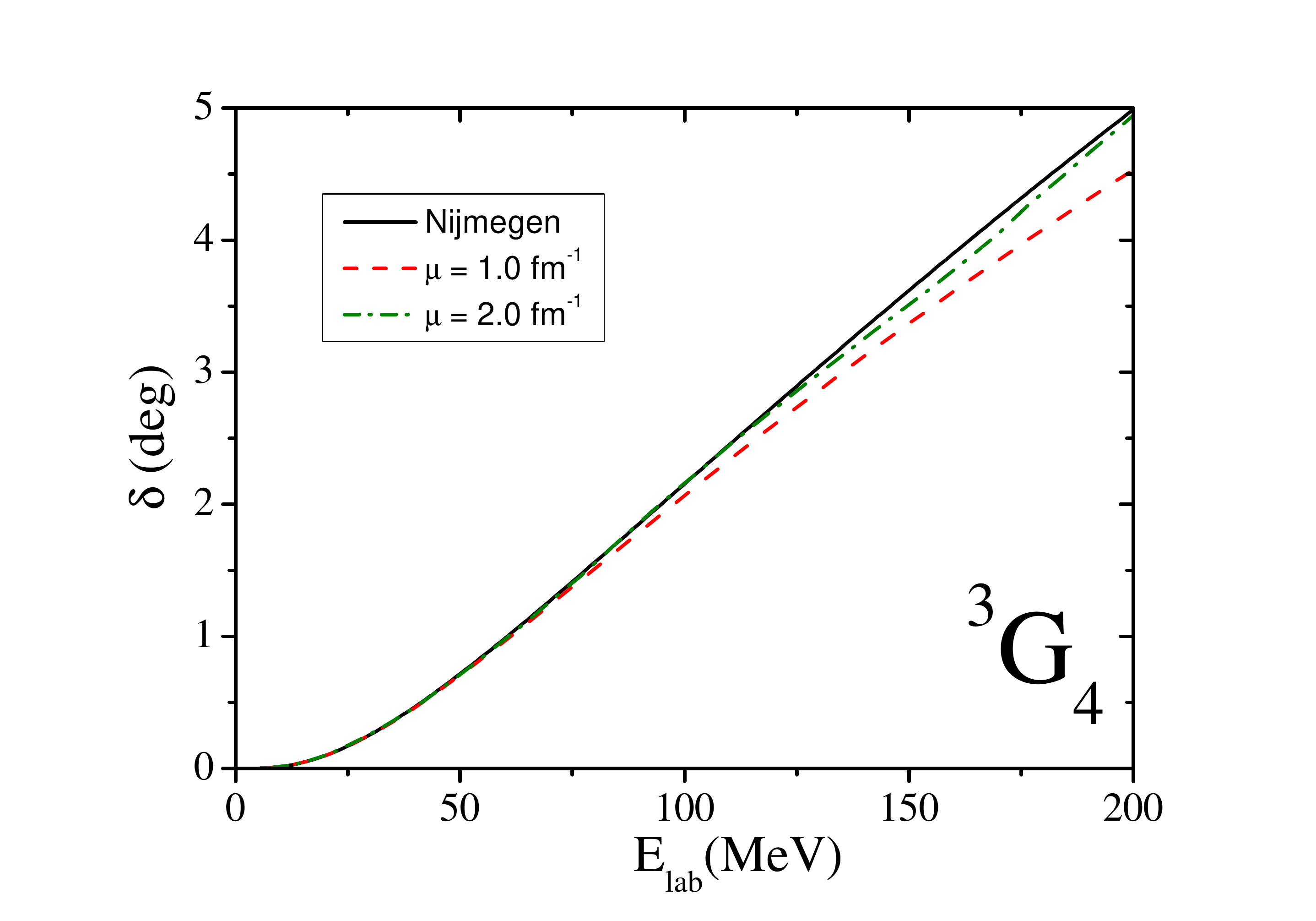}
\end{center}
\caption{(Color on-line) Phase-shifts in the $^1G_4$ and $^3G_4$ channels calculated from the solution of the subtracted LS equation for the $K$-matrix with five subtractions for the N3LO-EGM potential for several values of the renormalization scale compared to the Nijmegen partial wave analysis.}
\label{fig8}
\end{figure*}

\begin{figure*}[t]
\begin{center}
\includegraphics[scale=0.2]{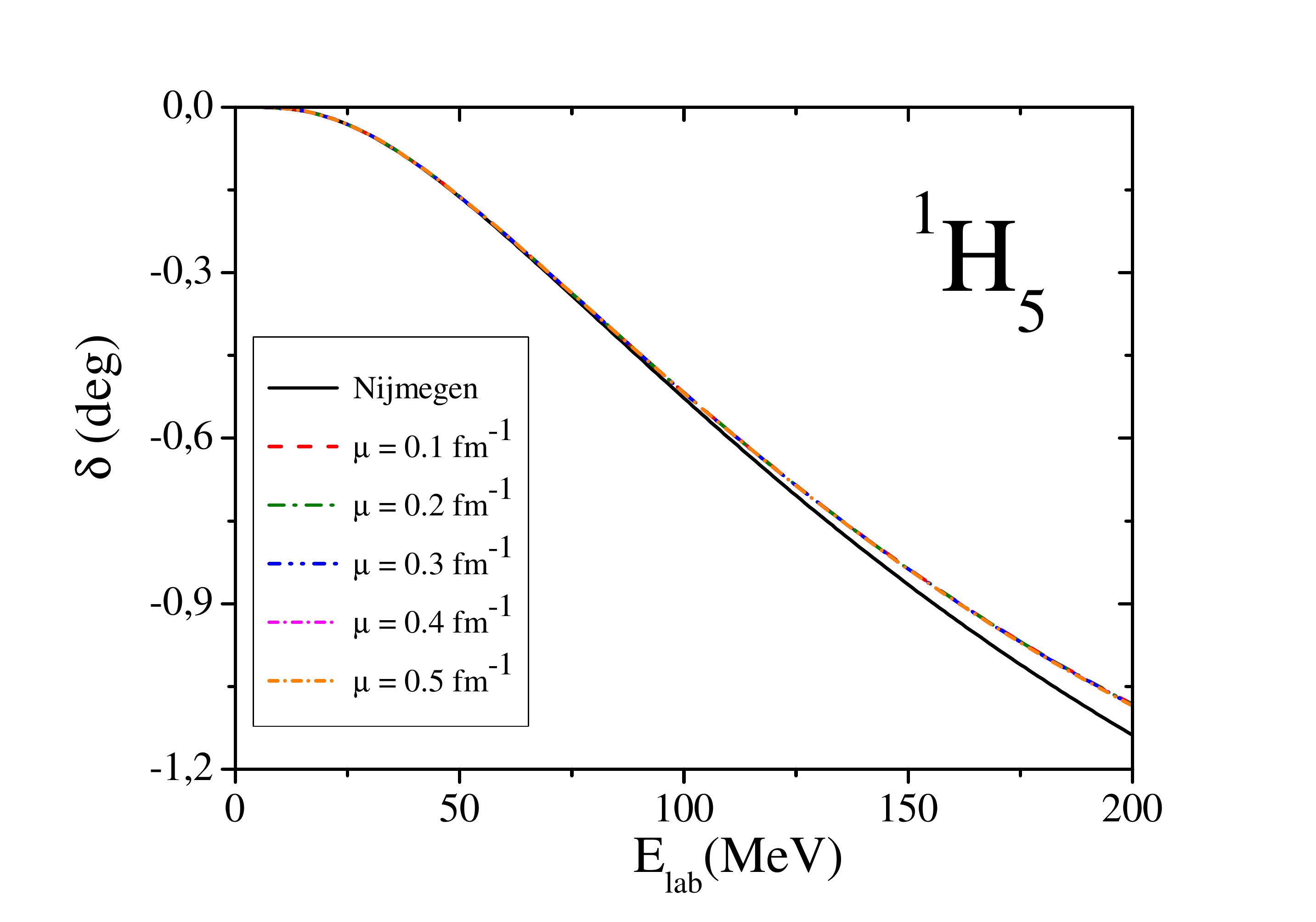}\hspace*{0.1cm}\includegraphics[scale=0.2]{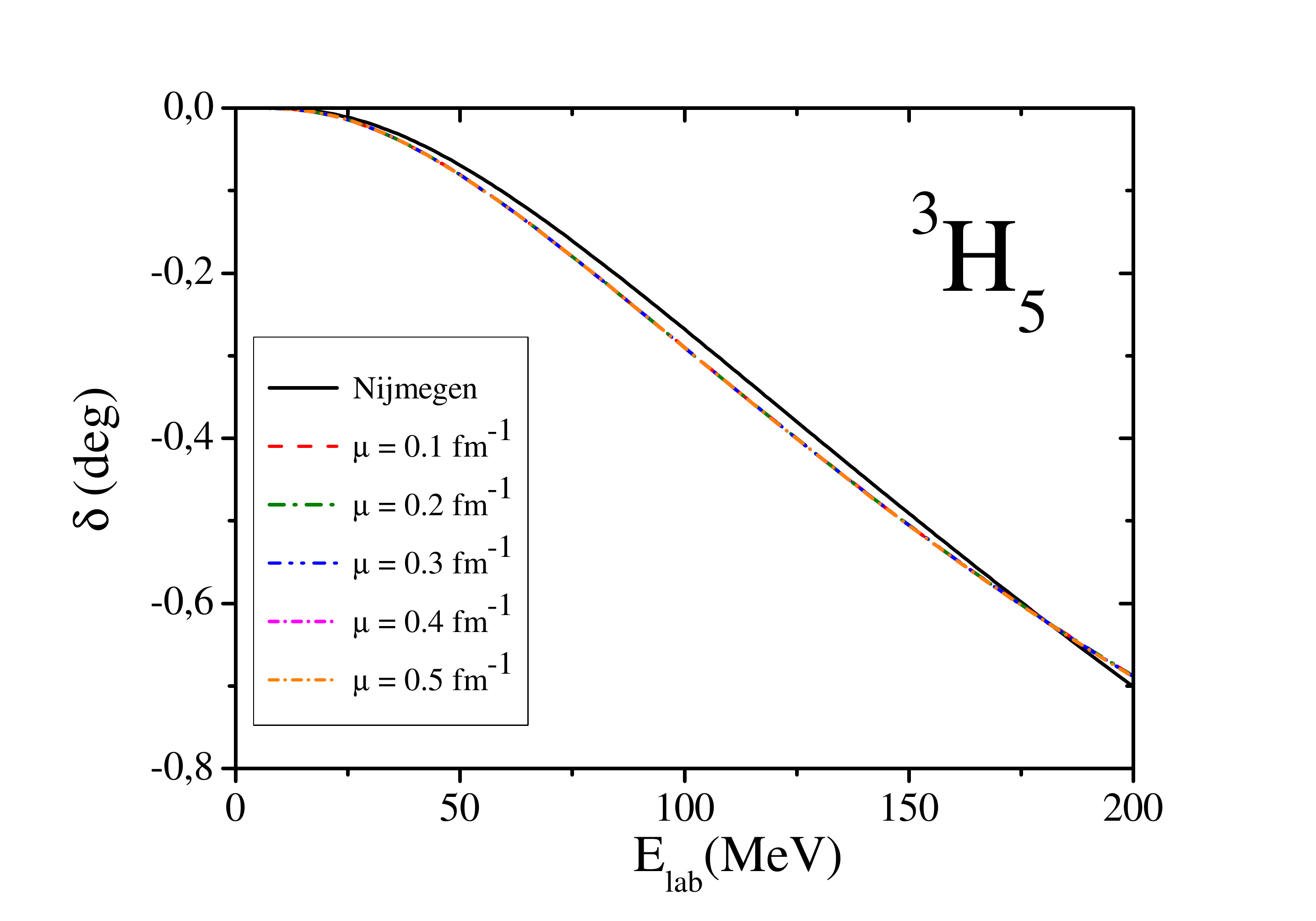} \\
\includegraphics[scale=0.2]{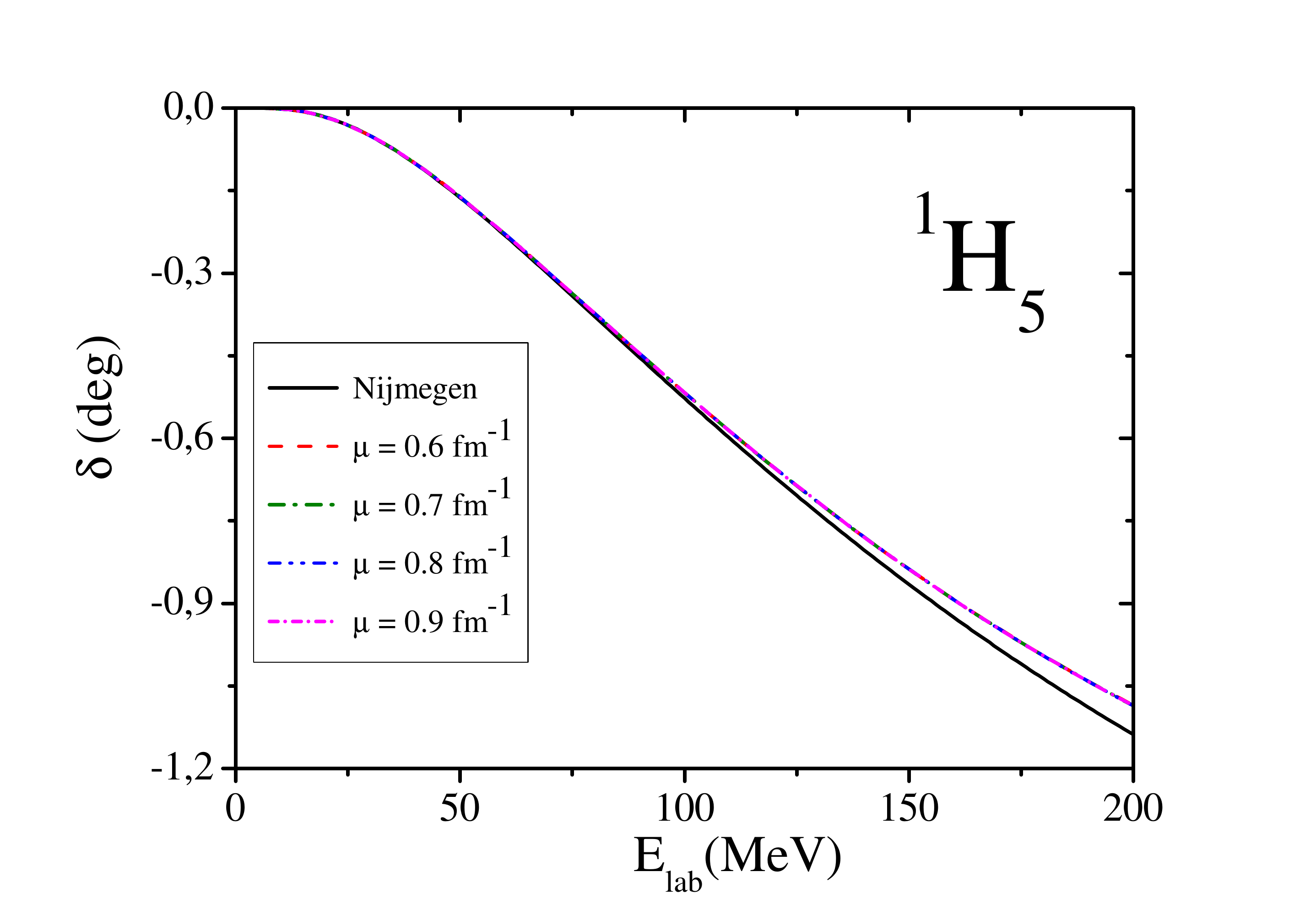}\hspace*{0.1cm}\includegraphics[scale=0.2]{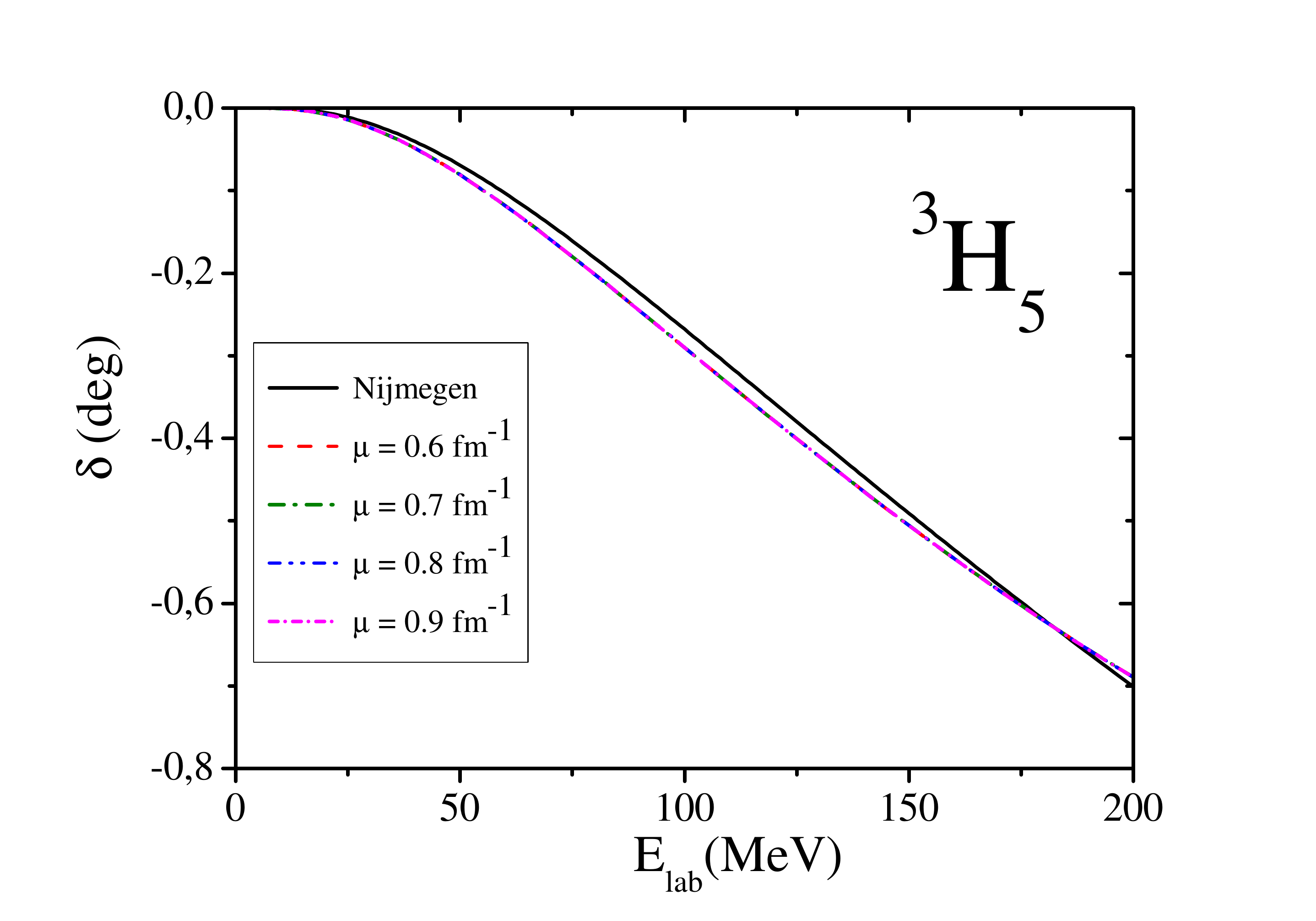} \\
\includegraphics[scale=0.2]{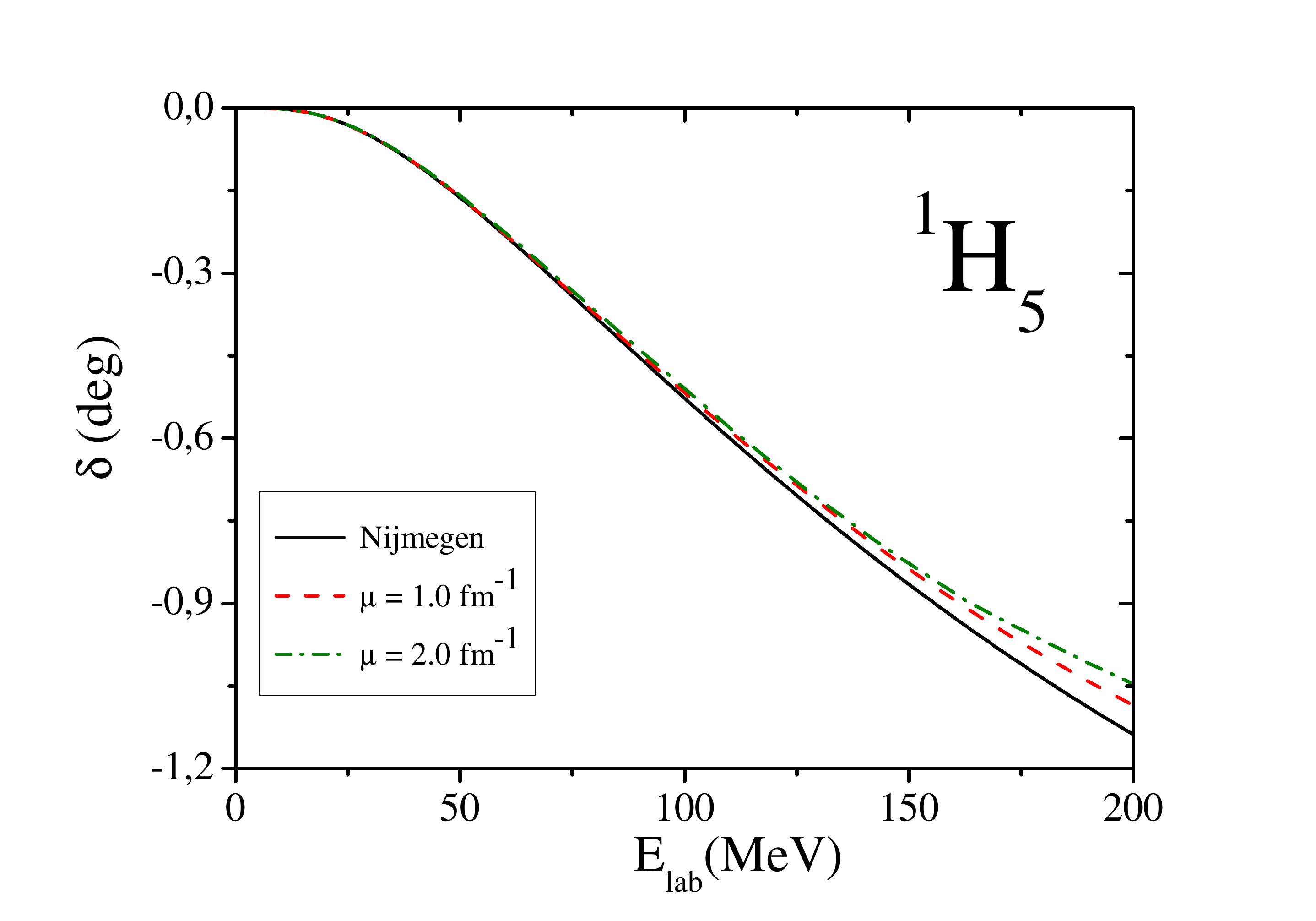}\hspace*{0.1cm}\includegraphics[scale=0.2]{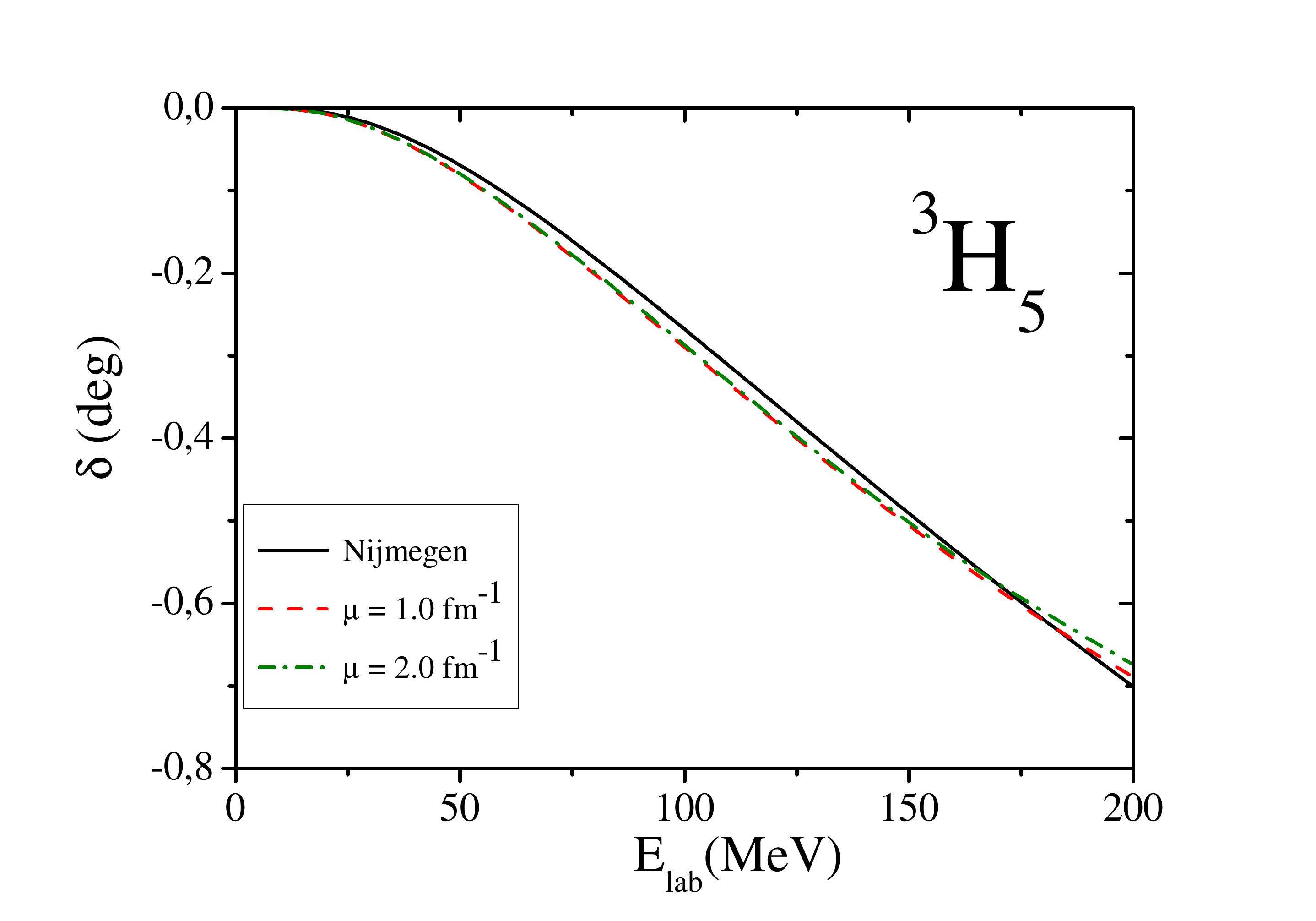} 
\end{center}
\caption{(Color on-line) Phase-shifts in the $^1H_5$ and $^3H_5$ channels calculated from the solution of the subtracted LS equation for the $K$-matrix with five subtractions for the N3LO-EGM potential for several values of the renormalization scale compared to the Nijmegen partial wave analysis.}
\label{fig9}
\end{figure*}

\begin{figure}[t]
\begin{center}
\includegraphics[scale=0.2]{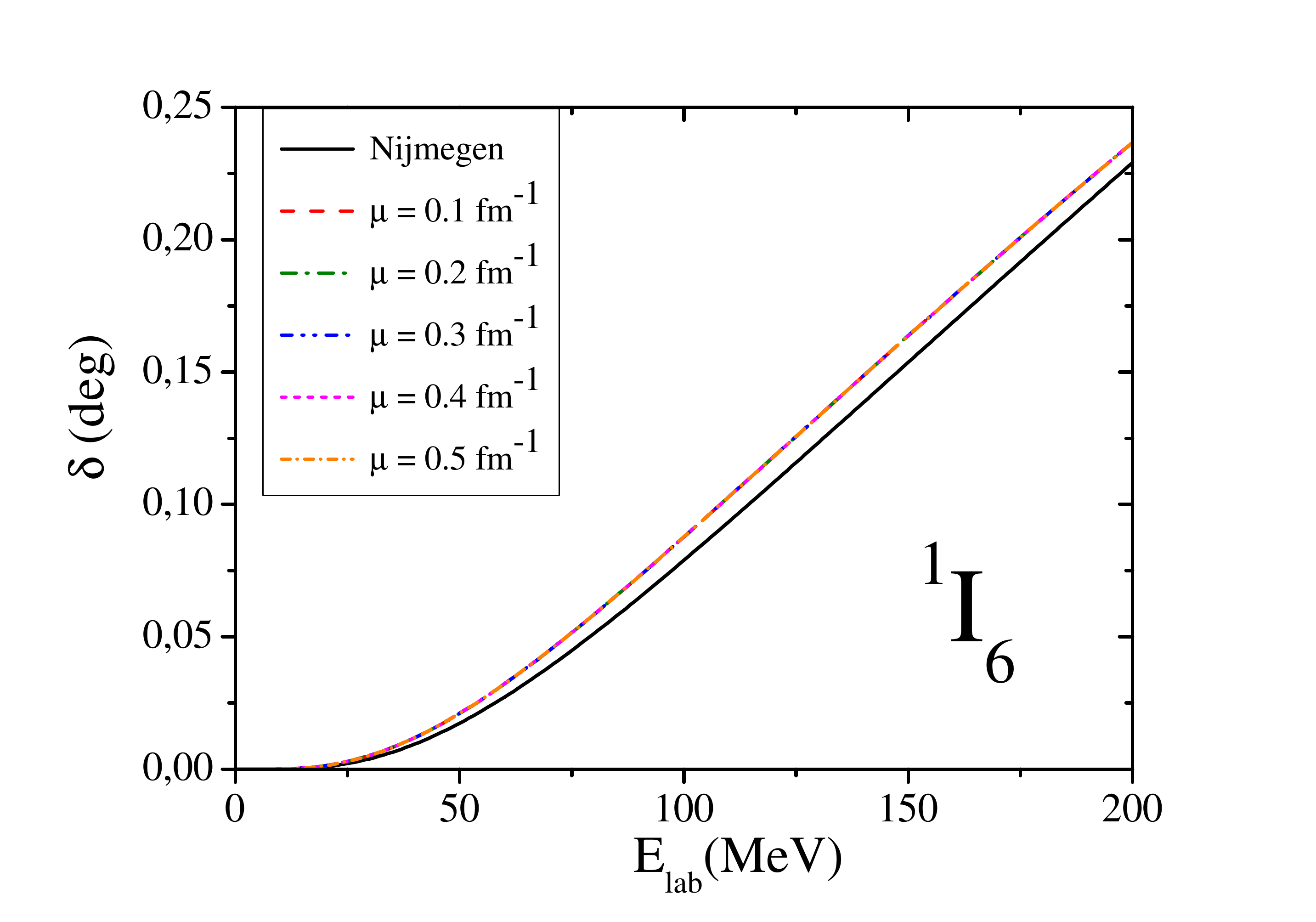}\hspace*{0.1cm}\includegraphics[scale=0.2]{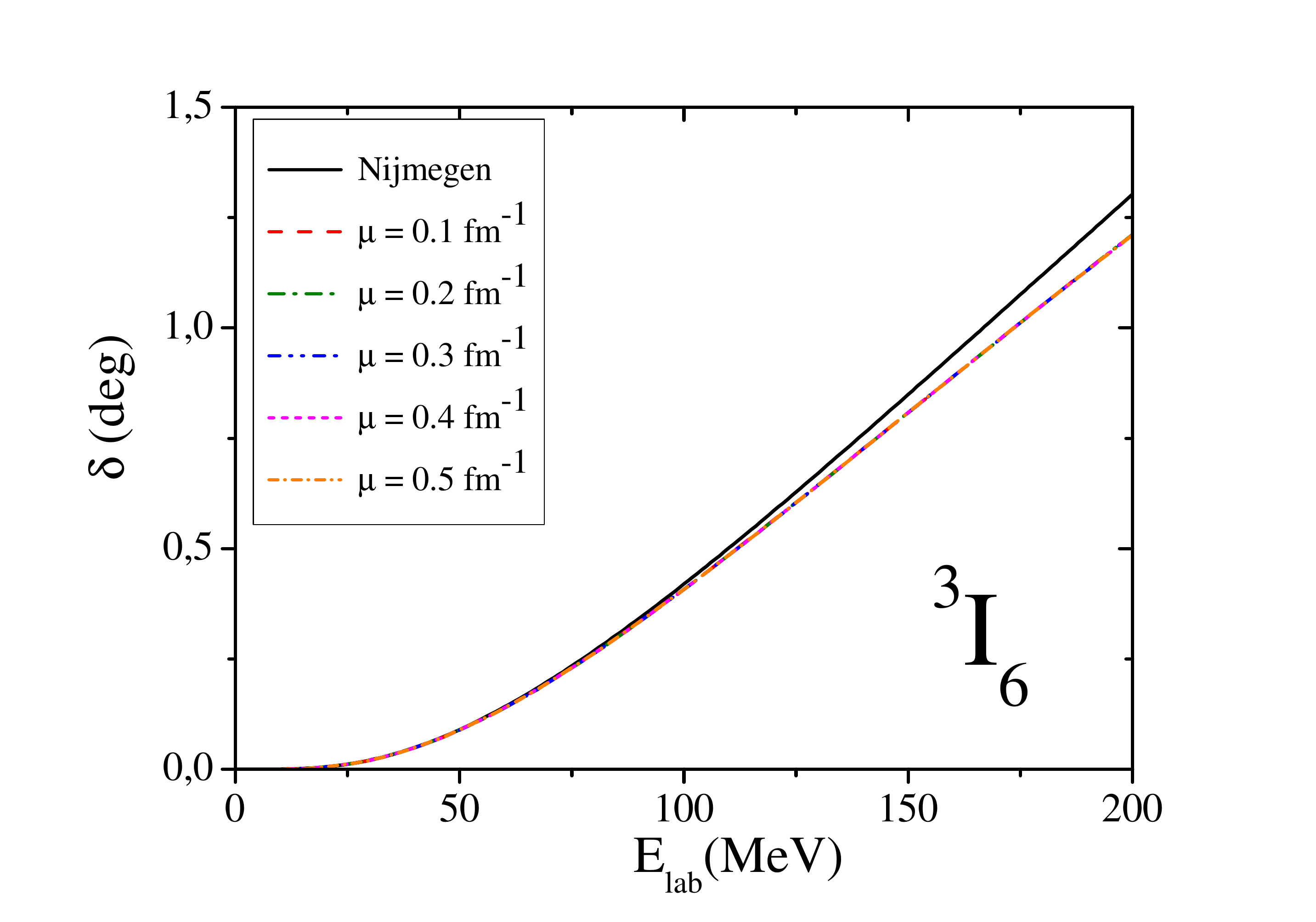} \\
\includegraphics[scale=0.2]{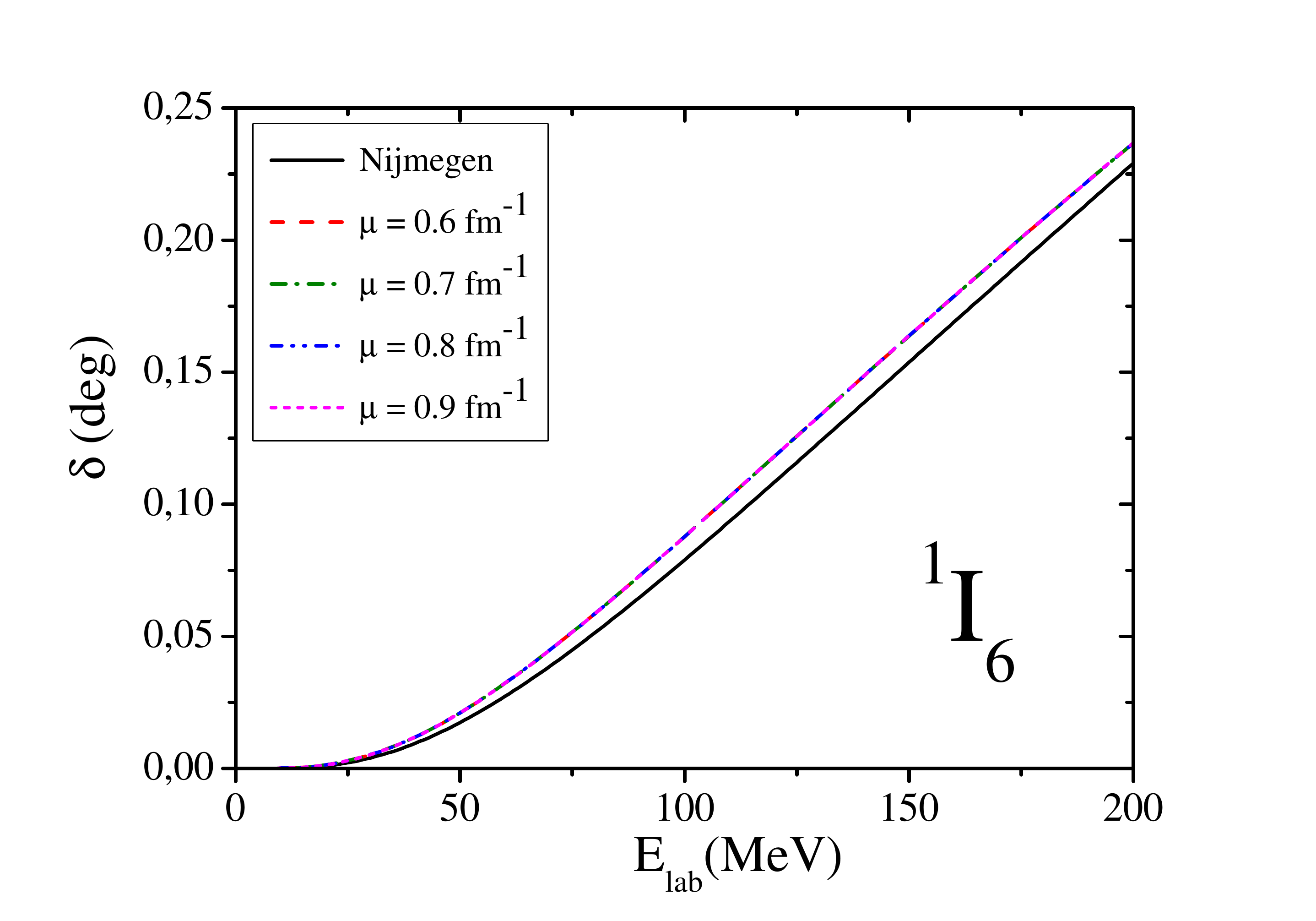}\hspace*{0.1cm}\includegraphics[scale=0.2]{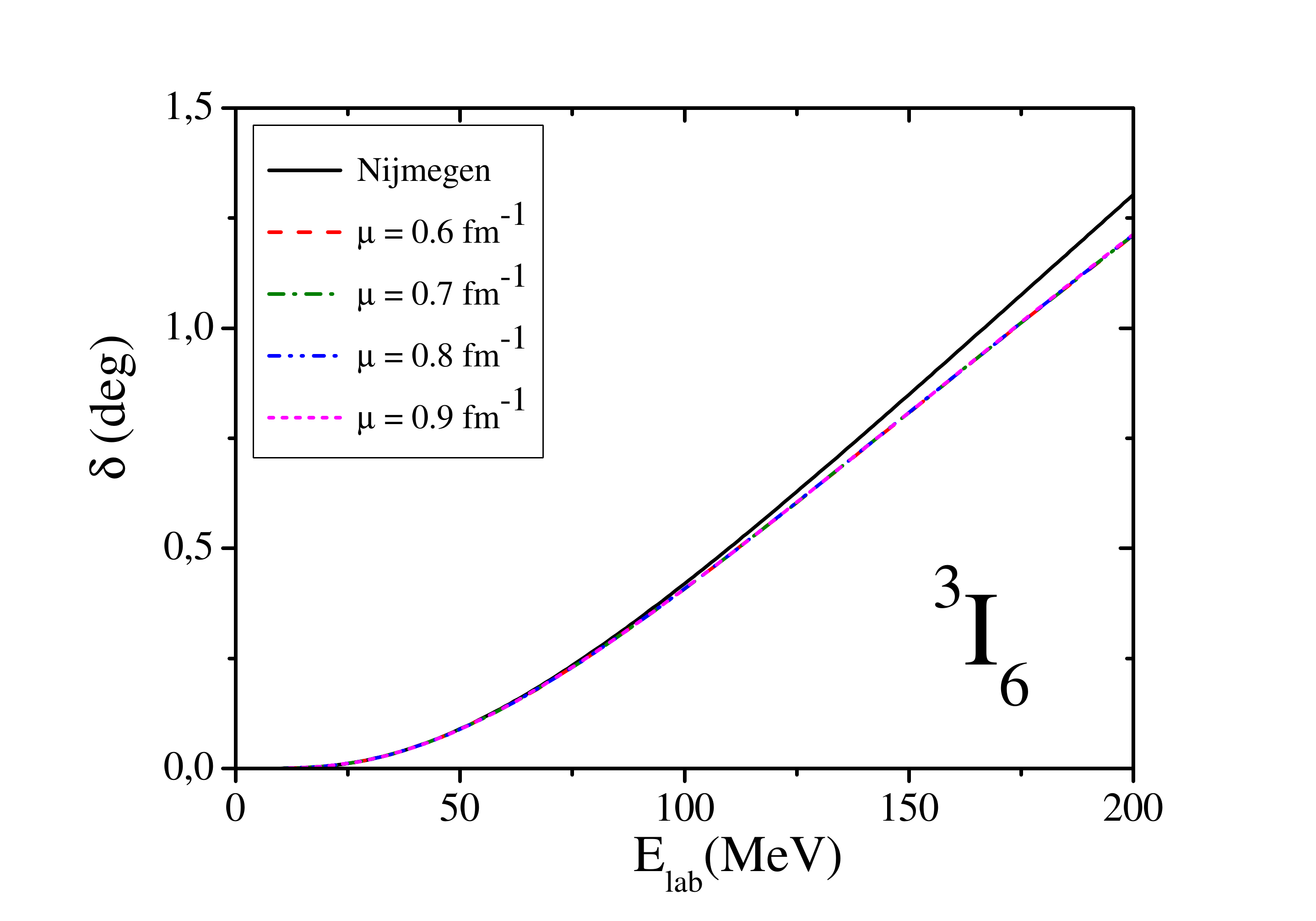} \\
\includegraphics[scale=0.2]{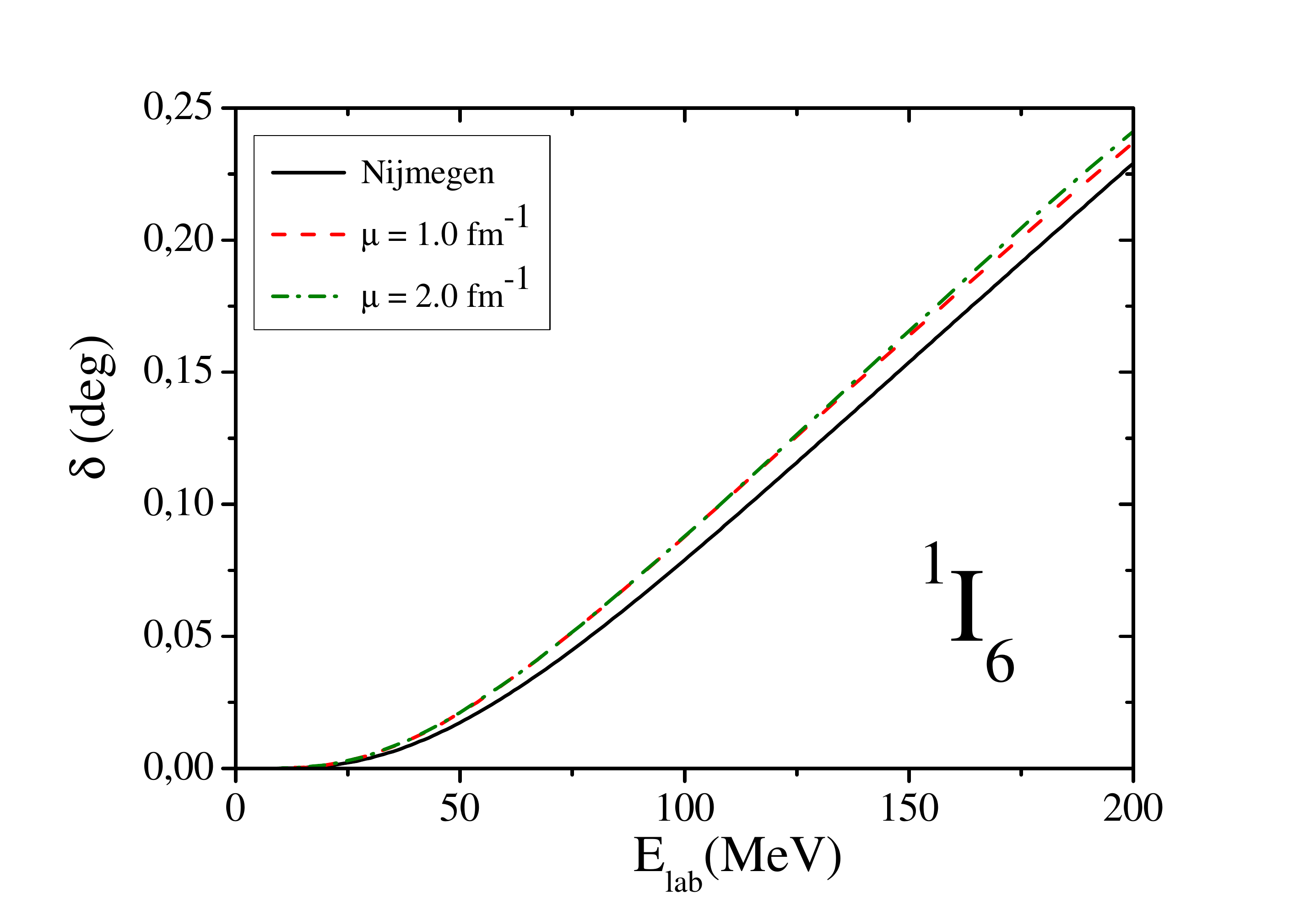}\hspace*{0.1cm}\includegraphics[scale=0.2]{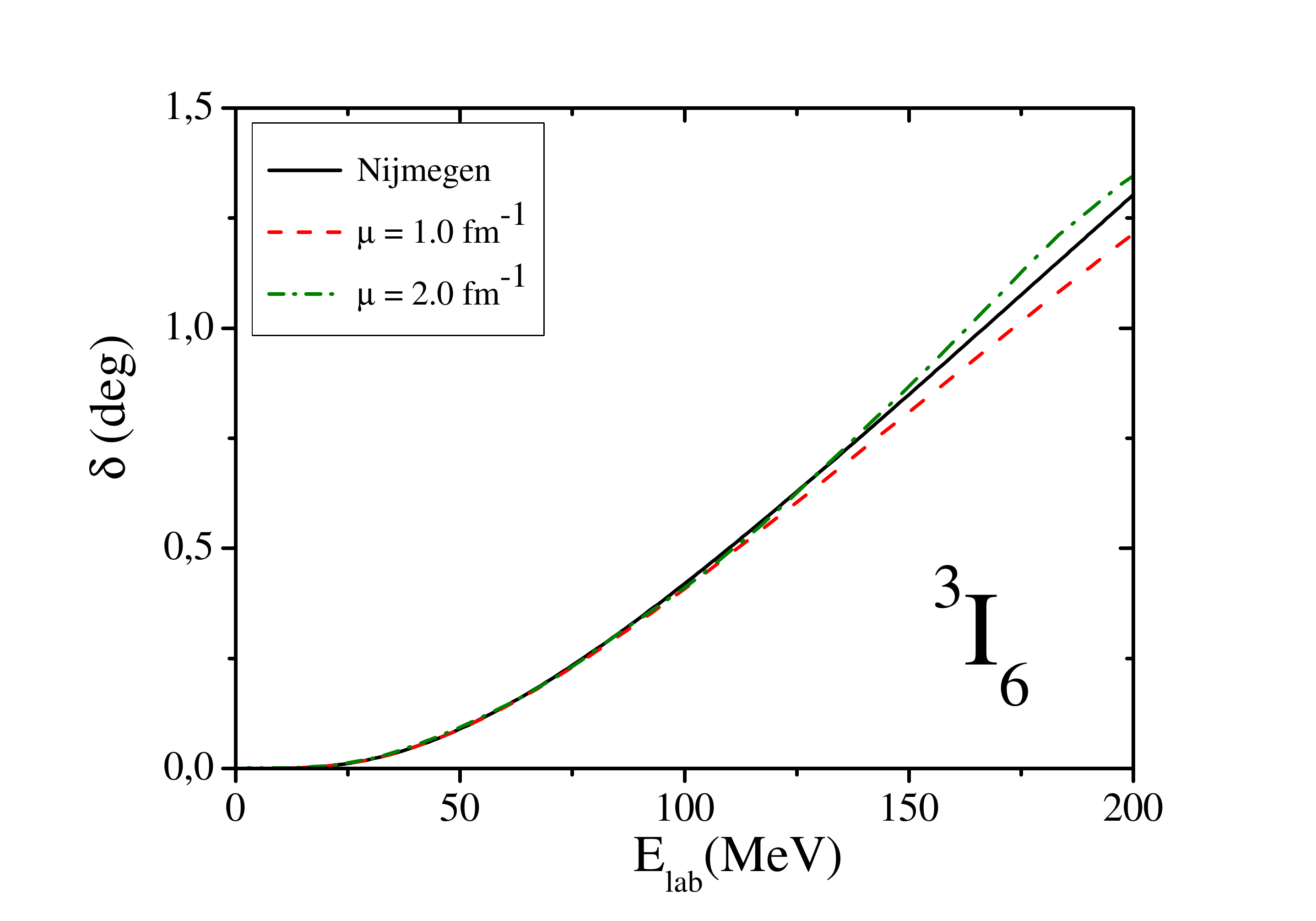} 
\end{center}
\caption{(Color on-line) Phase-shifts in the $^1I_6$ and $^3I_6$ channels calculated from the solution of the subtracted LS equation for the $K$-matrix with five subtractions for the N3LO-EGM potential for several values of the renormalization scale compared to the Nijmegen partial wave analysis.}
\label{fig10}
\end{figure}

\begin{figure}[t]
\begin{center}
\includegraphics[scale=0.17]{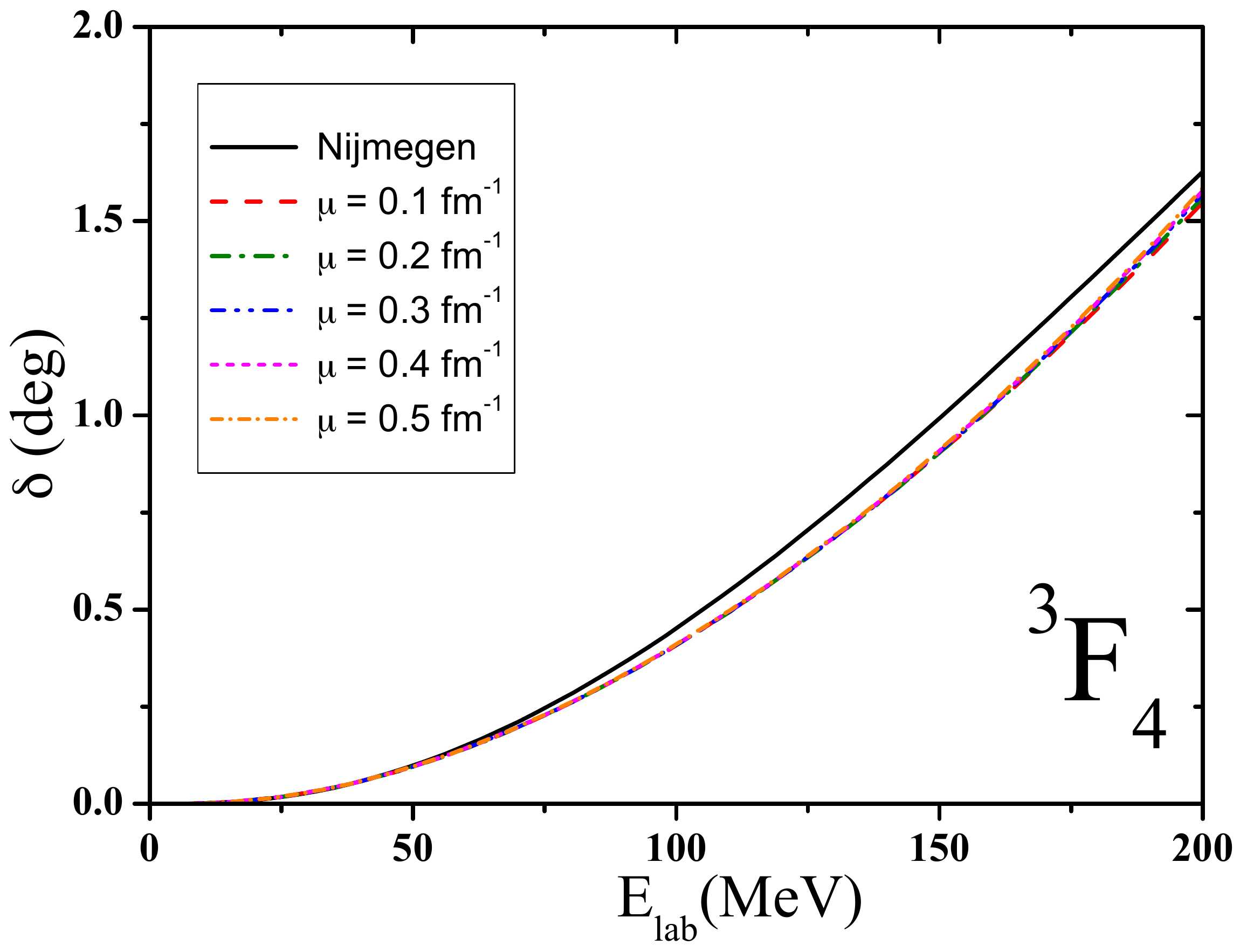}\includegraphics[scale=0.17]{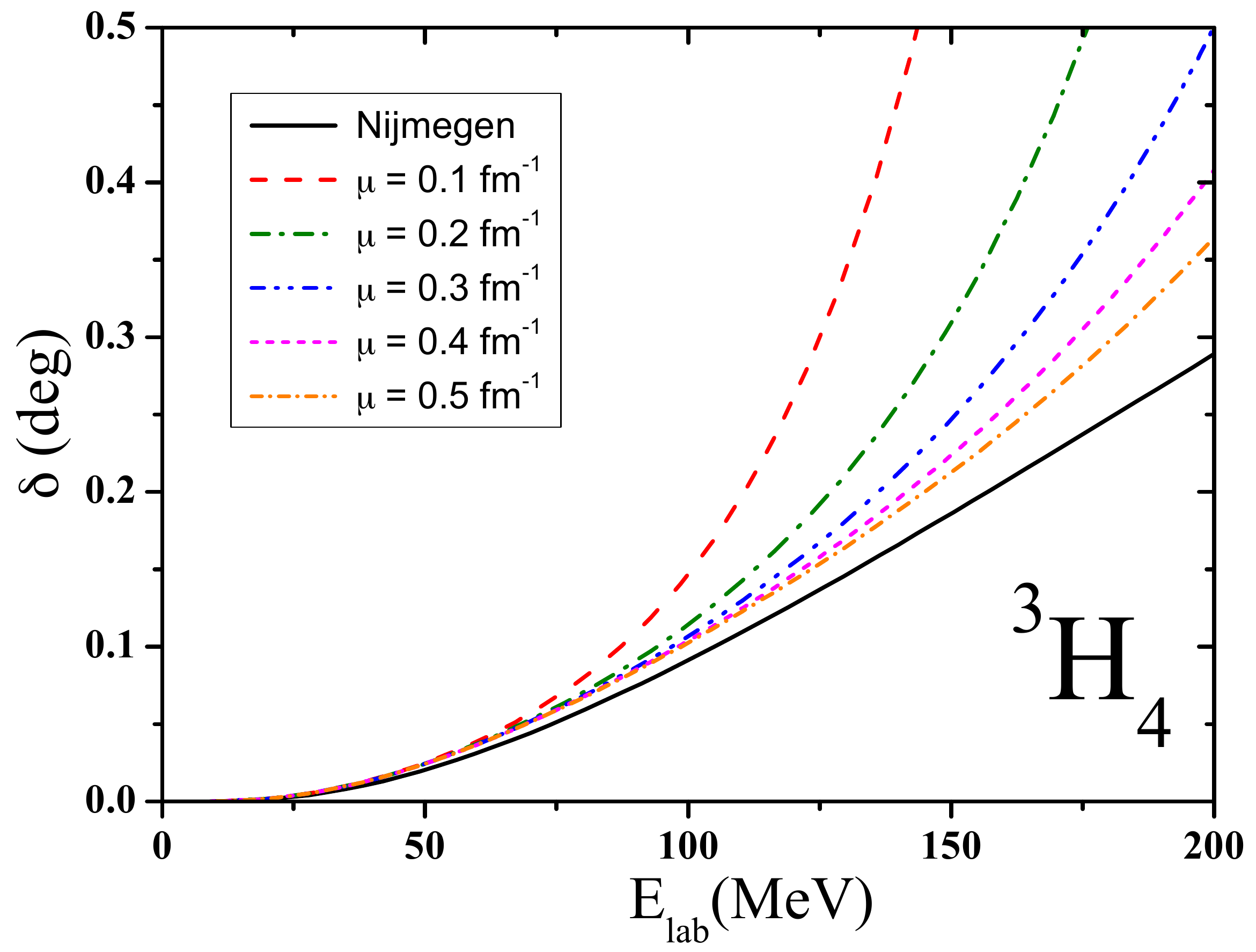}\includegraphics[scale=0.17]{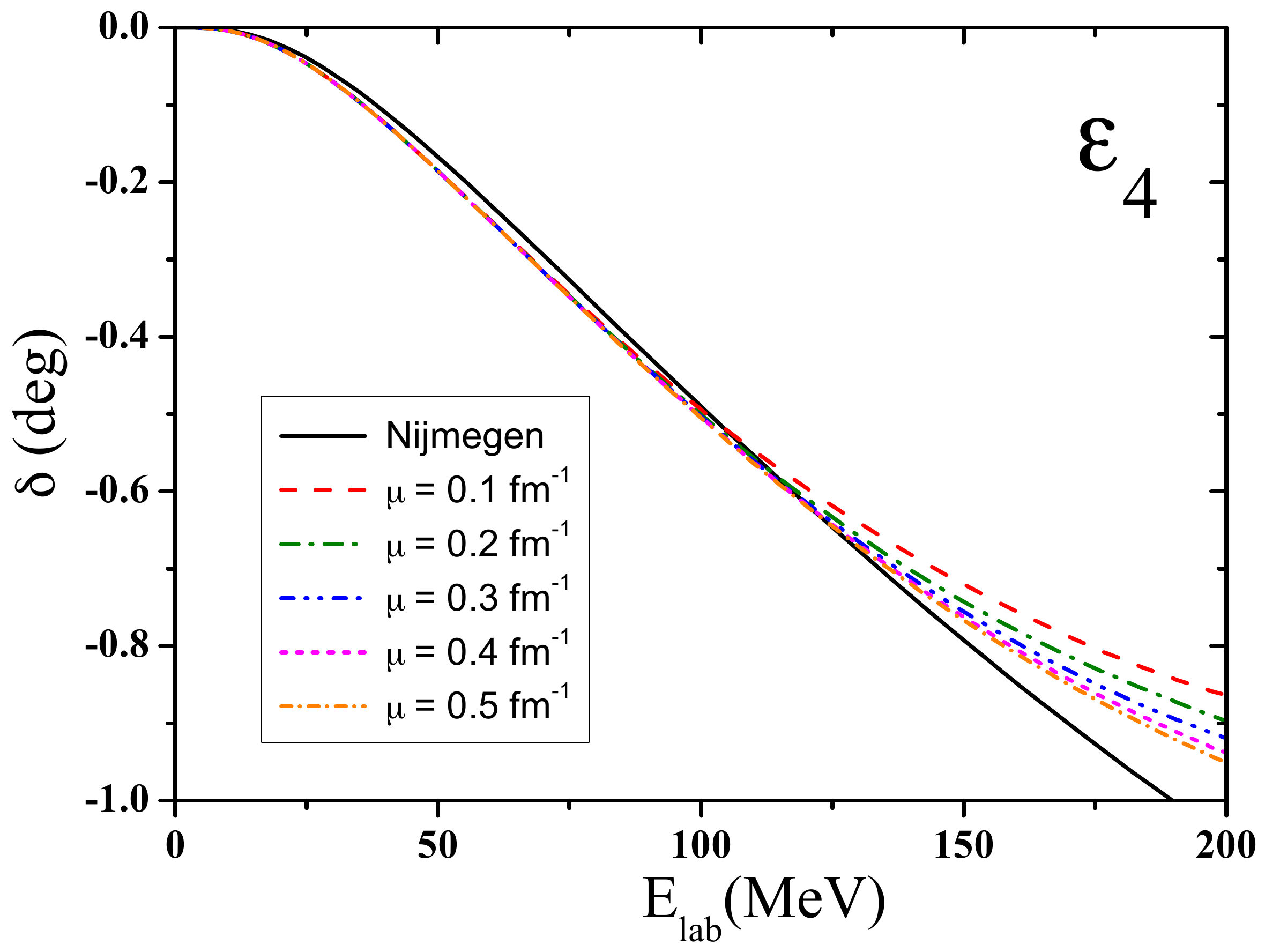} \\ \vspace*{1cm}
\includegraphics[scale=0.17]{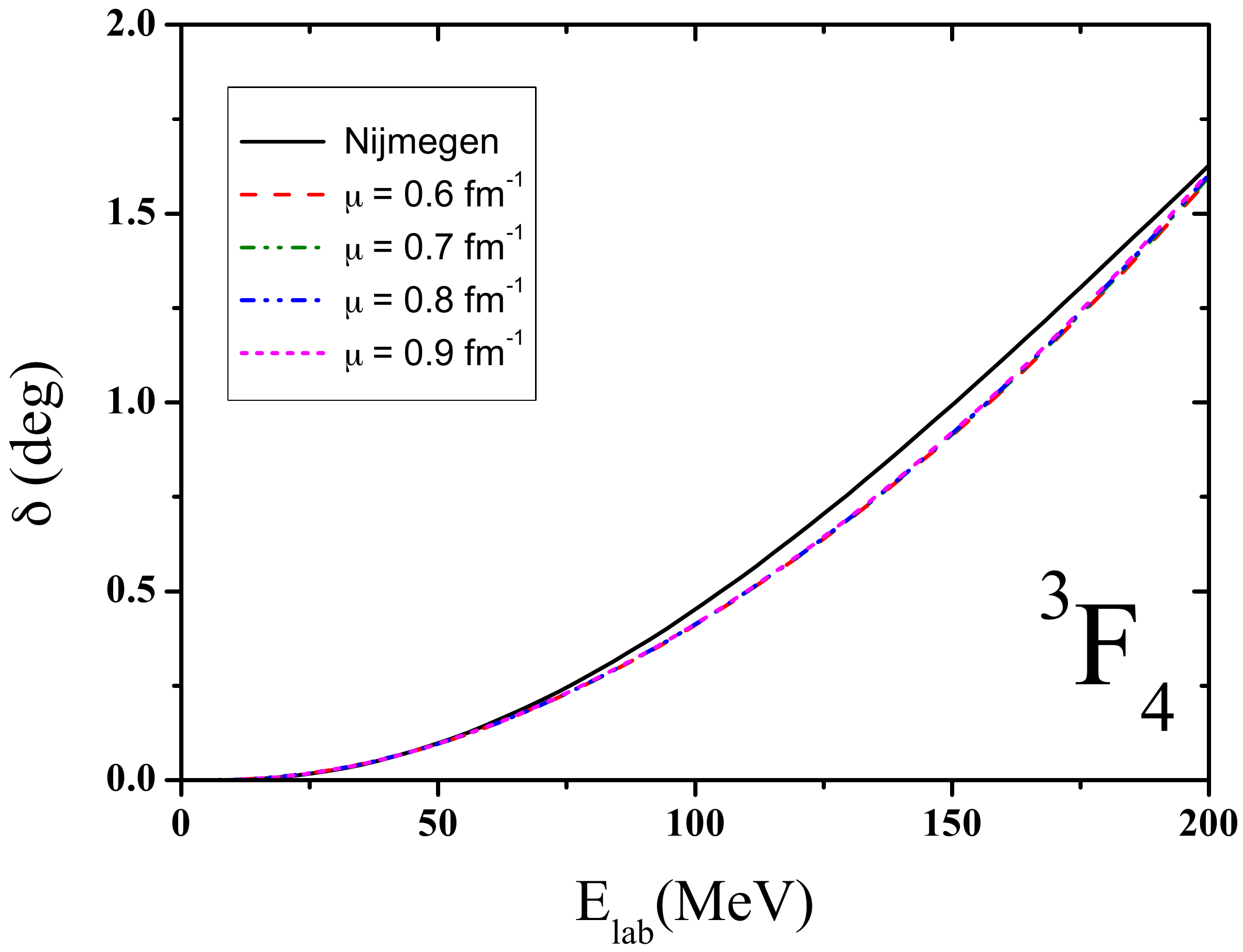}\includegraphics[scale=0.17]{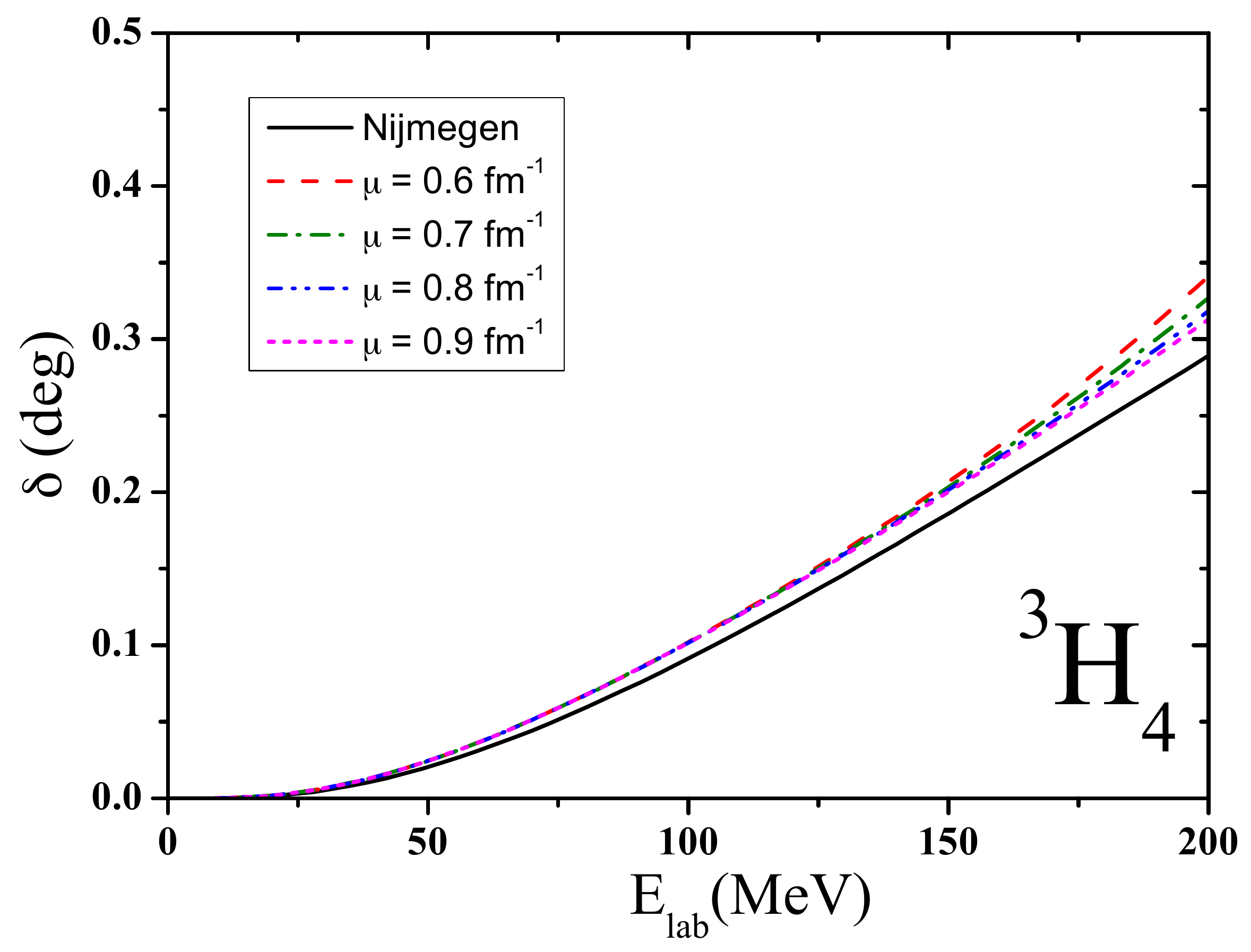}\includegraphics[scale=0.17]{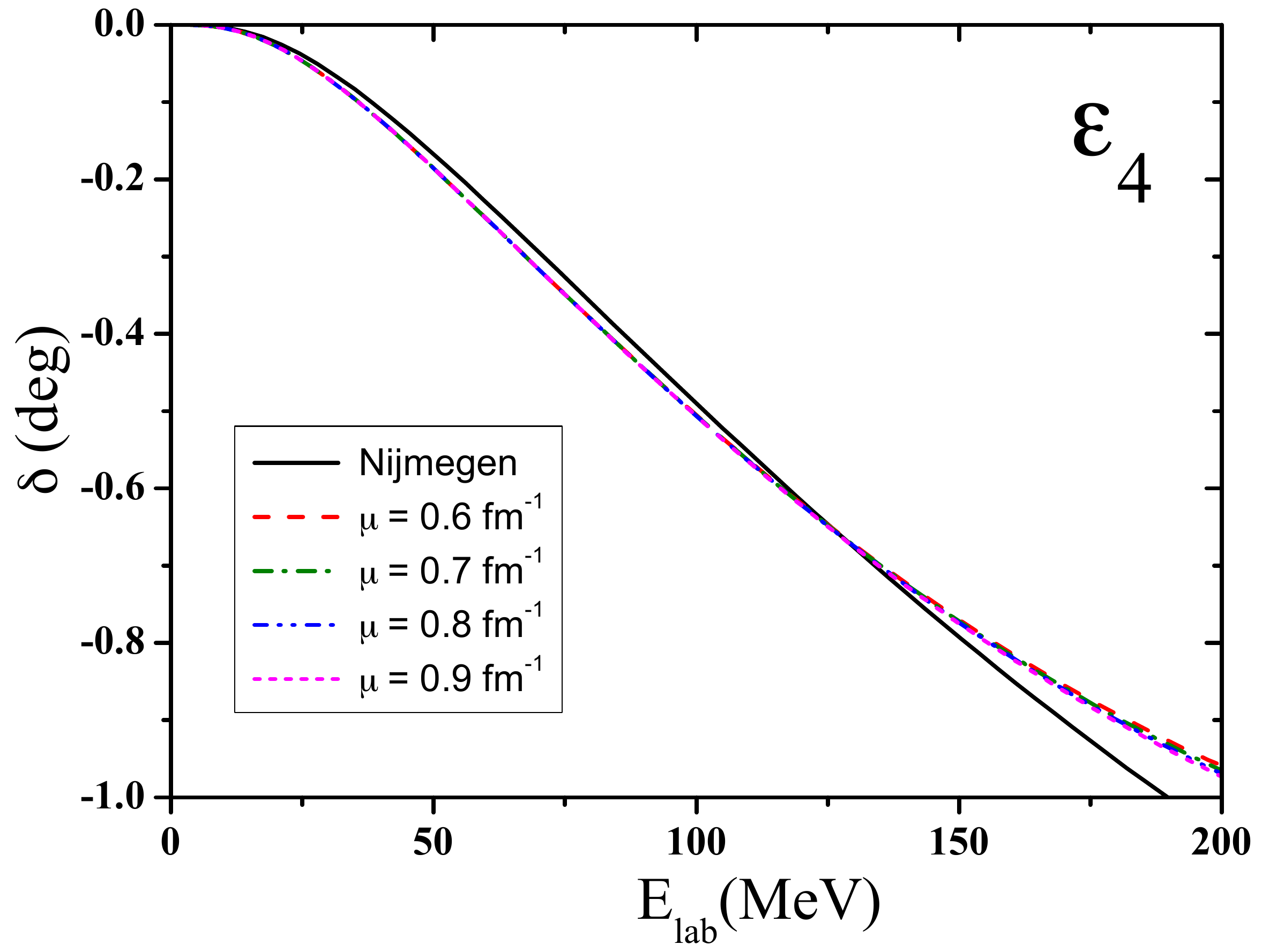}\\ \vspace*{1cm}
\includegraphics[scale=0.17]{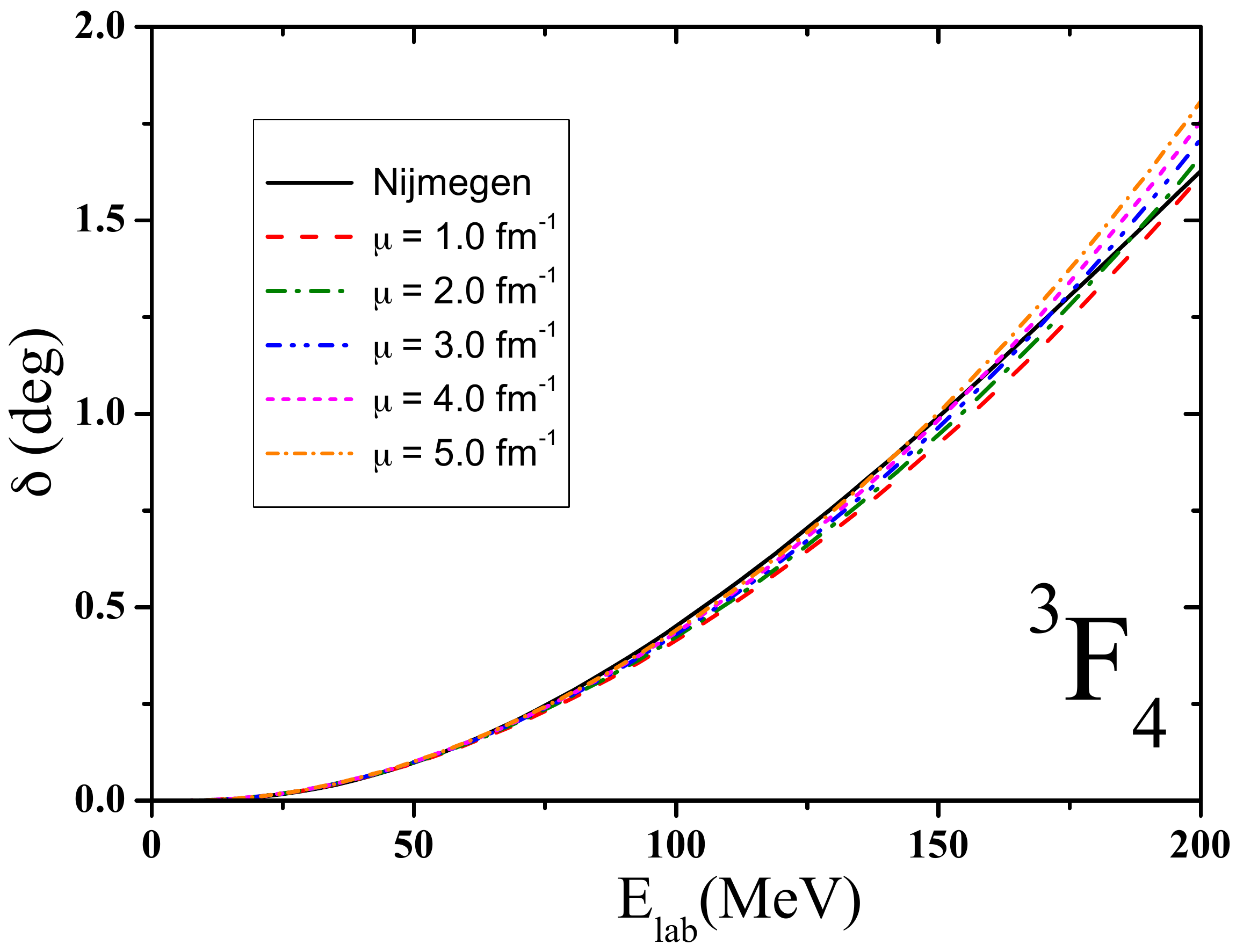}\includegraphics[scale=0.17]{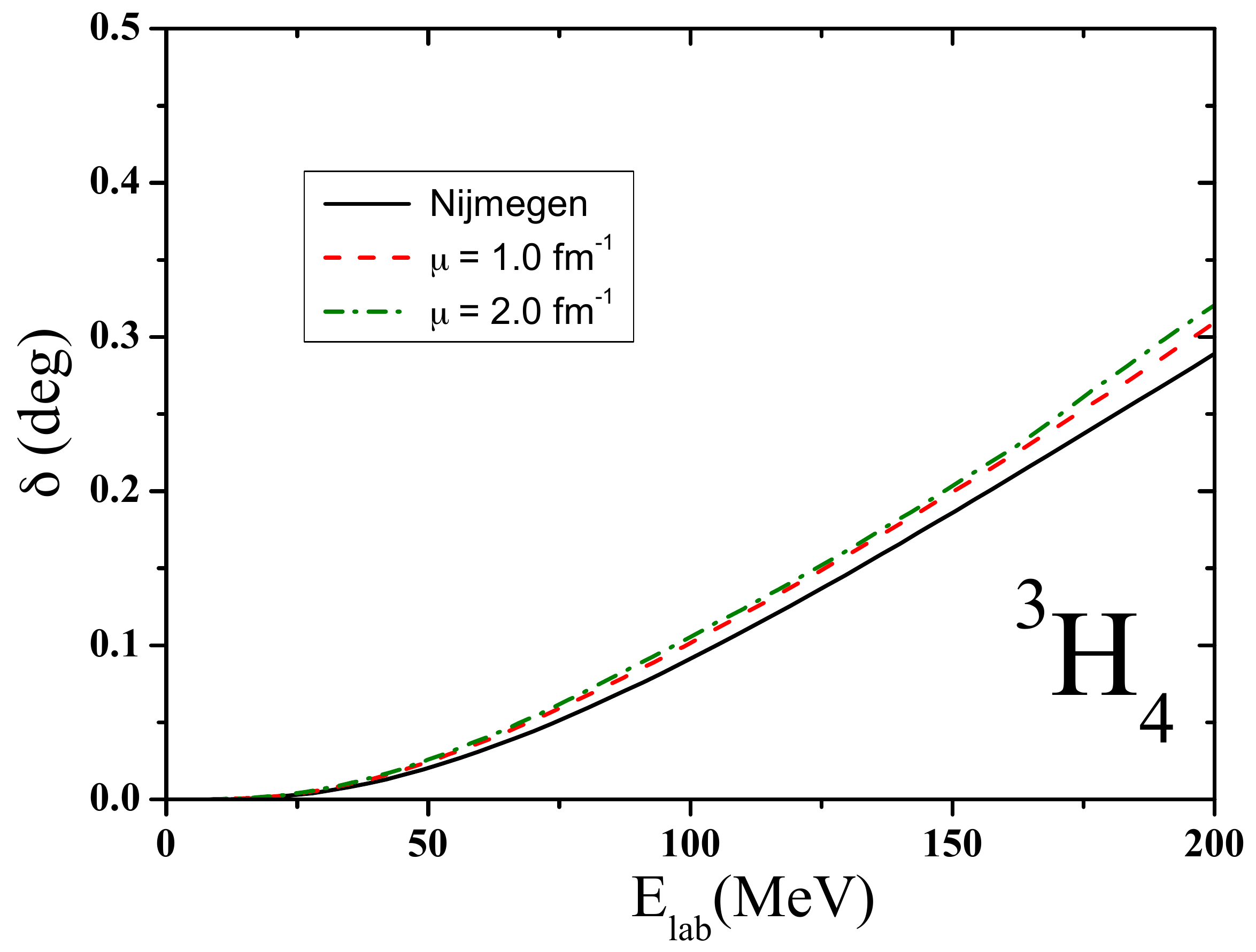}\includegraphics[scale=0.17]{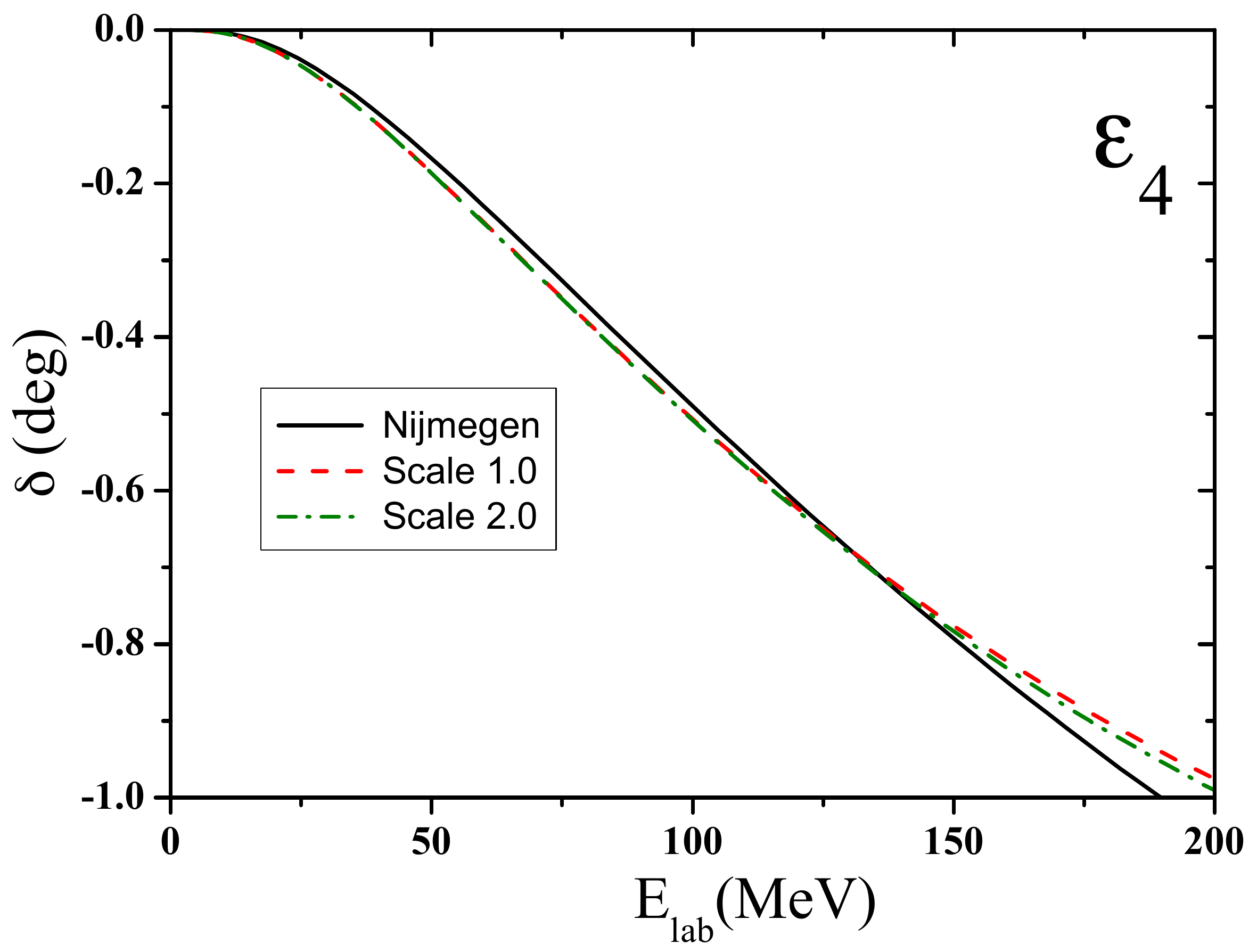} 
\end{center}
\caption{(Color on-line) Phase-shifts in the $^3F_4 - ^3H_4$ coupled channels calculated from the solution of the subtracted LS equation for the $K$-matrix with five subtractions for the N3LO-EGM potential for several values of the renormalization scale compared to the Nijmegen partial wave analysis.}
\label{fig11}
\end{figure}

\begin{figure}[t]
\begin{center}
\includegraphics[scale=0.17]{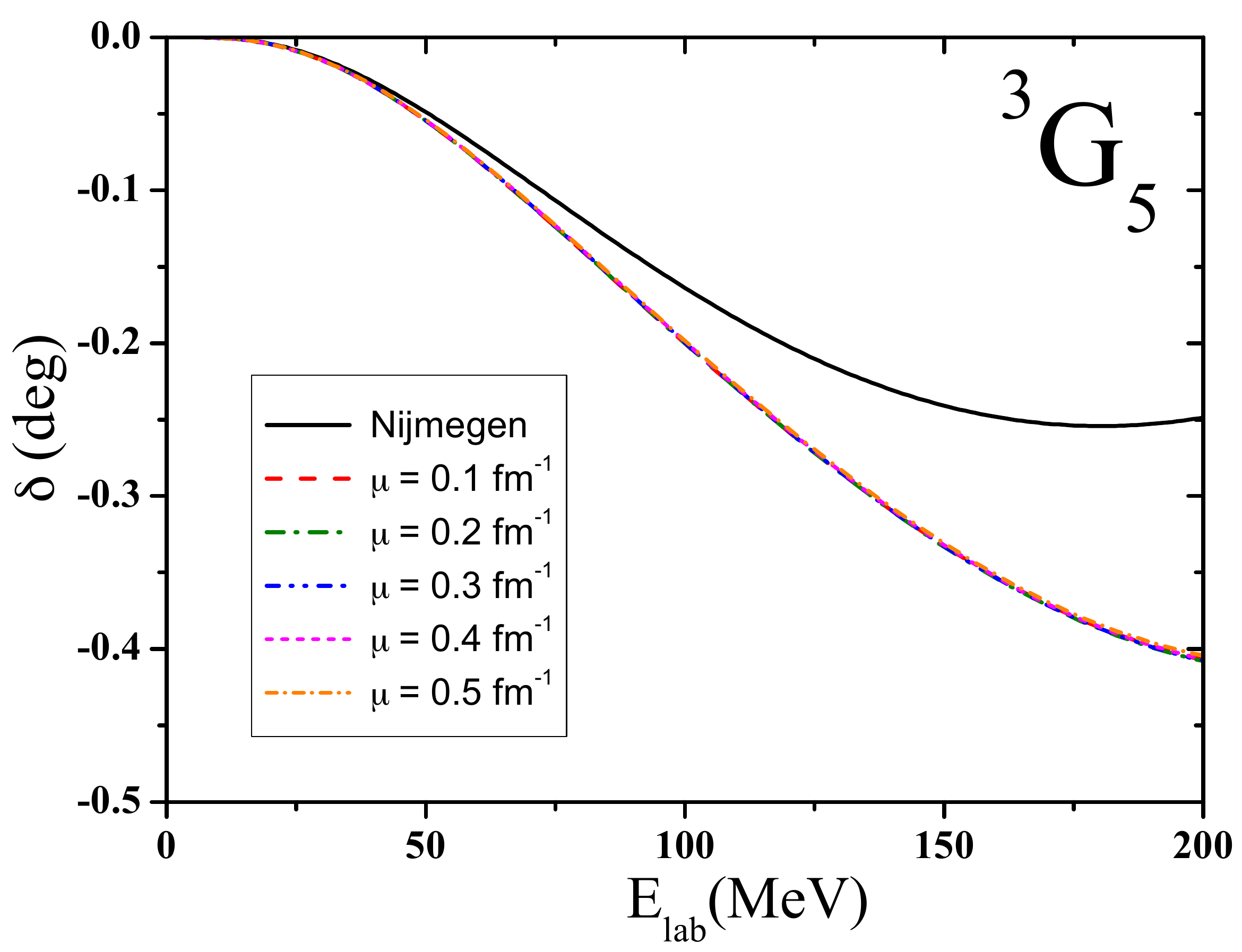}\includegraphics[scale=0.17]{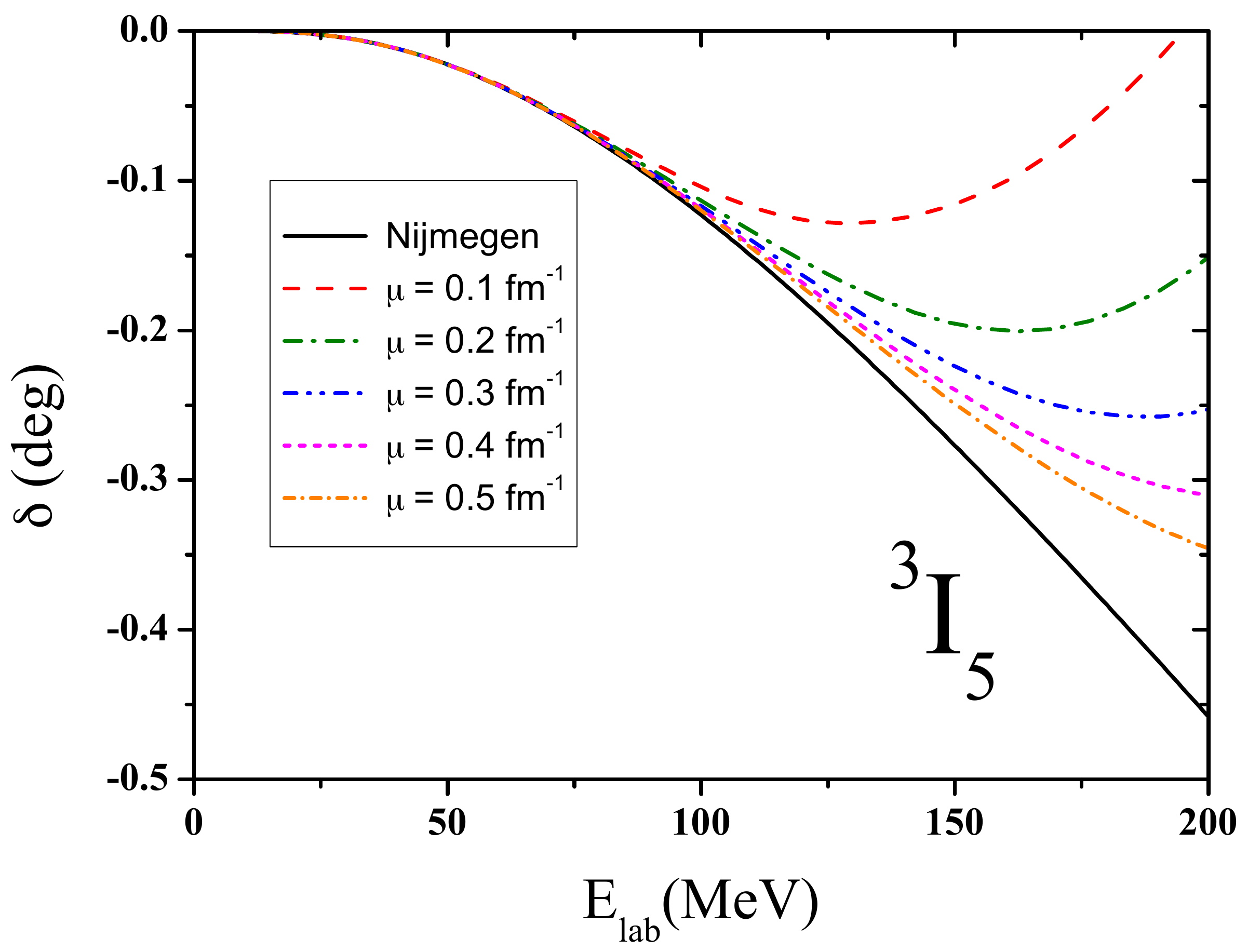}\includegraphics[scale=0.17]{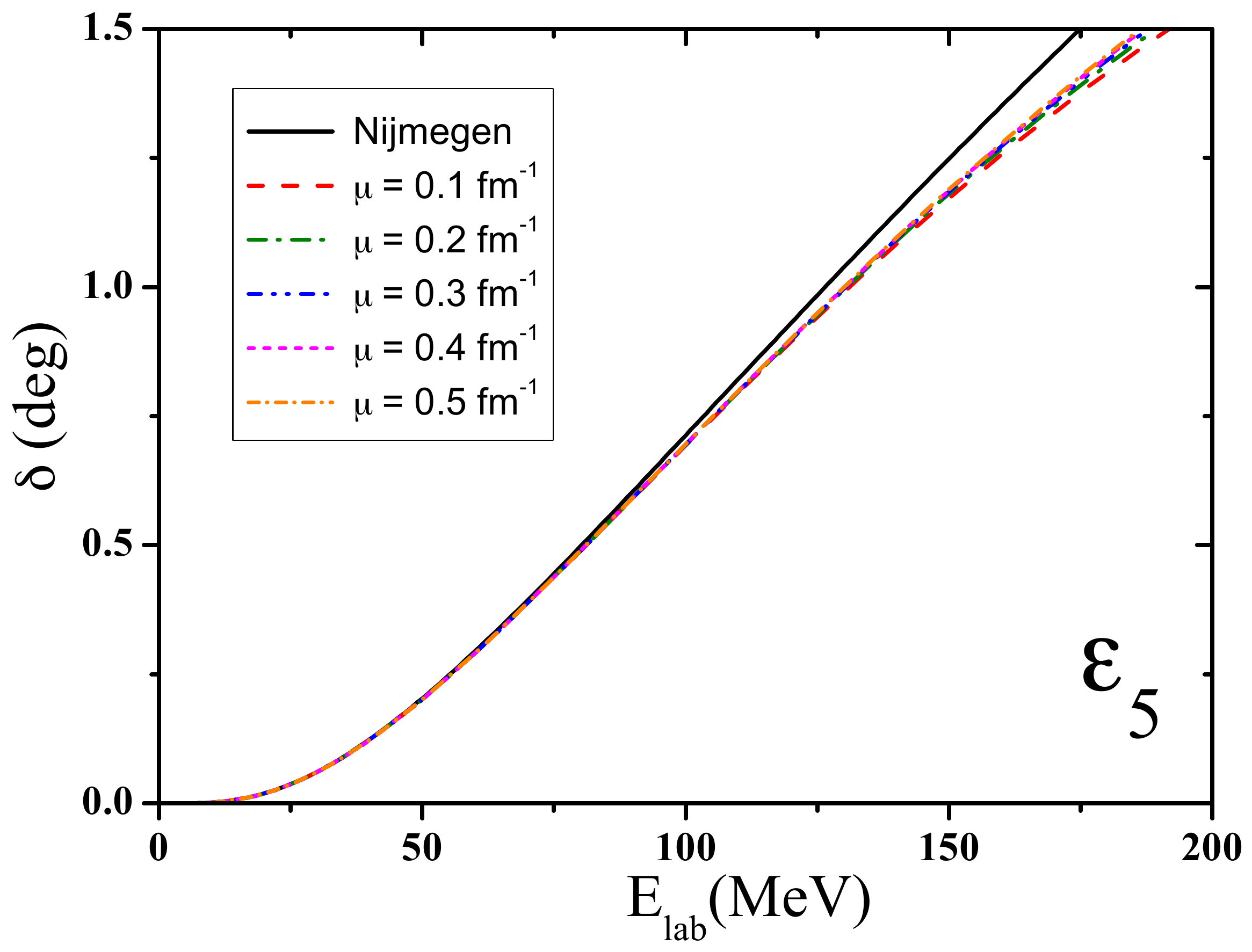} \\ \vspace*{1cm}
\includegraphics[scale=0.17]{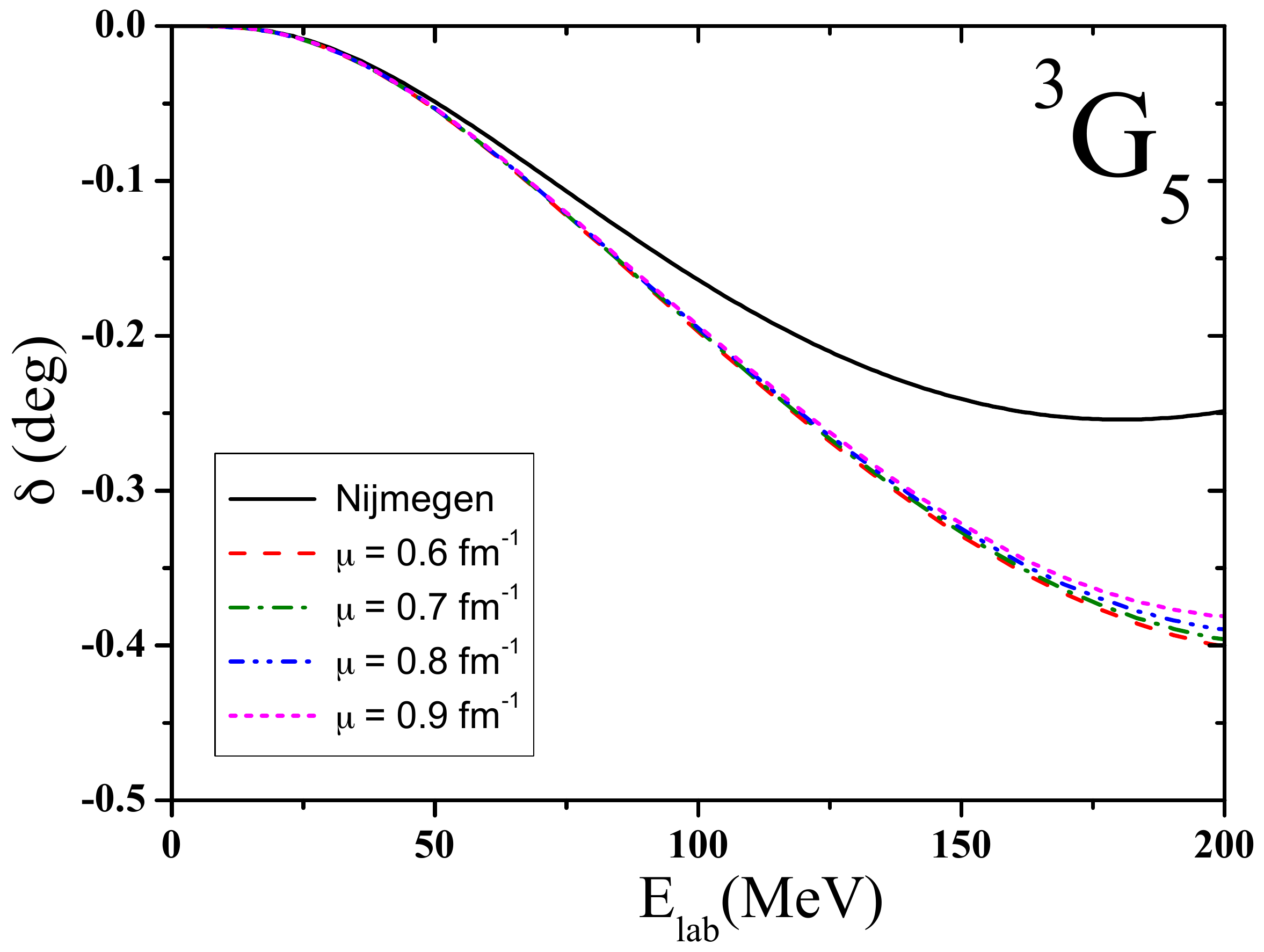}\includegraphics[scale=0.17]{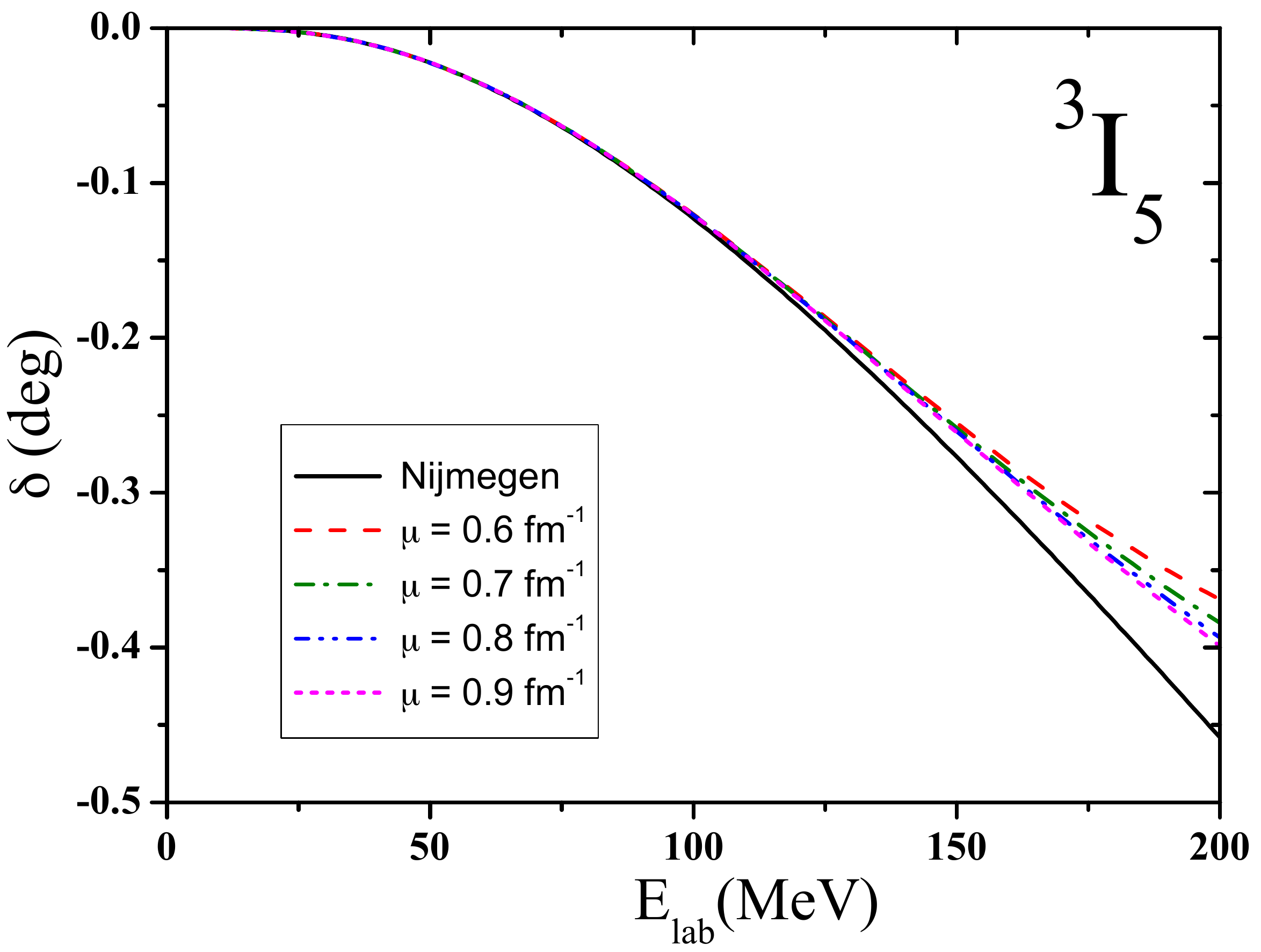}\includegraphics[scale=0.17]{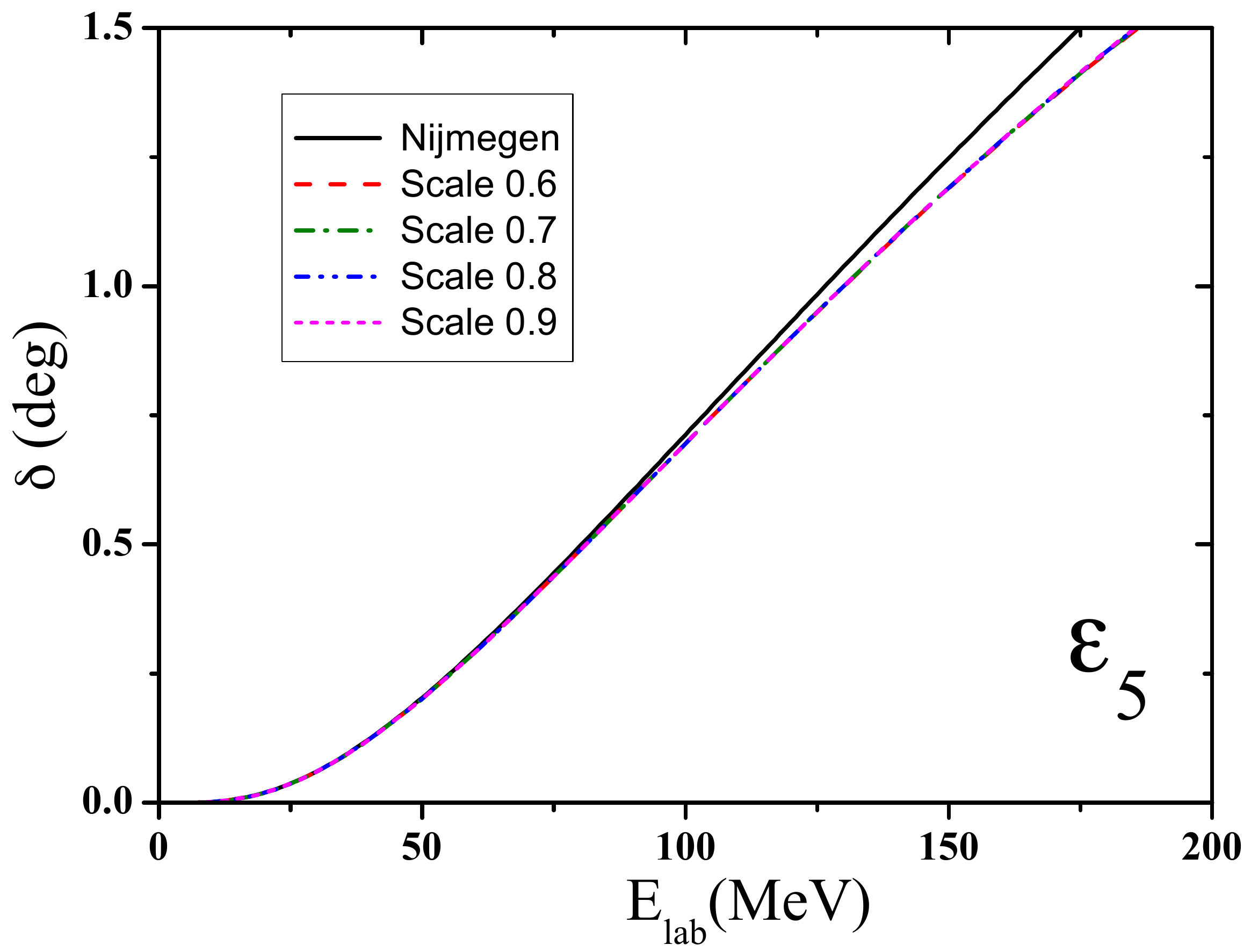}\\ \vspace*{1cm}
\includegraphics[scale=0.17]{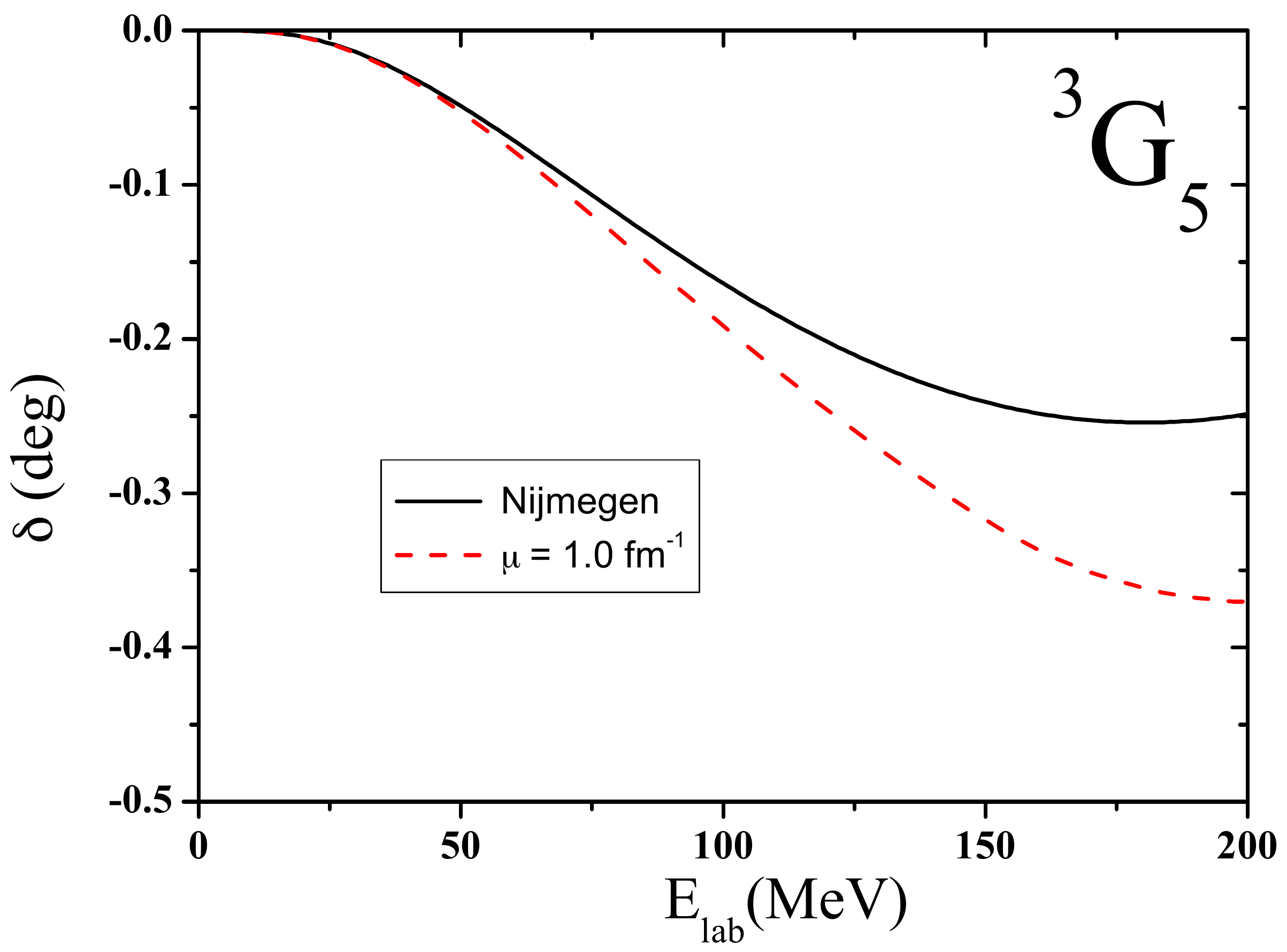}\includegraphics[scale=0.17]{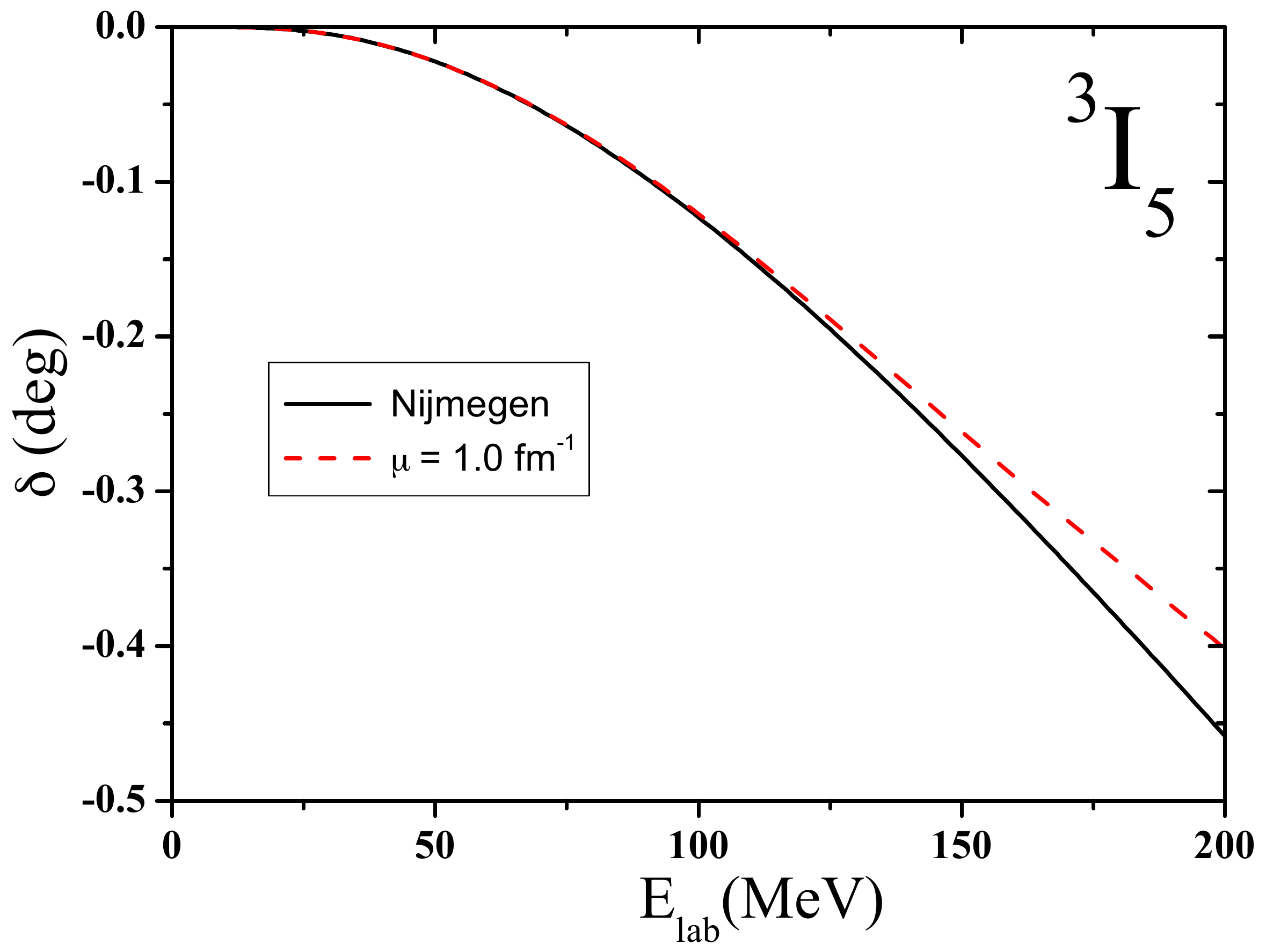}\includegraphics[scale=0.17]{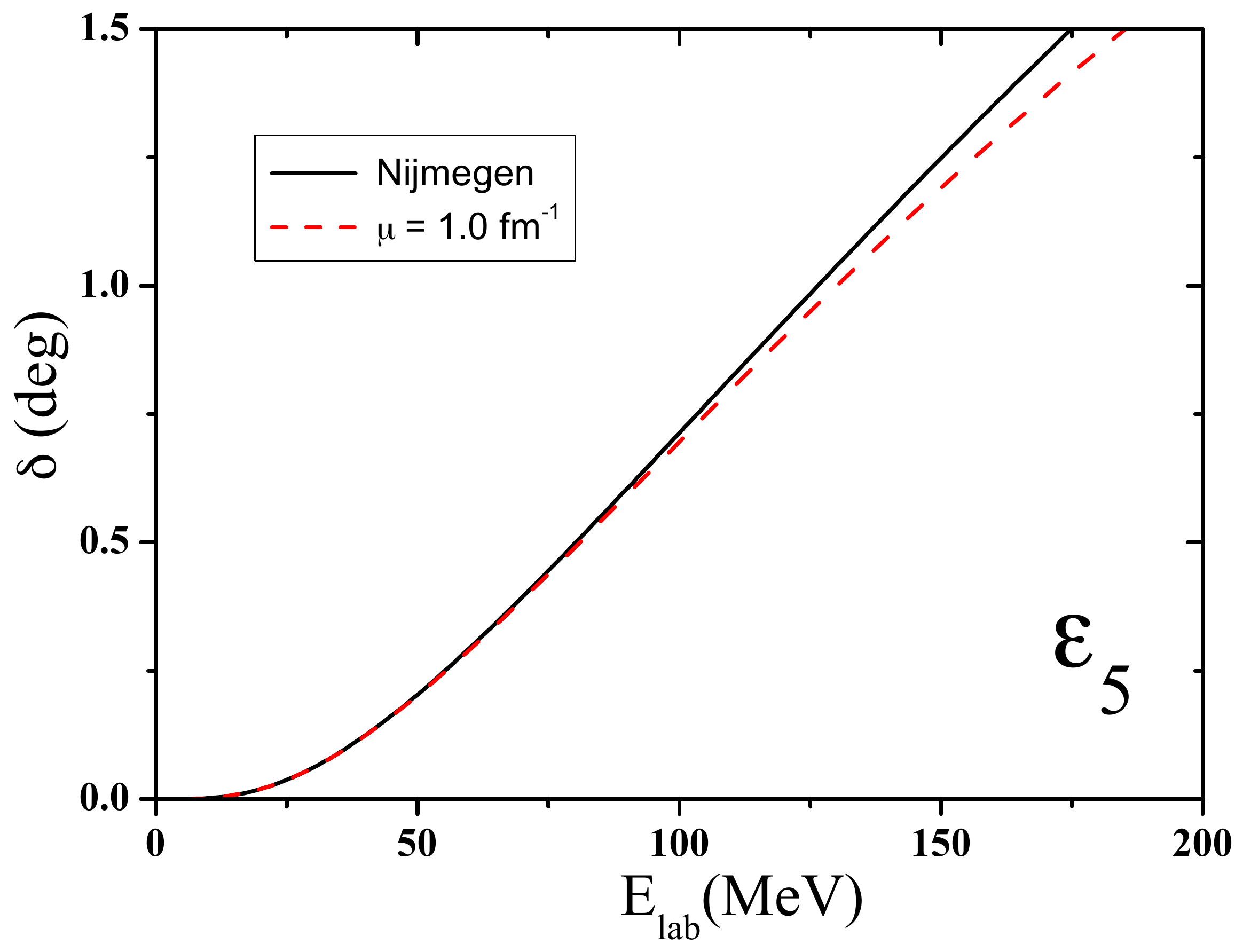}
\end{center}
\caption{(Color on-line) Phase-shifts in the $^3G_5 - ^3I_5$ coupled channels calculated from the solution of the subtracted LS equation for the $K$-matrix with five subtractions for the N3LO-EGM potential for several values of the renormalization scale compared to the Nijmegen partial wave analysis.}
\label{fig12}
\end{figure}

The results for the N3LO-EM potential are displayed in Figures \ref{fig1} to \ref{fig4} 
for the uncoupled channels and in Figures \ref{fig5} and \ref{fig6} for the coupled channels.
For the N3LO-EGM potential, the results are displayed in Figures \ref{fig7} to \ref{fig10} 
for the uncoupled channels and in Figures \ref{fig11} and \ref{fig12} for the coupled channels.
The phase-shifts are compared to the Nijmegen partial wave analysis \cite{Nij}. For un updated
high quality partial wave and error analysis, see the works from the Granada group. \cite{GRpwa}. 

Note that in the end we used different ranges of renormalization scales for EM and EGM and 
the reason is that the results for EGM converges faster as far as $\mu$ is concerned and there 
is no need to go above $2~{\rm fm}^{-1}$. In the case of EM, since there are still some variations 
in few waves for $\mu > 2~{\rm fm}^{-1}$ we extended the range of the subtraction point to 
$10~{\rm fm}^{-1}$. We believe this is related to the differences between EM and EGM that 
makes EGM softer than EM.

With few exceptions, in most of the channels we observe very small variation of the phases 
as we change the subtraction point, indicating that peripheral waves are nearly renormalization
group invariant or fixed-points of the subtractive renormalization group. The exceptions are
the $^3H_4$ and $^3I_5$ waves, which show slower convergence and the $^3G_5$ channel 
where the agreement with the partial wave analysis is only up to $\sim 50~{\rm MeV}$. Also, 
in the case of the EM potential, the phases for the $^3F_3$ wave present some oscillations 
when the renormalization scale approaches $\mu \sim 10 ~ {\rm fm}^{-1}$.

In this case of the coupled channels, the renormalization scheme is not failing but just requiring 
a slightly larger $\mu$ in the coupled channels, where we have the very singular tensor force. 
And this is what is different in the these channels compared to other F waves. The case of the 
uncoupled triplet 3F3 is somewhat different: the oscillation observed when going from 
$\mu = 6~ {\rm to}~ 9~{\rm fm}^{-1}$ is related to the TPE without SFR of the EM potential. 
This oscillation is not seen in the 3F3 wave when the EGM potential is used. An additional 
subtraction doesn’t modify the results and the  $\mu$ dependence in the these channels, 
so only the minimum number of subtractions need to be performed (five at N3LO). 

It is important to mention that the $\mu$-dependence of the phase-shifts is encoded in the
recursive driving terms $V^{(n)}_\mu$ since they all depend on the renormalization scale. 
In the case of cutoff regularization in configuration space \cite{mnl06}, the cutoff radius $R_S$
dependence has been traced to the most singular part of the interaction which can be attraction 
or repulsion, depending on the channel. Here we observe more dependence on the coupled
channels due to the very singular tensor force and the $F$-wave issue is due to the differences
between the TPE components of the EM and EGM potentials.  

\section{Conclusions and Outlook}

So far we have renormalized the N3LO interactions with multiple subtractions 
in peripheral channels considering an infinite cutoff $\Lambda = 30 ~ {\rm fm}{-1}$. 
Only pions contribute to the nuclear force in these waves and the results are parameter
free. 

The five subtractions performed in the kernel of the LS equation provide finite 
$K$-matrix in peripheral waves and the resulting phase shifts are rather independent 
of the subtraction point with the exception of the $^3H_4$ and $^3I_5$ waves where 
the fixed point is reached at approximately $\mu \sim 1 ~ {\rm fm}^{-1}$ after the slow 
convergence shown in Figs. \ref{fig5}, \ref{fig6}, \ref{fig11} and \ref{fig12}.

The oscillation in the $^3F_3$ channel suggests a closer look at this wave with 
the renormalization group flow equation that governs the driving terms (interactions) 
as the subtraction point is changed with the constraint of an invariant scattering amplitude. 

The advances given by our approach when compared to cutoff regularization are: the NN force 
doesn't have to be modified prior its insertion in the scattering equation; only the scattering equation 
is modified; the method is renormalization group invariant by construction and provides a non-relativistic 
flow equation for the driving terms that tell us how they change when the renormalization scale is modified 
in order to keep the amplitude invariant. Finally, we would like to point out that the cutoff looses any physical 
significance since our results were obtained with an extremely large value $\Lambda = 30 ~ {\rm fm}^{-1}$.

\section*{Acknowledgments}

We would like to thank financial support from FAPESP (grant 2016/07061-3), 
CNPQ (grant 306195/2015-1) and FAEPEX (grant 3284/16).



\end{document}